\documentclass[longauth]{aa}
\usepackage{graphicx}
\usepackage{txfonts}
\usepackage{gensymb}
\usepackage{threeparttable}
\usepackage{natbib,twoopt}
\usepackage{booktabs}
\usepackage{multirow}
\usepackage{upgreek}
\usepackage{placeins}
\usepackage{xcolor}
\usepackage[breaklinks=true]{hyperref}
\bibpunct{(}{)}{;}{a}{}{,}  
\makeatletter
  \newcommandtwoopt{\citeads}[3][][]{\href{http://adsabs.harvard.edu/abs/#3}
    {\def\hyper@linkstart##1##2{}
     \let\hyper@linkend\@empty\citealp[#1][#2]{#3}}}
  \newcommandtwoopt{\citepads}[3][][]{\href{http://adsabs.harvard.edu/abs/#3}
    {\def\hyper@linkstart##1##2{}
     \let\hyper@linkend\@empty\citep[#1][#2]{#3}}}
  \newcommandtwoopt{\citetads}[3][][]{\href{http://adsabs.harvard.edu/abs/#3}
    {\def\hyper@linkstart##1##2{}
     \let\hyper@linkend\@empty\citet[#1][#2]{#3}}}
  \newcommandtwoopt{\citeyearads}[3][][]
    {\href{http://adsabs.harvard.edu/abs/#3}
    {\def\hyper@linkstart##1##2{}
     \let\hyper@linkend\@empty\citeyear[#1][#2]{#3}}}
\makeatother

\begin{document}

    \title{Hot Rocks Survey III: A deep eclipse for LHS 1140c and a new Gaussian process method to account for correlated noise in individual pixels}
   
   \author{Mark Fortune \inst{1}\fnmsep\thanks{Corresponding author: fortunma@tcd.ie} \orcid{0000-0002-8938-9715},
           Neale P. Gibson \inst{1} \orcid{0000-0002-9308-2353},
           Hannah Diamond-Lowe \inst{2, 3} \orcid{0000-0001-8274-6639},
            João M. Mendonça \inst{3, 4, 5} \orcid{0000-0002-6907-4476},
            Amélie Gressier \inst{2} \orcid{0000-0003-0854-3002},
            Daniel Kitzmann \inst{6} \orcid{0000-0003-4269-3311},
            Natalie H. Allen \inst{7} \orcid{0000-0002-0832-710X},
            Prune C. August \inst{3} \orcid{0000-0003-3829-8554},
            Jegug Ih \inst{2} \orcid{0000-0003-2775-653X},
            Erik Meier Valdés \inst{8} \orcid{0000-0002-2160-8782},
            Merlin Zgraggen \inst{9} \orcid{0009-0001-6868-6171},
            Lars A. Buchhave \inst{3} \orcid{0000-0003-1605-5666},
            Brice-Olivier Demory \inst{6, 9, 10} \orcid{0000-0002-9355-5165},
            Néstor Espinoza \inst{2, 7} \orcid{0000-0001-9513-1449},
            Kevin Heng \inst{11, 10, 12, 13} \orcid{0000-0003-1907-5910},
            Kathryn Jones \inst{9} \orcid{0000-0002-2316-6850}, and
            Alexander D. Rathcke \inst{3} \orcid{0000-0002-4227-4953}
            }

   \institute{School of Physics, Trinity College Dublin, University of Dublin, Dublin 2, Ireland
        \and
        Space Telescope Science Institute, 3700 San Martin Drive, Baltimore, 21218, MD, USA
        \and
        Department of Space Research and Space Technology, Technical University of Denmark, Elektrovej 328, 2800 Kgs.\,Lyngby, DK
        \and
        Department of Physics and Astronomy, University of Southampton, Highfield, Southampton SO17 1BJ, UK
        \and
        School of Ocean and Earth Science, University of Southampton, Southampton, SO14 3ZH, UK
        \and
        Space Research and Planetary Sciences, Physics Institute, University of Bern, Gesellschaftsstrasse 6, 3012 Bern, Switzerland
        \and
        Department of Physics and Astronomy, Johns Hopkins University, 3400 N. Charles Street, Baltimore, MD 21218, USA
        \and
        Department of Physics, University of Oxford, Keble Road, Oxford, OX1 3RH, UK
        \and
        Centre for Space and Habitability, University of Bern, Gesellschaftsstrasse 6, 3012 Bern, Switzerland
        \and
        ARTORG Center for Biomedical Engineering Research, University of Bern, Murtenstrasse 50, CH-3008, Bern, Switzerland
        \and
        Ludwig Maximilian University, Faculty of Physics, Scheinerstr. 1, Munich D-81679, Germany
        \and
        University College London, Department of Physics \& Astronomy, Gower St, London, WC1E 6BT, United Kingdom
        \and
        University of Warwick, Department of Physics, Astronomy \& Astrophysics Group, Coventry CV4 7AL, United Kingdom
             }

    \titlerunning{Searching for an atmosphere on LHS 1140c}
    \authorrunning{M. Fortune et al.}
   \date{Received 20 February 2025 / Accepted 27 May 2025}

  \abstract
   {Time-series photometry at mid-infrared wavelengths is becoming a common technique to search for atmospheres around rocky exoplanets. This method constrains the brightness temperature of the planet to determine whether heat redistribution is taking place, which would be indicative of the presence of an atmosphere, or whether the heat is reradiated from a low-albedo bare rock. By observing at 15$\upmu$m, we are also highly sensitive to CO$_2$ absorption. We observed three eclipses of the rocky super-Earth LHS 1140c, using MIRI/Imaging with the F1500W filter. We found a significant variation in the initial settling ramp for these observations and identified a potential trend between the detector settling and the previous filter used by MIRI. We analysed our data using aperture photometry, however, we also developed a novel approach, which performs a joint fit of the pixel light curves using a shared eclipse model and a flexible multi-dimensional Gaussian process which can model changes in the PSF over time. Using simulated data, we demonstrate that our method has the ability to weight away from particular pixels that exhibit increased systematics, allowing for the recovery of eclipse depths in a more robust and precise way. Both methods, as well as an independent analysis, have detected the eclipse at $>\!5\sigma$, while recovering an eclipse depth consistent with a low-albedo bare rock. We measured a dayside brightness temperature of $T_\mathrm{day} = 561\pm44$ K, close to the theoretical maximum of $T_\text{day; max} = 537\pm9$ K. We rule out a wide range of atmospheric forward models to $>\!3\sigma$, including pure CO$_2$ atmospheres with surface pressure $\ge\!10$ mbar and pure H$_2$O atmospheres with surface pressure $\ge\!1$ bar. Our strict constraints on potential atmospheric composition, in combination with future observations of the exciting outer planet LHS 1140b, could provide a powerful benchmark for understanding atmospheric escape around M dwarfs.
   }


   \maketitle

\section{Introduction}
\label{sec:1}

    The search for atmospheres on rocky exoplanets is one of the major science goals defining the early JWST era. However, we still lack any definitive evidence of an atmosphere around any rocky planet, with only a few hints of signals reported to date (e.g. \citealt{Prune2024, 55cancrie, 55cancrie2, sulphur, Banerjee_2024}). At present, M-dwarf systems offer the best hope of characterising rocky exoplanet atmospheres using current facilities and they have therefore become the targets of numerous JWST programs (e.g. \citealt{2023Natur.618...39G, 2023Natur.620..746Z, Mansfield2024, Alam2025}). It remains an open question as to how many M-dwarf targets can retain atmospheres due to the significant flaring and intense EUV emission these stars are believed to exhibit in their first $\sim\!1$ Gyr \citep{2016PhR...663....1S}. However, there is a large degree of uncertainty around the extent of this activity, additionally, a planet might retain a secondary atmosphere after this active phase finishes \citep{Totton2024} or could even form a new secondary atmosphere from volcanic outgassing \citep{volcano_degas, Tian2024}. Ultimately, the only way we can answer this question is by observing a sample of M-dwarf targets and measuring whether or not they host atmospheres.

    There are multiple approaches to inferring the presence of an atmosphere, with different methods varying in observational cost. We used eclipse photometry over the 15$\upmu$m wavelength band offered by the MIRI F1500W filter. This approach assumes that the planet being studied is tidally locked to its host star (which is expected for our target from dynamical modelling of the LHS 1140 system; \citealt{modeleccentricity}) and asserts that the surfaces most likely to exist on rocky worlds within a given temperature range (300 K < $T_\mathrm{eq}$ < 880 K) would possess low surface albedos of $A < 0.2$ \citep{Koll2019, Mansfield2019}. As there should be insignificant heat redistribution on a planet without a substantial atmosphere \citep{Joshi1997, Selsis2011, Kollf}, we can expect this to result in a hot dayside, which would then produce strong thermal emission at mid-infrared (MIR) wavelengths. Alternatively, a planet with a substantial atmosphere should display more efficient heat redistribution to its nightside, cooling the dayside and reducing its thermal emission. In addition, carbon dioxide (CO$_2$) hosts a strong absorption band at 15$\upmu$m, which can further reduce this eclipse depth if present.
    
    The advantages of infrared eclipse photometry are that it can require fewer observing hours to identify an atmosphere compared to observing full phase curves \citep{Koll2019}. Unlike transmission spectroscopy, our observations are not affected by the transit light source effect (TLS; \citealt{2018ApJ...853..122R}) as we observe the planet as it passes behind its star, avoiding degeneracies in distinguishing a planetary atmosphere from stellar surface inhomogeneities. Compared with eclipse spectroscopy, photometry offers a higher signal-to-noise ratio (S/N), which helps reduce the observational cost -- albeit at the expense of increased degeneracy in our interpretations \citep{Hammond2024}.
    
    These benefits are the motivation behind the Hot Rocks Survey (GO 3730 PI: Diamond-Lowe, Co-PI: Mendonça), a large JWST Cycle 2 programme aimed at observing nine rocky planets with equilibrium temperatures ranging from $T_\mathrm{eq} \sim 420$ K - 910 K, using 15$\upmu$m eclipse photometry. Of these targets, LHS 1140c has the lowest equilibrium temperature (T = $422\pm7$ K; \citealt{2024ApJ...960L...3C}) and lowest instellation ($S \approx 5$ $S_{\mathrm{\oplus}}$; similar to Mercury at aphelion) of the sample. With a mass of $M = 1.91\pm0.06$ $M_{\mathrm{\oplus}}$ and radius of $1.272\pm0.026$ $R_{\mathrm{\oplus}}$ it can be considered a super-Earth. This gives it a reasonably high escape velocity of $v = 13.7\pm0.3$ km/s (using constraints from \citealt{2024ApJ...960L...3C}), which is larger than any Solar System rocky body. Combining these properties together makes it a high priority target as the cosmic shoreline model suggests atmospheres are more likely to be retained by planets with low instellation (in particular low integrated EUV flux) and high escape velocities \citep{2017ApJ...843..122Z}. Results from models of atmospheric escape suggests LHS 1140c might well have retained a secondary atmosphere \citep{volcano_degas, novelphysics} and observations from Swift \citep{2004ApJ...611.1005G} have shown the LHS 1140 system shows relatively low levels of present day NUV flux and estimated FUV flux for a star of this type \citep{NUV_obs}. However, we note recent work suggests mid-type M dwarfs such as LHS 1140 may have had longer early active periods than previous models predicted, making atmospheric retention more challenging \citep{Pass2025}.
    
    LHS 1140c was originally discovered by \citet{2019AJ....157...32M} and is the innermost known planet of the system. The only other confirmed planet in the system, LHS 1140b, lies within its star's habitable zone and has a low-density with a mass of $M = 5.60\pm0.19$ $M_{\mathrm{\oplus}}$ and radius of $1.730\pm0.025R_{\mathrm{\oplus}}$, potentially consistent with a water world or mini-Neptune \citep{2024ApJ...960L...3C}. Transit observations using NIRISS/SOSS \citep{2024ApJ...970L...2C} and NIRSpec \citep{Damiano2024} suggest that LHS 1140b is inconsistent with the mini-Neptune case due to the flat transmission spectrum recovered. This implies that it could be a water world with a high mean molecular weight atmosphere. The NIRISS/SOSS observations happened to capture a transit of our target LHS 1140c entirely within a transit of LHS 1140b, although there was a low S/N coming from just a single transit and the transmission spectrum was consistent with a flat line with no molecular features or scattering-slope from hazes detected. Additionally, it has previously been noted that there is a $4\sigma$ discrepancy between the transit depths of LHS 1140c from TESS observations and a single Spitzer observation at 4.5$\upmu$m. This result suggests possible absorption from CO$_2$ \citep{2024ApJ...960L...3C}. However, these recent NIRISS/SOSS observations appear to show that the TESS observations underestimated the transit depth, with the new observations lying between the TESS and Spitzer transit depths. Eclipse observations using the F1500W filter can further resolve this discrepancy as our observations are highly sensitive to the presence of CO$_2$.

    One of the lessons we have learned is that systematics affect at least some of the observations in our survey (e.g. \citealt{Prune2024,Meier2025}). In addition, there can be significant variation in the strength and sign of the initial settling ramp which covers at least the first $\sim 30 - 60$ minutes of our observations. We present a possible explanation for these variations related to the previous filter used by MIRI, as we note it remains in place throughout target acquisition. In addition, we present a new Gaussian process (GP) based approach which fits the eclipse depth at the level of individual pixels without performing an aperture extraction. We also retrieve information about how the point spread function (PSF) changes in time and characterise the strength of interpixel capacitance. Our method has the benefit of being able to identify and weight away from systematics found in individual pixels, which we demonstrate using simulated data. In one of our observations, this method identifies a possible persistence effect caused by a cosmic ray.
    
   Our work is laid out as follows: first we introduce our new pixel-based approach to account for systematics in Section~\ref{sec:2} and we demonstrate the advantages of this method using simulated data in Section~\ref{sec:3}. Those primarily interested in the observations may move straight to Section~\ref{sec:4} which presents the analysis of our observations using the standard approach of aperture photometry as well as using our pixel-based approach. It also includes our findings on MIRI detector settling. Section~\ref{sec:5}  describes our modelling of LHS 1140c and the interpretation of our recovered eclipse depth. Section~\ref{sec:6}  presents a discussion of the implications of our findings and elaborate on the strengths and weaknesses of our new method and where it may be extended. Our conclusions are outlined in Section~\ref{sec:7}.

    \section{Methods of fitting time-series photometry}
    \label{sec:2}

    \subsection{Motivation for a new pixel-fitting method}
    \label{sec:pixel_motivation}

    There have been a number of methods developed to analyse time-series photometric observations. After the data cleaning, a standard approach is to define an aperture region where flux from the target star is greatest and add up the flux within this region with a uniform weighting on each pixel. This classic approach is known as aperture extraction and can perform well for bright targets where any remaining background noise from imperfections in the background subtraction is minimal. It has a couple of limitations, the choice of the aperture region and shape is quite arbitrary, but choosing too wide an aperture reduces the S/N due to increased background noise while too small an aperture reduces S/N from missed flux. Typically an intermediate aperture radius is chosen which maximises S/N. However, applying a uniform weighting for each pixel is generally not optimal as brighter pixels will be less impacted by background noise than outer pixels and might ideally be more strongly weighted.

    Optimal extraction attempts to remedy these issues by fitting for the PSF shape and estimating the Poisson noise in each pixel \citep{optimal_extraction}. This makes it possible to weight each pixel in a way which maximises S/N. For observations which contain a lot of background noise, optimal extraction may result in a significant improvement in the S/N. This is because optimal extraction weights each pixel based on it's particular S/N value and can be mathematically shown to maximise the S/N in the presence of uncorrelated noise.

    However, the situation is more complicated in the presence of time-correlated noise. The Spitzer Space Telescope was known to suffer from systematics due to `intrapixel sensitivity', where the efficiency of each pixel varies across the surface of the pixel. In combination with instability in the telescope pointing and undersampling of the PSF, this resulted in systematics that could be different in each pixel \citep{Ingalls2012}. The first work to measure an exoplanet eclipse depth used detrending of the PSF centroid position to correct for this effect \citep{Charbonneau2005}. The technique of BiLinearly-Interpolated Subpixel Sensitivity (BLISS) mapping was later developed to account for this, directly mapping the sensitivity across each pixel's surface \citep{BLISS_mapping}. In addition, pixel-level decorrelation (PLD) also corrected for this issue by detrending the systematics in the extracted light curve against normalised pixel light curves \citep{2015ApJ...805..132D}. These techniques were used in early work characterising rocky exoplanet heat redistribution using phase curves from Spitzer \citep{Demory2016, 2019Natur.573...87K}. Other relevant techniques applied to Spitzer photometric time series include GPs which use centroid positions as input variables \citep{Evans2015}. Independent component analysis (ICA) -- introduced to exoplanet time series in \citet{Waldmann1, Waldmann2} -- has been applied to disentangle astrophysical and systematic effects and works by calculating statistically independent trends in the pixel time series \citep{Morello2014, Morello2015, Morello2016}. A separate approach used to account for systematics in Kepler time series is the `causal pixel model', which can distinguish systematics in each pixel by comparing them to multiple other stars on the detector \citep{2016PASP..128i4503W}. However, this approach is not applicable to our observations of a single star.

    Time-series photometry with the MIRI/Imaging mode is expected to suffer significantly less from intrapixel sensitivity variations compared to Spitzer due to the PSF being spread out over more pixels and the telescope pointing being more stable. There is also a difference in detector technology, the Si:As detectors on Spitzer were not significantly affected by intrapixel sensitivity while the InSb detectors were \citep{spitzer_intrapixelsensitivity}, MIRI uses Si:As detectors. We performed some simple tests simulating a PSF with a similar width to F1500W observations and modeling the pointing stability of JWST, which confirmed that effects such as intrapixel sensitivity and errors in flat fielding should be completely negligible compared to the photometric precision of these observations -- even when significant intrapixel or interpixel sensitivity variations were modelled (see Appendix~\ref{app:intrapixel_testing} for more details).
    
    However, at our early stage in understanding systematics from JWST instruments, we should be careful in assuming our observations are completely free from systematics. In addition, the more approaches we can develop to analyse these data, the less sensitive we are to the limitations of any individual method. With that in mind, we introduce our new pixel-fitting approach to analyse these data, which is radically different from existing approaches such as PLD or BLISS mapping. This method combines the benefits of optimal extraction by fitting for the shape of the PSF and accounting for the amplitude of Poisson noise in each pixel. However, it also fits for the amplitude of correlated noise in each pixel. This allows it to weight away from pixels that suffer from increased systematics, in addition to weighting based on reduced Poisson noise. This method is the first we are aware of which applies GPs to joint-fit pixel time series in astronomical data. Our approach can be more precise and robust when systematics are present in single pixels and can be used as an alternative to aperture photometry or optimal extraction for eclipse depth measurements. It also quantifies multiple detector effects in the process, such as constraining the amplitude of background noise and the amplitude of correlated Poisson noise resulting from interpixel capacitance.

    \subsection{Introduction to our new pixel-fitting method}
    \label{sec:pixel_intro}
    
    Gaussian processes are a method often used to fit correlated noise (also known as systematics) in exoplanet light curves (e.g. \citealt{2012MNRAS.419.2683G, Gibson2013, Evans2015, Mackey2017, 2023ARA&A..61..329A}). To apply them to an eclipse light curve, we must first model the eclipse signal using a deterministic transit model (e.g. using equations from \citealt{mandelagol}). This deterministic model is referred to as the `mean function', $\vec{\mu}$, and the underlying assumption of the method is that we can model the noise around this signal as a random draw from a covariance matrix built using a kernel function. That is to say, we model the noise as having a given amplitude or height scale, $h$, and the covariance in the noise between any two points is described by a kernel function which typically decreases as a function of the separation of the two points using a specific metric, such as time. For example, the exponential kernel combined with white noise is a common choice for a kernel function:
    \begin{equation}
    k(t, t') = h^2 \exp\left(-\frac{|t - t'|}{l}\right) + \sigma^2 \delta_{tt'},
    \label{eq:exp_kernel}
    \end{equation}
    where $t$, $t'$ could be the timestamps of two datapoints, $h$ is the height scale and $l$ the length scale of correlated noise, $\sigma$ is the amplitude of (uncorrelated) white noise, and $\delta_{tt'}$ is the Kronecker delta which equals unity when $t = t'$ and zero otherwise.

    If we want to fit multiple time series simultaneously (e.g. joint-fitting individual pixel light curves), then we can use a 2D GP model, where the dataset lies along a 2D grid of pixels-by-time. In this case, our kernel function can describe correlation as a function of separation in time and as a function of distance away on the detector. As the computational cost of GPs can scale as the cube of the total number of datapoints, namely, $\mathcal{O}(N_p^3 N_t^3)$ for $N_p$ pixels and $N_t$ time points, we need to exploit a particular GP optimisation to make this method computationally tractable for our datasets. We chose the optimisation originally identified in \citealt{Rakitsch2013}, and introduced to exoplanet time series by \citet{Fortune2024}, as it has sufficient flexibility in the choice of kernel functions we can use and it scales efficiently as $\mathcal{O}(2 N_p^3 + 2 N_t^3 + N_p N_t (N_p + N_t))$, making the analysis computationally feasible. The method calculates the exact log likelihood value without approximations and works by breaking up the expensive decomposition of the covariance matrix into separate covariance matrices for each dimension, see \citet{Fortune2024} for more details. It is implemented in an open-source Python package called \textsc{luas} which is available on GitHub\footnote{\href{https://github.com/markfortune/luas}{https://github.com/markfortune/luas}} along with documentation and tutorials \citep{Fortune2024}.

    \subsection{Pixel-fitting mean function}
    \label{sec:psf_fit}

    First, we need to model the PSF and account for how it will move in each integration due to telescope pointing instability. The PSF was not found to be well-characterised by simple parameterisations such as a 2D Gaussian profile or using the models generated by the package \textsc{webbpsf} \citep{webbpsf} and so instead we directly fit for the average flux in time for each pixel included in the fit. We also fit for a linear trend in flux for each pixel (with the slope denoted as $T_\mathrm{grad; (i, j)}$) as different pixels can show quite distinct slopes in time throughout the observations. To account for movements in telescope pointing, we fit for shifts in the PSF position along the $x$ and $y$ directions for each integration. Finally, the amplitude of the PSF as a function of time is fit with an eclipse model with a single shared set of eclipse parameters featuring the eclipse depth, $d = f_p/f_*$; period, $P$; system scale, $a/R_*$; radius ratio, $R_p/R_*$; impact parameter, $b$; central transit time, $T_{0}$; eccentricity, $e,$ and argument of periastron, $\omega$. We also included a shared exponential ramp model between all pixels with height scale, $h_\mathrm{ramp}$, and length scale, $l_\mathrm{ramp}$, to account for the sharp changes in flux typically seen in the first $\sim\!60$ minutes of the observations; however,  this ramp does appear quite different between pixels in the first $\sim\!30-45$ minutes of the observations (as  discussed in Section~\ref{sec:settling}). Thus, we avoided fitting the first 45 minutes of data with this method. Alternatively it may be possible to fit this with a separate ramp for each pixel but we did not explore this in this work.
    
    Combining the eclipse and settling effects, we model the light curve for each pixel $(i, j)$ using:
    \begin{align}
    f(i, j, t) &= f_\mathrm{eclipse}(T_0, P, a/R_*, R_p/R_*, b, d, \sqrt{e} \cos\omega, \sqrt{e} \sin\omega, t) \nonumber\\
    &*f_{(i, j)}(\vec{f}, \vec{\Delta x}(t), \vec{\Delta y}(t)) \nonumber\\
    &*\left[1 + T_\mathrm{grad; (i, j)}(t - t_\mathrm{mid}) + h_\mathrm{ramp} \exp\left(-\frac{t - t_\mathrm{init}}{l_\mathrm{ramp}}\right)\right],
    \label{eq:mean_function}
    \end{align}
    where $f_{(i, j)}(\vec{f}, \vec{\Delta x}(t), \vec{\Delta y}(t)$) is the flux of pixel $(i, j),$ as calculated by a 2D interpolating function which takes a grid of average flux values located at the central position of each pixel and linearly interpolates to a new grid offset by $\vec{\Delta x}(t), \vec{\Delta y}(t); $  namely, the $x$ and $y$ shift in PSF position for the integration at a time, $t$; $\vec{f}$ may be thought of as the baseline flux for each pixel light curve, while $\vec{\Delta x}(t)$ and $\vec{\Delta y}(t)$ fit for the time series of telescope pointing movements. Then, $\vec{f}$, $\vec{\Delta x}(t)$ and $\vec{\Delta y}(t)$ were all fit for as free parameters. We note we chose to fit for $\sqrt{e} \cos\omega$ and $\sqrt{e} \sin\omega$ instead of $e$ and $\omega$ as it puts a simple uniform prior on $e$ \citep{Eastman_2013}.
    
    \subsection{Pixel-fitting kernel function}
    \label{sec:kernel choice}

    There are multiple noise processes which are correlated between detector pixels that are not normally required to be accounted for when fitting an aperture extracted light curve. However, if we are trying to fit an eclipse depth directly using the pixel light curves, then we must account for the correlation in the noise between different pixels in order to avoid underestimating or overestimating our eclipse depth uncertainty. We present each noise process to be accounted for in this section.
    
We must account for how each noise process is correlated both between different pixels and also in time. To meet the restrictions of the 2D GP optimisation from \citet{Fortune2024}, we must assume that these correlations are separable and that the full kernel function of the dataset can be represented as a sum of two kernel functions separable as a product of pixel and time kernel functions:
\begin{equation}
    k(i, j, t, i', j', t') = k_{p}(i, j, i', j') k_{t}(t, t') + s_{p}(i, j, i', j') s_{t}(t, t').
    \label{eq:kernel_form}
\end{equation}
Here, we define two pixel kernel functions, $k_p$ and $s_p$, and two time kernel functions, $k_t$ and $s_t$. The terms in $k_p$ and $k_t$ describe the systematics which are correlated over time, while the terms in $s_p$ and $s_t$ describe white noise or correlations between pixels which are instantaneous in time. We introduce each noise process in turn over the rest of this section, but the overall form of our kernel function is
\begin{align}
    k(i, j, t, i', j', t') = &\left[ k_\mathrm{p; FCS} + k_\mathrm{p; IPS} + k_\mathrm{p; CS}\right] \exp\left(-\frac{|t - t'|}{l_{t}}\right) \nonumber\\
    +  &\left[s_\mathrm{p; wn+IPC}  + s_\mathrm{p; backgr}\right] \delta_{tt'},
    \label{eq:basic_kernel}
    \end{align}
where we have three time-correlated systematics processes. Due to the restrictions of this GP optimisation, they must all share the same time kernel function with a shared length scale (chosen to be the exponential kernel), but with different correlations as a function of separation of pixels on the detector. These correlations between pixels are denoted by the pixel kernel functions to be described in this section: $k_\mathrm{p; FCS}$, $k_\mathrm{p; IPS}$, and $k_\mathrm{p; CS}$. We also have multiple noise processes that we modeled as independent in time, including correlations associated with the background subtraction, as well as white noise which must be adjusted to account for interpixel capacitance (IPC). These are described by the pixel kernel functions $s_\mathrm{p; backgr}$ and $s_\mathrm{p; wn+IPC}$. Note that our kernel function is in the form of Equation~\ref{eq:kernel_form} and that none of the pixel kernel functions can have any dependence on time.

Depending on the noise process, it can either be simpler to describe the kernel function in terms of the covariance matrices that it builds or by writing out the kernel function explicitly; thus, we sometimes switch between these notations. Our kernel function restriction may equivalently be expressed with Equation~\ref{eq:kronsum}, where the full covariance matrix of the dataset must be expressible as the sum of two Kronecker products
\begin{equation}\label{eq:kronsum}
    \mathbf{K} = \mathbf{K}_{p} \otimes \mathbf{K}_{t} + \mathbf{\Sigma}_{p} \otimes \mathbf{\Sigma}_{t}.
    \end{equation}
Here, we have defined two pixel covariance matrices $\mathbf{K}_p$ and $\mathbf{\Sigma}_p$ (generated using the kernel functions $k_p$ and $s_p$) and two time covariance matrices $\mathbf{K}_t$ and $\mathbf{\Sigma}_t$ (generated using $k_t$ and $s_t$). The Kronecker product is an operation where each element in the matrix left of the $\otimes$ sign is multiplied by the full matrix on the right. For two matrices of shape $(N_p, N_p)$ and $(N_t, N_t),$ this will result in a matrix of shape $(N_p N_t, N_p N_t)$. Since our choice of GP optimisation only requires calculating the separate pixel and time covariance matrices, we avoid memory issues from building the full $(N_p N_t, N_p N_t)$ covariance matrix. In addition, instead of needing to decompose the full covariance matrix $\mathbf{K}$ (e.g. using Cholesky factorisation or eigendecomposition), we only need to separately decompose the pixel and time covariance matrices (specifically using eigendecomposition), significantly reducing the computational costs.

    \subsubsection{Flux-conserved systematics (FCS)}
    \label{sec:PSF_shape}

\begin{figure}
        \includegraphics[width=\columnwidth]{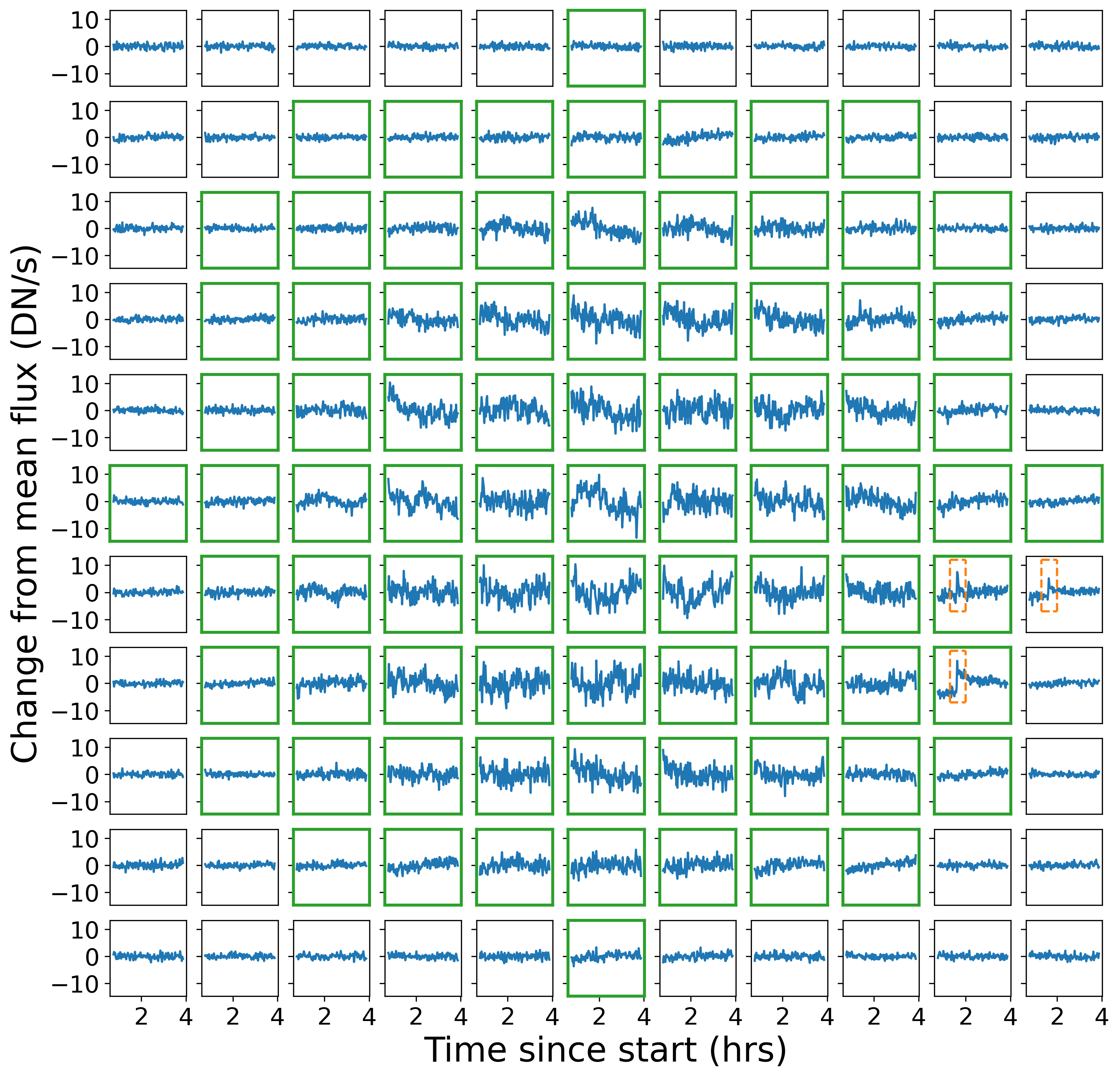}
        \caption{Pixel light curves centred on the PSF for the second eclipse, excluding the first 45 minutes dominated by settling and binned for clarity. A sharp persistence effect from a cosmic ray is highlighted with orange dashed boxes (see Appendix~\ref{app:cosmic_ray} for details). Pixels highlighted with green boundaries were fit using the pixel-fitting method except where specified we excluded the pixels containing the highlighted flux jump. Significant systematics are visible in the central light curves which may be flux-conserved between pixels or be independent for each pixel.}
        \label{fig:real_px_lightcurves}
    \end{figure}

    Examining the pixel light curves reveals significant systematics that do not appear obvious in the aperture extracted light curves (see Figure~\ref{fig:real_px_lightcurves}). They are particularly strong in the highest flux pixels at the centre of the PSF. This could be interpreted as the PSF continuously changing in time on timescales of $\sim\!30$ minutes, perhaps due to thermal distortion of mirrors on JWST \citep{JWST_mirrors} or alternatively may be due to charge migration between pixels. In principle, we could examine whether these systematics arise from a global process (e.g. from distortion of the mirrors) or from a local process (e.g. from charge migration) if we have another bright star on the detector to compare the pixel light curves to (similar to the `causal pixel model'; \citealt{2016PASP..128i4503W}). Unfortunately, we could not find an example of this in a MIRI/Imaging time-series dataset. The systematics not being apparent in the aperture extracted light curve suggests that they may be flux-conserving; namely, a bump up in one pixel may be paired with a bump down in a neighbouring pixel. The autocorrelation of neighbouring pixel light curves from our observations hints that this may be true (see Appendix~\ref{app:anticorr}). Regardless of the underlying cause, we modelled this using anti-correlated GPs between neighbouring pixels and we fit for the amplitude of this effect. This allows the model to fit these bumps in individual pixel light curves while keeping the overall aperture sum flux-conserved. We can then fit for additional non-flux-conserving systematics, which may arise from other causes that we might be able to weight away from.
    
    We used the exponential kernel to describe this correlation as a function of time and our pixel kernel function $k_\mathrm{p; FCS}$ to describe how these systematics are (anti-)correlated between pixels. To model the flux as conserved we assumed that the amount of flux `moving between' two neighbouring pixels, $i$ and $j$, is proportional to the geometric mean of the baseline flux falling on each pixel, $\sqrt{f_i f_j}$. We can then model the correlation in flux between both pixels with a covariance matrix as:
    \begin{equation}
   \mathbf{K}_p = 
    \begin{bmatrix}
    h_{FCS}^2 f_i f_j & -h_{FCS}^2 f_i f_j \\
    -h_{FCS}^2 f_i f_j & h_{FCS}^2 f_i f_j \\
    \end{bmatrix} = 
    h_{FCS}^2 f_i f_j\begin{bmatrix}
    +1 & -1 \\
    -1 & +1 \\
    \end{bmatrix}.
    \label{eq:2x2_FM}
    \end{equation}
This covariance matrix can describe correlated noise which is perfectly anti-correlated between two pixels, meaning that the sum of this correlated noise will be zero if the pixels are summed within an aperture extraction\footnote{Note: while this matrix is only positive semi-definite and not invertible, once white noise is included in the overall kernel then the full covariance matrix $\mathbf{K}$ will be invertible.}.

Each pixel actually ends up surrounded by eight neighbours, four pixels which share an edge and four pixels which share a corner. For simplicity, we might assume that flux movement occurs with equal amplitude in all directions including between adjacent and diagonal pixels. We can describe this by summing the covariance terms corresponding to flux movement between each neighbouring pair of pixels together. Given these assumptions, we can describe the pixel kernel function of flux-conserved systematics for all pixels using:
\begin{equation}
k_\mathrm{p; FCS}(i, j, i', j') = h_\mathrm{FCS}^2 f_{(i, j)} \left(f_\mathrm{neigh; (i, j)} \delta_{ii'} \delta_{jj'} - f_{(i', j')} n_{(i, j), (i', j')}\right),
\label{eq:general_FM}
\end{equation}
    where we define $h_\mathrm{FCS}$ as the amplitude of flux-conserving systematics, $f_{(i, j)}$ is the average flux on pixel $(i, j)$, $f_\mathrm{neigh; (i, j)}$ is defined as the sum of the fluxes of all neighbouring pixels to pixel $(i, j)$, and we define $n_{(i, j), (i', j')}$ analogously to the Kronecker delta but for neighbouring pixels:
    \begin{equation}
    n_{(i, j), (i', j')} =
    \begin{cases}
            1, &         \text{if } (i, j), (i', j') \text{ are neighbours},\\
            0, &         \text{otherwise.}
    \end{cases}
    \end{equation}
    We note that we chose not to fit all the pixels on the detector, which means we ended up fitting pixels that have neighbours not included in our fit. As the flux may still `move between' these pixels, we  still account for the additional variance experienced by these border pixels by including these pixel fluxes in the sum of neighbouring fluxes, $f_\mathrm{neigh; (i, j)}$. However, there are no corresponding covariance terms included in the covariance matrix. The consequence of this is that the flux moving between pixels being fit and pixels not being fit will not be conserved; although this only affects pixels at the edges of our boundary, which will typically experience low flux. This effect can be reduced by including more pixels in the fit, although this comes at an increased computational cost.

    \subsubsection{Independent pixel systematics}
    \label{sec:independent_pixel_systematics}

    Each pixel may also have systematics which are independent to neighbouring pixels, such as due to persistence from cosmic ray hits or unmasked bad pixels. For example, the pixel light curves from the second LHS 1140c eclipse appear to show a jump in flux in only a couple of pixels (as highlighted in Figure~\ref{fig:real_px_lightcurves}). For these systematics, we assume they can be modelled with the same kernel function in time $k_t$ as for flux-conserving systematics and have a pixel kernel function given by:
    \begin{equation}
    k_\mathrm{p; IPS}(i, j, i', j') = h_\mathrm{IPS; (i, j)}^2 f_{(i, j)}^2 \delta_{ii'} \delta_{jj'},
    \end{equation}
    where we refer to these as independent pixel systematics (IPS) with height scale, $h_\mathrm{IPS}$. Since the amplitude of systematics may vary significantly between pixels (e.g. due to persistence effects from a cosmic ray hitting specific pixels), we may choose to fit for the amplitude, $h_\mathrm{IPS; (i, j)}$, separately for each pixel. Checking which pixels recover large values of $h_\mathrm{IPS; (i, j)}$ could help identify pixels exhibiting increased systematics.
    
    \subsubsection{Common systematics between all pixels}
    \label{sec:common_systematics}

\begin{figure*}
        \includegraphics[width=\textwidth]{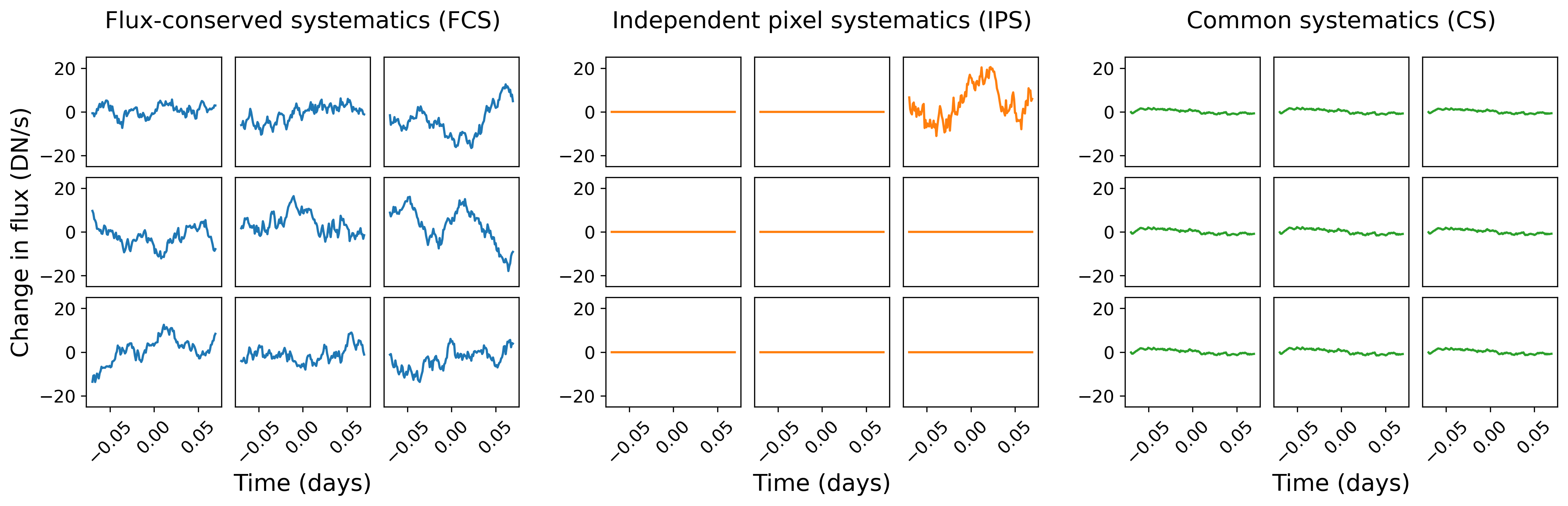}
        \caption{Three different random draws of time-correlated systematics for a grid of 3x3 pixels. Left grid shows flux-conserved systematics whose sum is zero after aperture extraction. The second grid shows systematics independent to a particular pixel. The third grid shows common systematics with the same shape and with an amplitude proportional to the flux on each pixel. While the amplitude of common systematics may appear small, their effect combines together over multiple pixels to reach a similar amplitude to the independent pixel systematics after aperture extraction.}
        \label{fig:systematic_visual}
    \end{figure*}

    We may also encounter systematics which have the same shape in all pixels and are proportional to the flux on each pixel. For example, if the target star flares, that could produce the same systematic shape in all pixels. These systematics may therefore represent stellar activity or alternatively could result from variations in the throughput of the detector which are correlated in time. We note that these systematics should be similar to fitting the aperture extracted light curve with a GP as they are not conserved and each pixel is treated similarly. Our pixel kernel function for these systematics is described by:
    \begin{equation}
    k_\mathrm{p; CS}(i, j, i', j') = h_\mathrm{CS}^2 f_{(i, j)}f_{(i', j')}.
    \label{eq:CS_term}
    \end{equation}
    We refer to these as common systematics (CS) with height scale $h_{CS}$. The limitations of our GP optimisation  require these systematics to also share the same time kernel function as the flux-conserved systematics and independent pixel systematics. A comparison between the three previous systematics described is shown in Figure~\ref{fig:systematic_visual}.

    \subsubsection{Interpixel capacitance (IPC)}
    \label{sec:interpixel_capacitance}

    Interpixel capacitance is an effect which leads to correlated Poisson noise between neighbouring pixels. It is not to be confused with charge migration which is when charge carriers generated in one detector pixel are read by a neighbouring pixel. Charge diffusion does not result in a correlation in Poisson noise because the charge is exclusively read by the pixel the charge migrated to. Interpixel capacitance is where the potential difference on a given detector pixel is coupled to neighbouring pixels due to a small amount of capacitance between these pixels. This results in an observed correlation in the noise between neighbouring pixels \citep{2006OptEn..45g6402M}.

    This effect can be modelled by taking the image we expect to receive (including Poisson noise) and using a 3x3 convolution kernel, $H$ (not to be confused with a GP kernel function), which blurs the signal received in each pixel into its neighbouring eight pixels. The kernel will always sum to unity which results in no change to the overall flux received. We can parameterise this convolution kernel using Equation~\ref{eq:ipc_kernel} with $\alpha_x$ the coupling between neighbouring pixels to the left or right, $\alpha_y$ for pixels above and below and $\alpha_{xy}$ for diagonal pixels. More complex coupling is also possible but this parameterisation is common in the literature \citep{2006OptEn..45g6402M, 2023PASP..135g5004M} as follows:
    \begin{equation}
    H = 
    \begin{bmatrix}
    \alpha_{xy} & \alpha_{y} & \alpha_{xy} \\
    \alpha_{x} & (1 - 2\alpha_{x} - 2\alpha_{y} - 4\alpha_{xy}) & \alpha_{x} \\
    \alpha_{xy} & \alpha_{y} & \alpha_{xy}
    \end{bmatrix}
    \label{eq:ipc_kernel}
    .\end{equation}
    \citet{2023PASP..135g5004M} suggested that for the MIRI detector $\alpha_x \approx \alpha_y \approx 3\%$ while $\alpha_{xy}$ is significantly smaller yet still non-zero. The exact extent of IPC is difficult to quantify and is claimed to vary between observations so we  rely on fitting for $\alpha_x, \alpha_y$ and $\alpha_{xy}$ for each observation.

    We need to account for both the amplitude of white noise for each pixel and the effect of IPC in our combined pixel kernel function $s_\mathrm{p; wn+IPC}$. However, as IPC can be modelled using linear transformations, it is more natural to describe this by the covariance matrix generated by this kernel function, $\mathbf{S}_\mathrm{p; wn+IPC}$. Suppose each pixel $(i, j)$ observes Poisson noise with variance $\sigma_{(i, j)}^2$ - which in the absence of IPC is independent between pixels. This can be described by a diagonal covariance matrix, $\mathbf{S}_\mathrm{p; wn}$, which contains the variance of each pixel $\sigma_{(i, j)}^2$ along its diagonal (the pixels must be sorted into some arbitrary order). We can simulate IPC by taking a random draw $\vec{z}$ from $\mathbf{S}_\mathrm{p; wn}$ and then convolving with our IPC kernel $H$ from Equation~\ref{eq:ipc_kernel}. Note we can construct a matrix $\mathbf{T}_\mathrm{IPC}$ which performs this convolution for each pixel via the matrix multiplication: $\vec{z}' = \mathbf{T}_\mathrm{IPC} \vec{z}$, where $\vec{z}'$ is our random draw of white noise with the effect of IPC simulated. The covariance matrix of $\vec{z}'$ is then given by:
    \begin{align}
    \mathbf{S}_\mathrm{p; wn+IPC} = \mathbb{E}((\vec{z}' - \mathbb{E}(\vec{z}'))(\vec{z}' - \mathbb{E}(\vec{z}'))^T) &= \mathbb{E}(z'z'^T), \\
    &= \mathbb{E}(\mathbf{T}_\mathrm{IPC} \vec{z} \vec{z}^T \mathbf{T}_\mathrm{IPC}^T), \\
    &= \mathbf{T}_\mathrm{IPC} \mathbf{S}_p \mathbf{T}_\mathrm{IPC}^T,
    \label{eq:IPC_cov}
    \end{align}
    where we have used the fact that $\mathbb{E}(\vec{z}') = 0$, which follows from $\mathbb{E}(\vec{z}) = 0$. For notating our overall kernel function, we define the value of our pixel kernel function $s_\mathrm{p; wn+IPC}$ calculated between points located at $(i, j)$ and $(i', j')$ as $\mathbf{S}_\mathrm{p; wn+IPC; (i, j), (i', j')}$, the element of the covariance matrix corresponding to these two pixels.

    \subsubsection{Background flux and systematics along rows and columns}
    \label{sec:background_scatter}

    There is significant background flux in these observations, due to a variety of sources including thermal emission from JWST itself \citep{background}. The exact level of background flux will vary in time, resulting in a correlation in the scatter between all pixels on the detector which can be partially corrected for using a background subtraction. In addition, due to effects such as the reset anomaly and the intricacies of the way pixels are read out by the MIRI instrument, there could be a correlation in the noise between pixels which share the same row or column.
    
    We account for this correlation using our pixel kernel function $s_\mathrm{p; backgr}$. We assume any residuals from the background subtraction will be independent in time and so these processes can share the same time kernel function $s_t = \delta_{tt'}$ as for white noise.
    
    We can describe the pixel covariance between pixels located at $(i, j)$ and $(i', j')$ using:
    \begin{equation}
    s_\mathrm{p; backgr}(i, j, i', j') = h_\mathrm{back}^2 + h_\mathrm{row}^2 \delta_{ii'} + h_\mathrm{col}^2 \delta_{jj'},
    \label{eq:background_corr}
    \end{equation}
    where $h_\mathrm{back}$ is the amplitude of overall background scatter, $h_\mathrm{row}$ is the amplitude of correlated scatter in each row and similarly $h_\mathrm{col}$ for each column. By fitting for the amplitude of $h_\mathrm{back}$, $h_\mathrm{row}$ and $h_\mathrm{col}$, we can constrain how much remaining background scatter is left by the background subtraction. We could potentially use this to compare different background subtraction procedures. We note that this does not give us the ability to detect whether the background has been over-subtracted or under-subtracted. Instead, it  simply allows us to see whether there is a lot of correlated scatter between rows, columns, or all pixels.

    \subsubsection{Combined kernel}
    \label{sec:combined_kernel}
    Our full kernel function for the pixel-fitting method is therefore given as:
    \begin{align}
    k(i, j, t, &i', j', t') = \left[ h_\mathrm{FCS}^2 f_{(i, j)}\left(f_\mathrm{neigh; (i, j)} \delta_{ii'} \delta_{jj'} - f_{(i', j')} n_{(i, j), (i', j')}\right) \right. \nonumber\\
    &+\left. h_\mathrm{IPS; (i, j)}^2 f_{(i, j)}^2 \delta_{ii'} \delta_{jj'} + h_\mathrm{CS}^2 f_{(i, j)}f_{(i', j')}\right] \exp\left(-\frac{|t - t'|}{l_{t}}\right) \nonumber\\
    &+  \left[\mathbf{S}_\mathrm{p; wn+IPC; (i, j), (i', j')}  + h_{back}^2 + h_{row}^2 \delta_{ii'} + h_{col}^2 \delta_{jj'}\right] \delta_{tt'}
    \label{eq:full_kernel}.
    \end{align}

\section{Testing on simulated data}
\label{sec:3}

    To better examine how the pixel-fitting method compares to aperture photometry, we generated synthetic observations for which the true eclipse depth was known. This then allows us to examine how precisely and robustly each method recovers the eclipse depth in different circumstances. We examined how well each method performed when there were only flux-conserved systematics present, when systematics were added in only a single pixel light curve, and when there were systematics present across all pixels with the same shape (e.g. from a flare).

    \subsection{Simulating the target signal}
    \label{sec:mf}
    
    The simulations were based on the first eclipse of LHS 1140c, which consisted of 1,262 integrations and covered $\sim$233 minutes in total (for more details see Section~\ref{sec:4}), although we assume the first 30 minutes have been clipped. To minimise the computational cost, we simulated eight times fewer integrations. This permitted each pixel-fitting analysis to run in $\sim\!$40 minutes on a CPU cluster node, and we could run eight of these analyses in parallel. As we were simulating observations with eight times the integration time, we reduced the amplitude of Poisson noise by a factor of $\sqrt{8}$ to maintain similar constraints on the eclipse depth.
    
    The PSF model was taken from the median frame of the first LHS 1140c eclipse after clipping the first 30 minutes. This PSF model was then interpolated in $x$ and $y$ position for each integration with a change in position distributed as $\mathcal{N}(0, 0.005)$ pixels around the central location. No detector settling or linear slope in flux was introduced meaning that the baseline flux in each pixel was constant.
    
    The eclipse model was generated using the package \textsc{jaxoplanet} \citep{jaxoplanet}, which implements the equations of \citet{mandelagol} in \textsc{JAX} \citep{JAX}. The eclipse model was fixed for all simulations with the mean values of $P$, $a/R_*$, $R_p/R_*$, $b$ from literature values of LHS 1140c (as listed in Table~\ref{tab:priors}). The central eclipse time was fixed to the midpoint of the time series and the eccentricity was fixed to zero. For all fits of the simulated data, the eclipse depth was the only eclipse model parameter being fit for with every other eclipse parameter held fixed to the true values.

    \subsection{Simulating white noise and systematics}
    \label{sec:noise}
    
    White noise was generated with the amplitude for each pixel determined by taking the standard deviation of each pixel time series in the first LHS 1140c eclipse. This noise was then convolved with the IPC convolution kernel from Equation~\ref{eq:ipc_kernel} using values of $\alpha_x = 3\%$, $\alpha_y = 2\%$ and $\alpha_{xy} = 1\%$, before then being added to the simulated signal. Background noise was also introduced to simulate residual background scatter not removed by a background subtraction. Each integration had a constant value added to all pixels, drawn from $\mathcal{N}(0, 0.1)$ DN/s, which (given the flux in the target signal) works out to adding scatter with amplitude of $\approx\!135$ ppm for a 5px radius circular aperture. We also added scatter which was constant along each row and column but independent between each row and column. This was added by drawing from $\mathcal{N}(0, 0.4)$ DN/s for the rows and $\mathcal{N}(0, 0.3)$ DN/s for the columns. Our analysis of the real observations did not constrain the presence of these row or column systematics but it was a possibility we wanted to include which both aperture photometry and the pixel-fitting method should be able to account for.
    
\begin{figure}
        \includegraphics[width=\columnwidth]{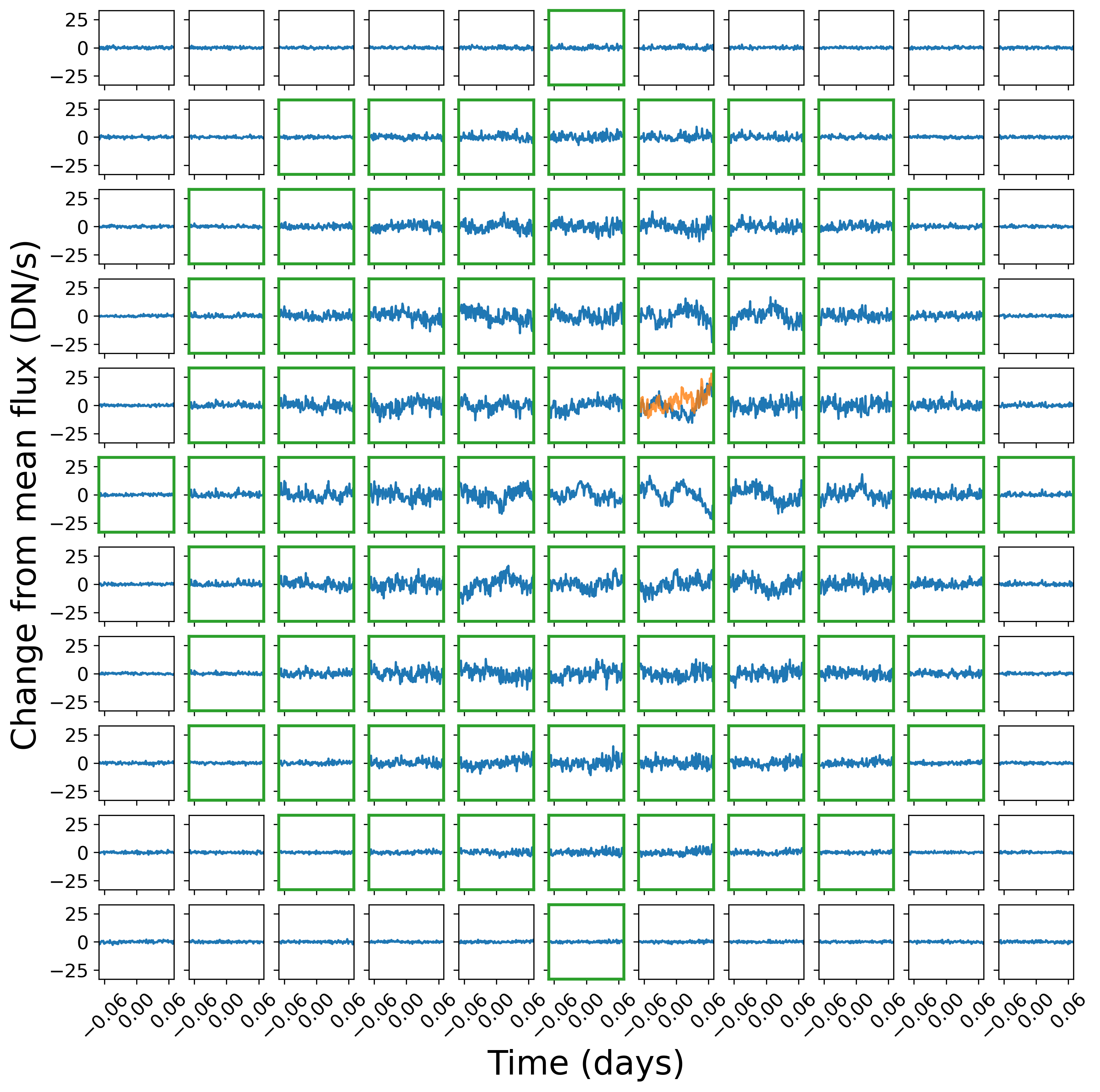}
        \caption{Pixel light curves for the first simulation in the flux-conserved systematics (FCS) and independent pixel systematics (IPS) scenarios. The only difference is the pixel light curve in orange which had independent systematics added to it. The pixels included within the pixel-fits are highlighted with green boundaries.}
        \label{fig:sim1and2visual}
    \end{figure}

    We simulated the flux-conserved systematics by taking a random draw from a GP with an exponential kernel and a length scale of 35 minutes and adding it to one pixel and subtracting it from a neighbouring pixel. This was performed for each pair of neighbouring pixels (i.e. 8 times per pixel) and the amplitude of the draw was scaled to be the geometric mean of the flux on each of the pair of pixels multiplied by the constant $h_\mathrm{FCS} = 990$ ppm. By adding and subtracting the same signal, it means that these systematics are flux-conserving with the only non-flux-conserving effects due to the finite area of the aperture (i.e. there is `flux movement' across the fixed aperture boundary), although the amplitude of this effect was small. For a pixel surrounded by eight pixels with the same flux values, the amplitude of systematics in this pixel would work out to$\sqrt{8} h_{\mathrm{FCS}} = 2800$ ppm. We note that $h_\mathrm{FCS}$ was meant to be set to 700 ppm - consistent with the values recovered in the observations - but the addition and subtraction was accidentally performed twice for each pair of pixels, effectively increasing the amplitude of this effect by$\sqrt{2}$. We generated 200 of these simulations and we refer to them as the flux-conserved systematics (FCS) scenario.
The other two sets of simulations included additional systematics which were not flux-conserved. The second set of simulations -- called the independent pixel systematics (IPS) scenario -- took each simulation from the FCS scenario and injected a $h_\mathrm{IPS} = 5000$ ppm height scale systematic into a single pixel time series. The pixel was randomly chosen for each simulation to be one of the nine brightest pixels at the centre of the PSF. As these pixels tend to make up about $3\% - 4.5\%$ of total flux, this results in a $\approx 150-225$ ppm systematic in the aperture extracted light curve. While aperture extraction with this pixel masked out may be a good approach, the flux-conserved systematics make it difficult to distinguish these additional systematics from flux-conserved systematics. This can be seen in Figure~\ref{fig:sim1and2visual}, which features a plot of the pixel light curves centred on the PSF for the first simulated dataset in the FCS scenario and the IPS scenario.

    \begin{figure}
        \includegraphics[width=\columnwidth]{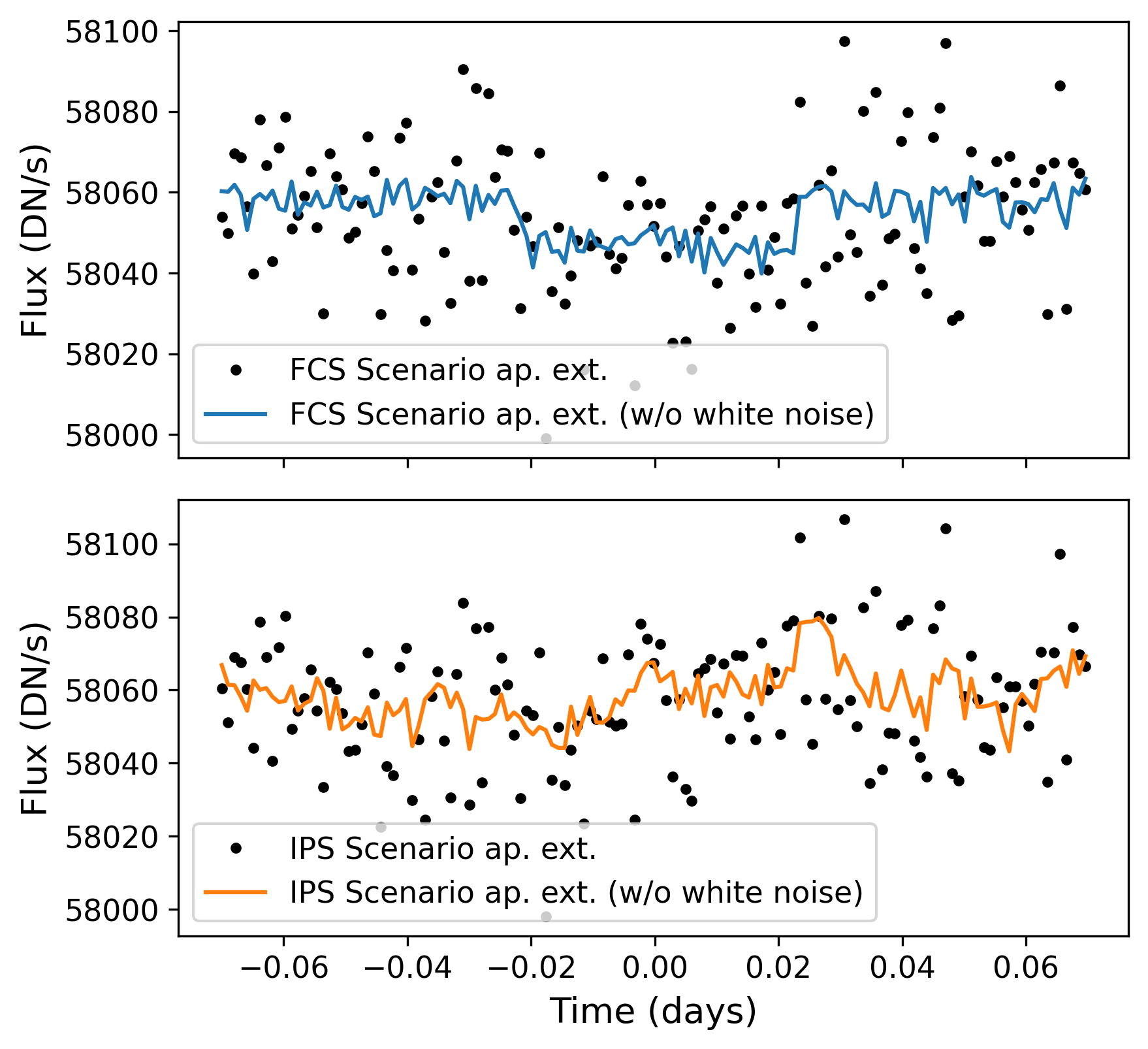}
        \caption{Aperture extracted light curves from simulated datasets shown in Figure~\ref{fig:sim1and2visual}. Aperture extractions were also performed without including white noise to visualise systematics more clearly (some jitter remains from imperfections in centroiding). The systematics injected into one pixel has dramatically affected the aperture extracted light curve.}
        \label{fig:sim1and2ap}
    \end{figure}
    
    The third set of simulations added a common systematic shape to all pixels at an amplitude equal to $h_\mathrm{CS} = 200$ ppm multiplied by the flux on each pixel. Here, this  is referred to as the common systematics (CS) scenario. This could represent real astrophysical systematics such as a flare or instrumental systematics which affect the throughput of the observation. The height scale was chosen as it resulted in a similar amplitude of systematics in an aperture extracted light curve as the IPS scenario, but with the systematics spread out over all pixels instead of originating entirely from a single pixel.

    For each scenario, the synthetic data were aperture extracted with a radius of 5 pixels centred on the best-fit centroid for each integration -- as performed for the real observations. This radius minimised scatter in the aperture extracted light curves compared to radii of 4.5 or 5.5 pixels, matching the real observations. In Figure~\ref{fig:sim1and2ap}, we give an example of a simulated dataset from the FCS and IPS scenarios after aperture extraction. The pixel-fitting method only included pixels with centres within 5 pixels from the PSF centre, resulting in 81 pixels being included, which is similar to the 78.5px$^2$ area for a circular aperture with a radius of 5 pixels.

    \subsection{Different fitting approaches tested}
    \label{sec:sim_methods}

    We used MCMC to recover the eclipse depth with each method, modelling the posterior as Gaussian and using the mean and standard deviation of the chains. For each method, we ran four chains of No U-Turn Sampling (NUTS; \citealt{Hoffman2011}) as implemented in \textsc{PyMC} \citep{Salvatier2015} -- with each chain having 1000 tuning steps and 1000 draws. We also used slice sampling \citep{Neal2003} in combination with NUTS, by performing blocked Gibbs sampling \citep{blocked_Gibbs} on some parameters close to their prior bounds (e.g. $h_\mathrm{back}$, $h_\mathrm{row}$, $h_\mathrm{col}$, $l_t$, $h_\mathrm{CS}$, $h_\mathrm{IPS}$). This was performed because it was found that running NUTS alone could sometimes result in chains getting stuck on these parameters. The Gelman-Rubin statistic \citep{1992StaSc...7..457G} was used to measure convergence of the chains. A slightly higher than normal threshold of $\hat{r}\leq1.02$ was used to determine convergence for each simulation in order to reduce computational costs arising from the large number of MCMCs performed.
    
    For each set of simulations, we tested aperture extraction with and without a GP. These MCMCs fit for the eclipse depth, $d$, the flux out-of-(secondary) transit, $F_\mathrm{oot}$, the white noise amplitude, $\sigma$, and in the case of the GP fits also the height scale, $h$, and length scale, $l_t$, of the GP. We also tested using a GP with $l_t$ fixed to the true simulated value of 35 minutes, as this may be a fairer comparison to the pixel-fitting method -- which obtains extra information about the time length scale as it fits for flux-conserved systematics with the same time length scale.

    For all pixel-fitting approaches, we fit for the eclipse depth, $d$, baseline flux for each pixel, $f_{(i, j)}$ (for 81 pixels), white noise amplitude for each pixel $\sigma_{(i, j)}$, change in centroid position at each integration $\vec{\Delta x}$ and $\vec{\Delta y}$ (for 137 integrations), IPC parameters $\alpha_x$, $\alpha_y$, and $\alpha_{xy}$, background scatter parameters, $h_\mathrm{back}$, $h_\mathrm{row}$ and $h_\mathrm{col}$, height scale and time length scale of flux-conserving systematics $h_\mathrm{FCS}$ and $l_t$. This describes the `only flux-conserving systematics' approach (abbreviated as `only FCS') and includes 445 free parameters which were all fit for. We also included a height scale for systematics with a common shape in all pixels $h_{CS}$ for the `common systematics' approach and height scales for independent systematics in each pixel $h_\mathrm{IPS; i}$ for the `independent systematics' approach. Finally, as this independent systematics approach turned out to give quite conservative predictions, we also reran each of these approaches where only the pixels recovered to have large height scales (specifically where the $3^{rd}$ percentile value was greater than a threshold of $10^{-4.8}\approx16$ ppm) were fit with independent height scales and the remaining pixels all shared a joint height scale $h_\mathrm{IPS; shared}$. We note these pixel systematics were still treated as independent but with a shared amplitude (i.e. they could still have completely different shapes). This is listed in tables as the `shared independent systematics' approach. We also included methods which could account for both common and independent pixel systematics as these were robust to a range of systematics.

    \subsection{Comparing the performance of each method}
    \label{sec:metrics}
    
    We measured the precision of each method using the root mean square error (RMSE) of the mean eclipse depth for each simulation. The smaller the RMSE, the more precise the eclipse depth measurement is. If the method is robust, then the average standard deviation of the eclipse depth measurement, $\bar{\sigma}_d$, should approximately match the RMSE. If the standard deviation is too big then our results are overly conservative and if it is too small then the results are unreliable. This can be summarised with the reduced chi-squared statistic $\chi_r^2$, which should equal unity when the uncertainties are correctly estimated, less than one for uncertainties which are too large/conservative and greater than one for uncertainties which are too small/unreliable\footnote{This assumes the posteriors for the eclipse depths are Gaussian}. For each scenario (consisting of 200 simulations), it is calculated as follows:
    \begin{equation}
    \label{chi2}
    \chi_r^2 = \frac{1}{200}\sum_{i=1}^{200}\left(\frac{d_i - d_\mathrm{true}}{\sigma_i}\right)^2
    .\end{equation}
    Here, $d_\mathrm{true} = 200$ppm is the true simulated eclipse depth and $d_i$, $\sigma_i$ are the mean and standard deviation of the recovered eclipse depth for the $i^\mathrm{th}$ simulation.

    Finally, we examined whether any method was biased towards measuring eclipse depths which were too large or too small by calculating whether the mean eclipse depth across all 200 simulations was consistent with $d_\mathrm{true} = 200$ppm. We found that all methods tested were within $2.5\sigma$ of the expected mean values across each of the scenarios. Given that this metric has little impact on the comparison between methods, and is typically very similar for methods within the same scenario, we chose to exclude it from each table of results.

    \subsection{Flux-conserved systematics (FCS) scenario results}
    \label{sec:sim1}

    This scenario only contained flux-conserved systematics, therefore, aperture extraction should result in very clean light curves and a GP should not be required to robustly analyse these observations. However, for real observations we can never be sure that the light curves are completely free from systematics and so we are interested in how well each method performs in this scenario. The methods that marginalise over systematics will naturally tend to give more conservative constraints on the eclipse depths (even when systematics are not present); this is true both for aperture extraction with a GP and for various pixel-fitting approaches. However, this will be balanced out by their more reliable performance when systematics are present.

    The results in Table~\ref{tab:sim_1} demonstrate this very clearly. The most precise methods (i.e. smallest RMSE) are aperture extraction without a GP and the pixel-fitting approach which only accounts for flux-conserved systematics. They both exhibit $\chi_r^2$ values close to unity across these simulations, showing their uncertainty estimates are reliable. Aperture extraction with a GP has a slightly worse precision and larger uncertainties. The significantly smaller value of $\chi_r^2 = 0.74$ shows that this method is quite conservative. Pixel-fitting only accounting for common systematics performs similarly.
    
\begin{figure}
        \includegraphics[width=\columnwidth]{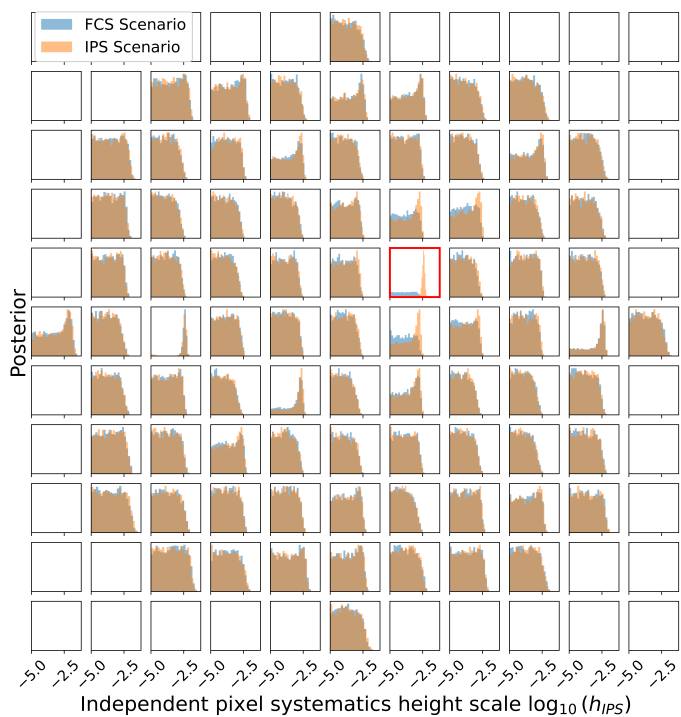}
        \caption{Constraints on the Independent Pixel Systematics (IPS) height scale for each pixel for the first simulation in the FCS and IPS scenarios (shown in Figure~\ref{fig:sim1and2visual}). Highlighted in red is the one pixel which had independent pixel systematics injected into it for the IPS scenario. This pixel has been correctly identified by the pixel-fitting method and would be weakly weighted in the eclipse fit.}
        \label{fig:sim1and2_h_PSS}
    \end{figure}

    However, we see that the other pixel-fitting approaches which can account for independent systematics in each pixel have worse accuracy and even larger uncertainty estimates. This is similar to aperture extraction with a GP and in general we do expect a more flexible systematics model to result in larger errorbars as it marginalises over a wider space of models \citep{Gibson2014}. The pixel-fitting method is particularly flexible as it can fit for potential systematics in every pixel. Despite the vast majority of these pixel systematics height scales being consistent with the minimum prior bound, they still marginalise over a range of amplitudes which are consistent with the data. See Figure~\ref{fig:sim1and2_h_PSS} for the posterior of each independent pixel systematics height scale, $h_\textrm{IPS}$, for the first simulation in the FCS and IPS scenarios.
    
    We see that sharing the amplitude of the smallest independent pixel systematics height scales does reduce how conservative the method is, with smaller average uncertainties and a $\chi_r^2$ value closer to one. Accounting for common systematics as well as independent pixel systematics offers the most conservative result; however this method is the most flexible and can be robustly applied to all three sets of simulations generated.

    \begin{table}
        \caption{Results of the 200 simulations in the Flux-Conserved Systematics (FCS) scenario.}
        \centering
        \begin{tabular}{llll}
                \hline
                Method & RMSE & $\bar{\sigma}_d$ & $\chi_r^2$ \\
                \hline
            Ap. Ext. w/o GP & 52 & 53 & 0.98\\
Px-fit (only FCS) & 51 & 52 & 0.94\\
Ap. Ext. w/ GP & 54 & 64 & 0.74\\
Ap. Ext. w/ GP (fixed $l_t$) & 55 & 66 & 0.73\\
Px-fit (common systematics) & 53 & 64 & 0.7\\
Px-fit (indep. systematics) & 63 & 81 & 0.59\\
Px-fit (shared indep. sys.) & 57 & 63 & 0.82\\
Px-fit (common+indep. sys.) & 68 & 88 & 0.57\\
Px-fit (common+shared indep. sys.) & 64 & 73 & 0.71\\
                \hline
        \end{tabular}
        \label{tab:sim_1}
    \end{table}

    \subsection{Independent pixel systematics (IPS) scenario results}
    \label{sec:sim2}

    The IPS scenario simulations were the exact same as the FCS scenario but with a single GP draw added to one of the brightest central nine pixels for each simulation. In this case, the pixel-fitting method has the potential to outperform aperture extraction as it can weight away from the pixel with systematics injected into it. However, it must be able to constrain the presence of independent systematics on top of the flux-conserved systematics. Since the method can fit for independent pixel systematics in all of the pixels, the results can still be a bit conservative given that the systematics are only present in a single pixel.

    \begin{figure*}
        \includegraphics[width=\textwidth]{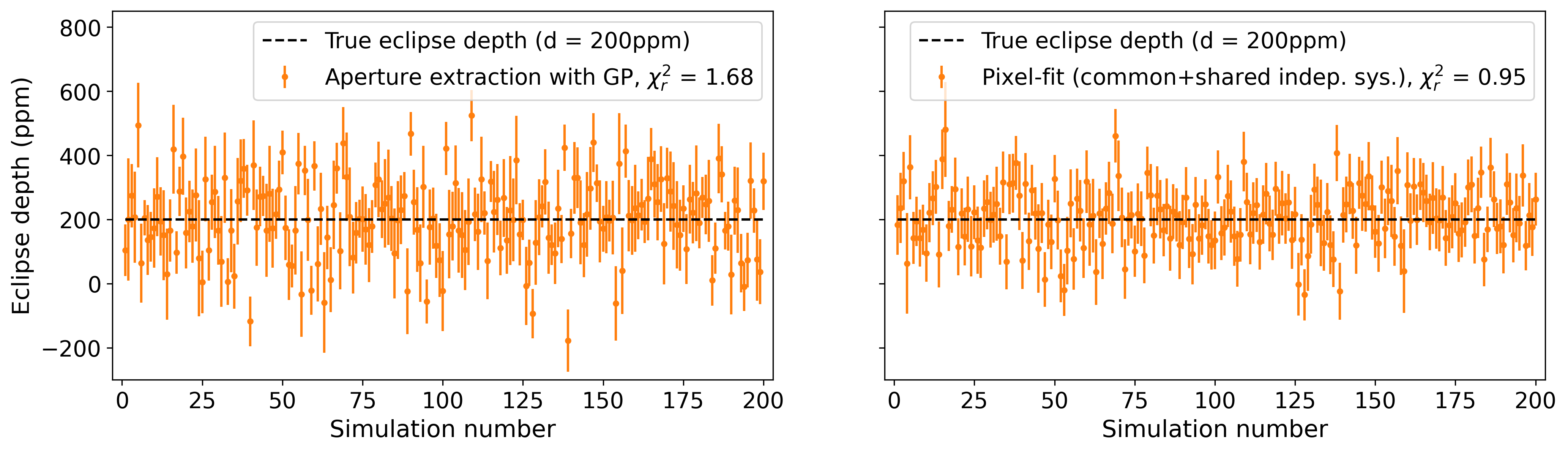}
        \caption{Eclipse depth constraint for all simulations in the IPS scenario, which each had a single pixel contaminated with time-correlated systematics. The left plot shows the results from using aperture extraction with a GP. The right plot shows pixel-fitting accounting for both independent pixel systematics (with a shared height scale for cleaner pixels) and common systematics. Pixel-fitting outperforms aperture extraction here in both precision (smaller RMSE) and reliability ($\chi_r^2$ closer to unity) as it can weight away from the single pixel contaminated by systematics.}
        \label{fig:PSS_comparison}
    \end{figure*}

    We see from Table~\ref{tab:sim_2} that all of the pixel-fitting approaches outperform aperture extraction with a GP on precision (RMSE), have tighter constraints (smaller $\bar{\sigma}_d$) and have more reliable constraints ($\chi_r^2$ closer to unity). This difference is also visualised in Figure~\ref{fig:PSS_comparison}, which shows the eclipse depth constraint for all 200 simulations in the IPS scenario for both aperture extraction with a GP and pixel-fitting (accounting for common systematics and shared independent systematics). Aperture extraction without a GP performs particularly poorly, with the lowest accuracy and with uncertainties which are far too small ($\chi_r^2 = 6.07$) because it cannot account for the systematics injected into the data. Aperture extraction with a GP does a reasonable job of accounting for the systematics but with quite low precision compared to pixel-fitting, as it has no way to weight away from the pixel with injected systematics. The chi-squared value of $\chi_r^2 = 1.68$ is also quite high, with a slightly better value of $\chi_r^2 = 1.46$ when the length scale in time is fixed to the true simulated value.
    
    For aperture extraction with a GP, the height scale of systematics $h$ is not constrained to be greater than the minimum prior bound in many of these simulations. This means that varying the location of this prior bound will increase or decrease the weighting of the area of the posterior where systematics are negligible. For example, we get $\chi_r^2 = 1.14$ if we increase the minimum bound to $h > 10^{-4}$ and if we instead lower the minimum bound to $h > 10^{-6}$, we get $\chi_r^2 = 1.88$. Increasing the minimum prior bound on the height scale can be interpreted as a more conservative analysis where the presence of systematics is more likely, and this would help lower the $\chi_r^2$ here. However, setting $h > 10^{-4}$ also lowers the $\chi_r^2$ value for the first set of simulations to just $\chi_r^2 = 0.46$, which was already overly conservative at $h > 10^{-5}$. Overall, for our normal prior bound of $h > 10^{-5}$, the $\chi_r^2$ from combining the FCS and IPS scenarios is $\chi_r^2 = 1.21$ when fitting for $l_t$ and $\chi_r^2 = 1.10$ when fixing it; thus, the method performs well over an even mix of clean and systematics-contaminated datasets.
    
    The pixel-fitting methods which account for common systematics on top of accounting for independent pixel systematics have slightly lower precision and larger uncertainties. However, they are accounting for additional systematics that could potentially exist in real data -- such as those from variations in throughput and detector settling -- so this is a more conservative approach that might make sense when we do not know the origin of systematics in the data.

    \begin{table}
        \caption{Results of the 200 simulations in the independent pixel systematics (IPS) scenario.}
        \centering
        \begin{tabular}{llll}
                \hline
                Method & RMSE & $\bar{\sigma}_d$ & $\chi_r^2$ \\
                \hline
            Ap. Ext. w/o GP & 143 & 59 & 6.07\\
Ap. Ext. w/ GP & 122 & 107 & 1.68\\
Ap. Ext. w/ GP (fixed $l_t$) & 119 & 112 & 1.46\\
Px-fit (indep. systematics) & 87 & 93 & 0.89\\
Px-fit (shared indep. sys.) & 84 & 81 & 1.12\\
Px-fit (common+indep. sys.) & 92 & 101 & 0.82\\
Px-fit (common+shared indep. sys.) & 89 & 92 & 0.95\\
                \hline
        \end{tabular}
        \label{tab:sim_2}
    \end{table}

    \subsection{Common systematics (CS) scenario results}
    \label{sec:sim3}

    Finally, the simulations in the common systematics scenario differ from the IPS scenario in that the systematics were injected into all pixels with a common shape instead of into a single pixel. The amplitude of these systematics was scaled by the flux in each pixel, so this could represent systematics inherent from the signal, such as stellar granulation or flaring. It could also arise from factors affecting the throughput which vary over time. The amplitude of this was such that aperture extraction with a GP performs similarly to the IPS scenario but with a different systematics origin. In the IPS scenario, the pixel-fitting approaches could outperform aperture extraction by weighting towards the cleaner pixels, however, this will not work when all the pixels suffer from a similar proportion of systematics. Pixel-fitting should therefore perform similarly to aperture extraction with a GP.

    We can see from Table~\ref{tab:sim_3} that all methods have low precision and most have large uncertainties. Aperture extraction without a GP has a similarly poor performance as it had for the IPS scenario, with uncertainties which are too small as it cannot robustly account for the systematics present. Any of the pixel-fitting methods which can account for common systematics approximately match the performance of aperture extraction with a GP. Similar to the IPS scenario, aperture extraction with a GP and also the pixel-fitting methods have quite high $\chi_r^2$ values. This could also be explained by the trade-off between performance on clean data and systematics-contaminated data when the systematics are not fully constrained to be present. The slightly more robust $\chi_r^2$ values when independent pixel systematics are also accounted for may arise from the method just being generally more conservative across all the simulations rather than it more closely modelling the systematics.

    \begin{table}
        \caption{Results of the 200 simulations in the common systematics (CS) scenario.}
        \centering
        \begin{tabular}{llll}
                \hline
                Method & RMSE & $\bar{\sigma}_d$ & $\chi_r^2$ \\
                \hline
            Ap. Ext. w/o GP & 155 & 60 & 6.86\\
Ap. Ext. w/ GP & 133 & 116 & 1.73\\
Ap. Ext. w/ GP (fixed $l_t$) & 133 & 120 & 1.54\\
Px-fit (common systematics) & 127 & 118 & 1.54\\
Px-fit (common+indep. sys.) & 128 & 121 & 1.24\\
Px-fit (common+shared indep. sys.) & 128 & 120 & 1.37\\
                \hline
        \end{tabular}
        \label{tab:sim_3}
    \end{table}

    \subsection{Summary of the simulations}
    \label{sec:sim_summary}
    
    \begin{figure}
        \includegraphics[width=\columnwidth]{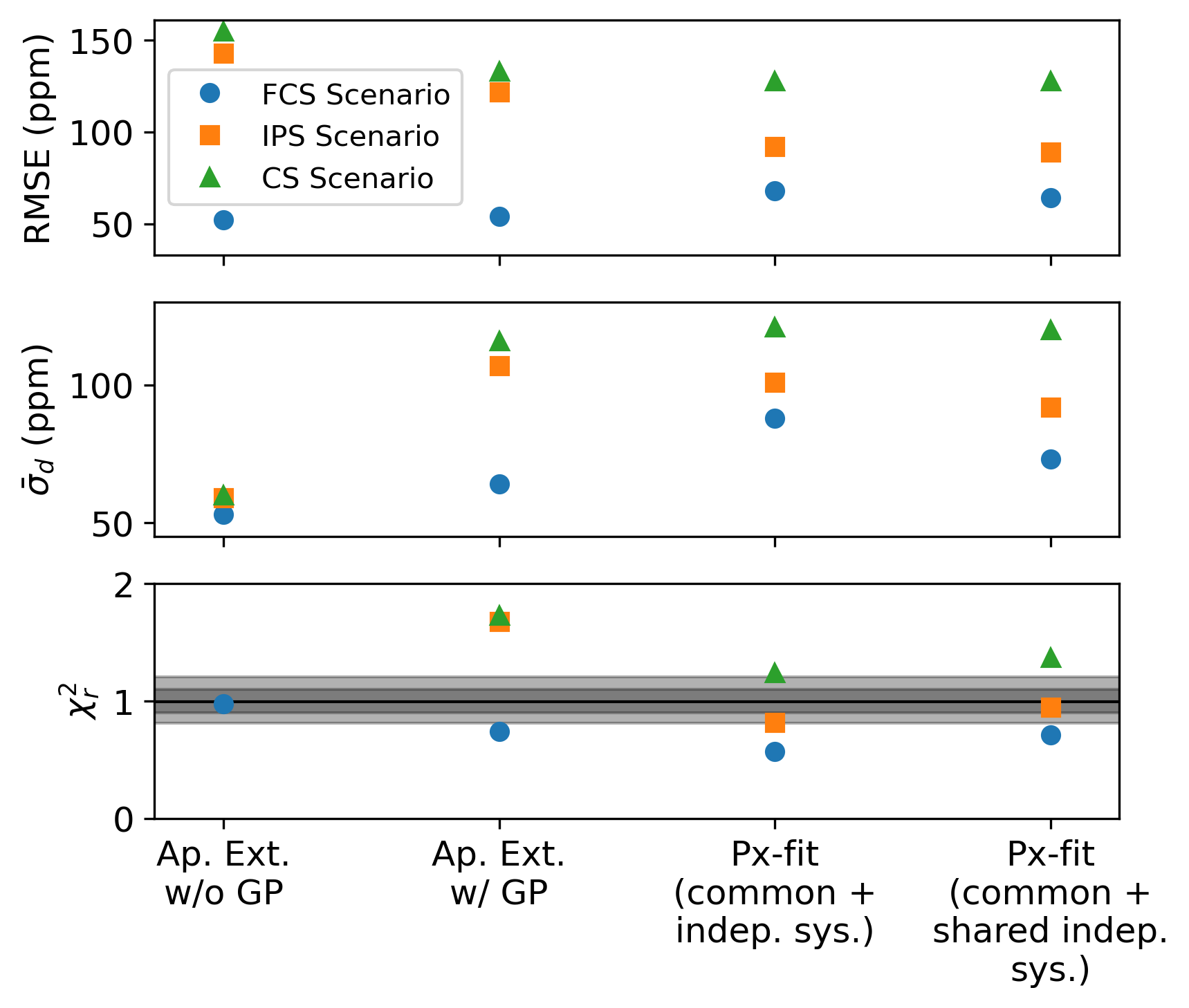}
        \caption{Results from Tables~\ref{tab:sim_1}, \ref{tab:sim_2}, and \ref{tab:sim_3} comparing the RMSE, mean uncertainty in eclipse depth, and $\chi_r^2$ of recovered eclipse depths for various methods. Aperture extraction with a GP is shown to perform similarly for the IPS and CS scenarios as the aperture extracted light curves contain similar amplitude systematics. Pixel-fitting achieves a better performance for the IPS scenario as it can weight away from the contaminated pixel.}
        \label{fig:sim_summary}
    \end{figure}

    The results demonstrate that pixel-fitting can outperform aperture extraction when systematics are large in a particular pixel, but the method is overly conservative when only flux-conserving systematics are present. It also matches the performance of aperture extraction with a GP when systematics have a common shape across all pixels. See Figure~\ref{fig:sim_summary} for a summary of the results in Tables~\ref{tab:sim_1}, \ref{tab:sim_2}, and \ref{tab:sim_3} for aperture extraction with and without a GP as well as for the general pixel-fitting approach of accounting for common systematics across all pixels and systematics in individual pixels. We can see that compared with aperture extraction with a GP, pixel-fitting performs very similarly when systematics are common between all pixels but much better when there are systematics present in specific pixels; for instance, after a cosmic ray hit (as shown in Figure~\ref{fig:real_px_lightcurves}).

\section{Observations}
\label{sec:4}

    The three eclipses of LHS 1140c were observed on $27^\mathrm{th}$ November 2023, $7^\mathrm{th}$ July 2024, and $19^\mathrm{th}$ July 2024. Each eclipse used the SUB256 subarray, read 36 groups for each integration and had 1,262 integrations. The integrations had an 11.1s cadence, with each observation lasting $\sim$233 minutes. This followed a strategy of leaving 60 minutes of flexible start time plus 30 minutes of settling time followed by observing for two eclipse durations, with one eclipse duration ($\approx\!67$ minutes) before and after expected mid-eclipse for a circular orbit. There was also an additional 9 minutes of padding to help catch slightly more eccentric orbits.

    \subsection{Primary data reduction}
    \label{sec:reduction}

    Our reduction began with a download of the uncalibrated `uncal' FITS files for our three eclipses from the Mikulski Archive for Space Telescopes (MAST), which had already been pre-processed by version 2024\_1a of the Science Data Processing (SDP) system. We ran the JWST calibration pipeline version 1.15.1 with the default settings for steps 1 and 2, except we changed the jump detection threshold to 7$\sigma$, excluded both the first and last groups from each integration when reading up-the-ramp (the default is to only exclude the last group) and skipped the \textsc{photom} step in stage 2 to keep the flux units in data numbers per second (DN/s). We excluded the first and last groups to avoid a known issue where these groups can include difficult to correct reset effects \citep{Morrison_2023}. The jump detection threshold was decided upon by calculating the average amplitude of white noise recovered across all three eclipse fits when only white noise and an exponential ramp + linear model was fit. It was found that 7$\sigma$ minimised this out of integer thresholds from $4 - 10\sigma$.
    
    One issue we encountered was that running the pipeline separately on each uncal data segment resulted in offsets in the flux time series between segments, despite these segments not representing any actual change in the observation but instead are simply a way of reducing the total filesize of any given file. To remedy this, for all our reductions we first combined the uncal segments together into a single FITS file and then ran the JWST pipeline on this combined uncal file, which avoided these offsets.

    Once we had our stage 1 and 2 calibrated `calints' files, we applied a median clipping filter to remove any additional outliers. A background subtraction was then performed on each integration by taking the median flux values from an annulus region centred around the target PSF with lower and upper radii of 50 to 80 pixels from the annulus centre. This annulus region was chosen as there still appeared to be a gradient in flux as far out as about 50 pixels from the centre while the flux appears to remain constant from 50 to 80 pixels.

    \subsubsection{Primary reduction: Aperture extraction analysis}
    \label{sec:ap_ext_fitting}

\begin{figure}
        \includegraphics[width=\columnwidth]{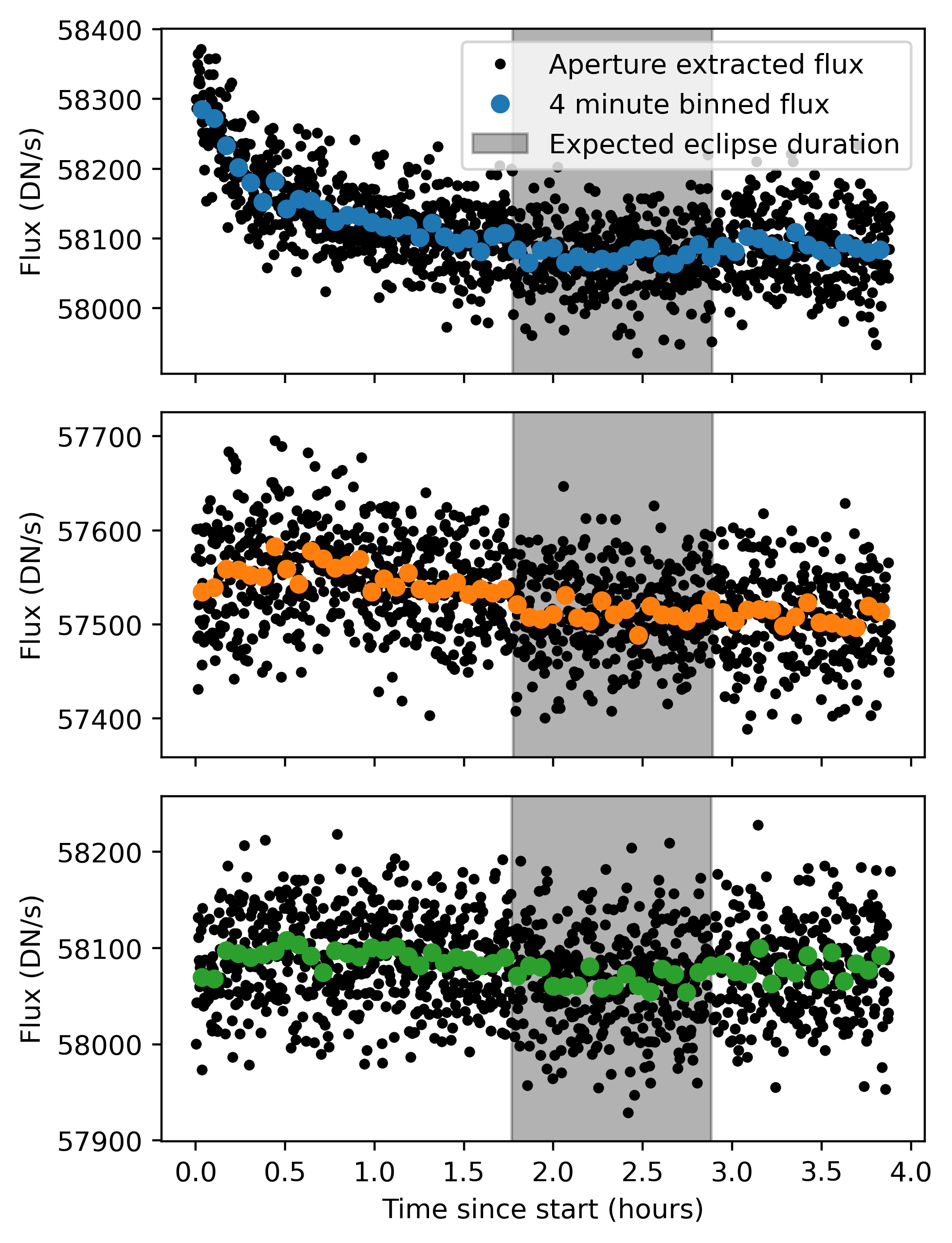}
        \caption{Three aperture extracted eclipse light curves from the primary reduction. Observations are binned into four minute bins for clarity. Expected eclipse location for a circular orbit is shaded in gray. There are differences in initial detector settling in the first half hour, where the first eclipse shows a strong decrease in flux, while the other two eclipses increase in flux.}
        \label{fig:three_eclipses}
    \end{figure}

    Aperture extraction was performed with a circular aperture using the package \textsc{photutils} \citep{photutils}. For each integration, the centroid position was fit for using a best-fit to a 2D Gaussian. A range of radii from 3.0px to 7.0px were used in 0.5px steps to extract the light curves and it was found a 5px radius minimised the amplitude of scatter in the light curves. The 5px aperture extracted flux for each light curve is shown in Figure~\ref{fig:three_eclipses} with only the first integration of each eclipse removed as these show significantly lower flux (consistent with the `first exposure effect'; \citealt{Morrison_2023}).

    We performed a joint fit of the three eclipses using priors on the central (primary) transit time, $T_0$; system scale, $a/R_*$; impact parameter, $b$; and radius ratio, $R_p/R_*$, from the NIRISS/SOSS transit observation of LHS 1140c, as analysed in \citet{2024ApJ...970L...2C}. As the NIRISS/SOSS analysis assumed a circular orbit with fixed period, we used priors for the period, $P$, and eccentricity parameters, $\sqrt{e}\cos\omega$ and $\sqrt{e}\sin\omega$, from the joint fit of radial velocity and previous transit observations (performed before the JWST transit observation) in \citet{2024ApJ...960L...3C}. Including the JWST transit observation from \citet{2024ApJ...970L...2C} within a joint fit of the RV data could help to tighten these constraints further but is beyond the scope of this paper. Table~\ref{tab:priors} shows a summary of these priors. As the time of primary transit is tightly constrained with this prior, the central eclipse time is largely determined by the eccentricity parameters (see Appendix~\ref{app:eccentricity}).

    To deal with detector settling, we clipped between 30-60 minutes (162-324 integrations) from the start of each observation, with all joint-fits clipping 45 minutes (243 integrations). We experimented with including an exponential ramp in addition to a linear slope, modelling the baseline flux as:
    \begin{equation}
    f_\mathrm{base}(t) = F_\mathrm{oot} + T_\mathrm{grad}(t - t_\mathrm{mid}) + h_\mathrm{ramp} \exp\left(-\frac{t - t_\mathrm{init}}{l_\mathrm{ramp}}\right),
    \label{eq:baseline}
    \end{equation}
    where $t_\mathrm{init}$ is the time stamp of the first integration being fit and $t_\mathrm{mid}$ the midpoint of the observation, $F_\mathrm{oot}$ is the baseline flux out-of-(secondary) transit, $T_\mathrm{grad}$ is a linear slope in baseline flux, $h_\mathrm{ramp}$ is the amplitude of the settling ramp and $l_\mathrm{ramp}$ is the lengthscale of the ramp. When fitting for a combined exponential ramp and linear slope, there were issues with degeneracies between long exponential ramps with large $l_\mathrm{ramp}$ values and the linear slope $T_\mathrm{grad}$; thus, in practice, we also subtracted the tangent line to the exponential ramp at the last integration, which helped reduce this degeneracy when $l_\mathrm{ramp}$ was large.

    When fitting the aperture extracted light curve with a GP, we chose an exponential kernel plus white noise:
    \begin{equation}
        \mathbf{K}_{ij} = h^2 \exp\left(-\frac{|t_i - t_j|}{l_{t}}\right) + \sigma^2 \delta_{ij},
        \label{eq:exp_kernel2}
    \end{equation}
    where $h$ is the height scale, $l_t$ the length scale of correlated noise, and $\sigma$ is the amplitude of white noise.

    \subsubsection{Primary reduction: Pixel-fitting analysis}
    \label{sec:px_fitting}

    For the pixel-fitting method, the aperture extraction was not needed and the reduced data were used directly. The same eclipse model and priors were used as for the aperture extraction method.
    
\begin{figure}
        \includegraphics[width=\columnwidth]{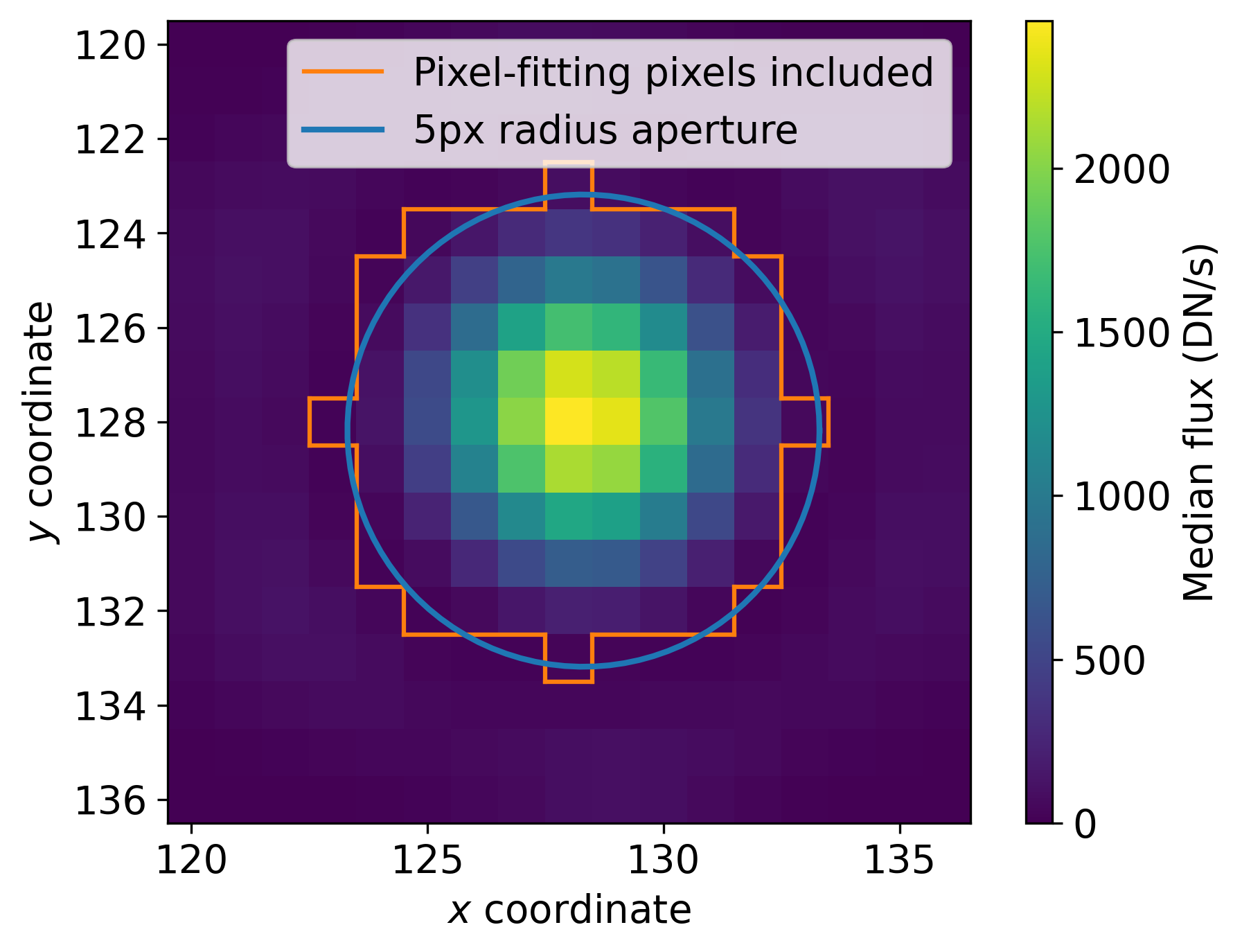}
        \caption{Median frame of the first observation centred on the target PSF. Aperture extraction was performed with a 5px circular aperture in blue, centred for each integration. Pixels used for the pixel-fitting are enclosed in orange.}
        \label{fig:PSF_visual}
    \end{figure}

    Figure~\ref{fig:PSF_visual} shows the choice of pixels within 5px of the PSF centre for the pixel-fitting method overlaid over the median frame of the PSF. It is also shown how it compares to the 5px circular aperture used for aperture extraction. We note that unlike aperture extraction, pixel-fitting can weight away from lower S/N pixels similar to optimal extraction, so we can choose to include more pixels to try improve our constraints. However, including more pixels becomes more computationally expensive particularly as there are many more parameters to recover, and the change in constraints was found to be quite negligible when pixels 8.5px away from the centre were included (covered in Appendix~\ref{app:extra_eclipse_depths}).
    
    The computational cost of joint-fitting the three eclipse light curves using pixel-fitting was quite high and took approximately 20 hours for each analysis when running on one NVIDIA A100 GPU. This was largely due to the relatively large number (1,019) of integrations fit for each eclipse which required the decomposition of a large time covariance matrix $\mathbf{K}_t$ in each MCMC step. In addition, there was a huge number of free parameters to be fit, especially because each integration in all three datasets had two parameters which marginalised over the uncertainty in the change in centroid position (resulting in over 6,000 free parameters). As these parameters were all well constrained and had Gaussian-like posteriors, they were efficiently sampled using NUTS and all had an $\hat{r} \le 1.01$. The computational cost could have been reduced substantially by binning in time, parallelising each MCMC chain over more GPUs or running on each dataset separately. Running the method on a single dataset and parallelising over 4 GPUs on a GPU cluster took approximately one hour, even without binning in time. This means the method is better suited to examining pixel systematics in individual eclipses but poorly suited to joint-fitting large numbers of eclipses at high time-resolution.

    \subsection{Secondary data reduction}
    \label{sec:sec_reduction}
    We performed a secondary reduction using a completely independent pipeline, which has been previously used in \citet{Prune2024} and will be described in more detail in the upcoming paper (Gressier et al., in prep.). This pipeline employs the \texttt{transitspectroscopy} \citep{espinoza_nestor_2022} wrapper for the initial stages of the \texttt{jwst} pipeline, skipping the reference pixel correction and incorporating a custom jump correction. The identified 10 Hz heater noise is corrected. Once the groups are converted to slopes, we proceed to Stage 2. We compute a median frame, which is then used for the following step. To estimate the background shape, we mask pixels within a box centered around the source. The masked pixels are interpolated using a linear interpolation with neighboring pixels. This frame is then smoothed using a Gaussian filter with a standard deviation of 7 pixels. The resulting median background image is subtracted from all frames. Pixel light curves are extracted, and outliers are identified using a median filter. These outliers are replaced with the median value of each pixel. Once we obtain a background corrected, cleaned image, we proceed to flux extraction. 
    
    We fit a 2D Gaussian to each frame to determine the centroid position and then apply a circular mask with a 10 pixel aperture to isolate the flux. Flux extraction is performed using two techniques: optimal extraction and classic aperture photometry with a 5 pixel radius aperture. For optimal extraction, the PSF is modelled by fitting a 2D Gaussian to the background-corrected cleaned median frame. The normalized PSF (scaled to unity) is used as weights to extract the flux within the 10 pixel circular mask. The uncertainties in the flux are estimated through error propagation.  We apply the same reduction process to all three observations.
    
    \subsubsection{Secondary reduction analysis}
    \label{sec:secondary_fitting}
    
    We use \texttt{juliet} \citep{Espinoza_2019} to fit the extracted light curves from this secondary reduction. The three observations are jointly fitted using five detrending methods: a linear slope in time (\texttt{L}), a single exponential (\texttt{E}), a linear slope with a single exponential (\texttt{L+E}), a linear slope with a GP (\texttt{L+GP}), and a linear slope with a single exponential and a GP (\texttt{L+E+GP}). We use the Matérn 3/2 kernel, which is described by:
    \begin{equation}
        k_\mathrm{sec.}(t, t') = \xi^2\left(1 + \sqrt{3} \frac{|t - t'|}{\rho}\right) \exp{\left(-\sqrt{3}\frac{|t - t'|}{\rho}\right)} + \sigma^2\delta_{tt'},
    \end{equation}
    with $\xi$ the height scale of systematics, $\rho$ the length scale in time and $\sigma$ the amplitude of white noise.
    
    We trim the first 200 integrations of each observation, corresponding to 37 minutes. Each detrending model is applied independently to each observation, while the eclipse depth and orbital parameters are jointly fitted. Table\,\ref{tab:priors:juliet} presents the priors and distributions of the planetary parameters included in the joint fit. We provide details on the parameters of the detrending methods and their corresponding priors.

    \subsection{Eclipse depth results}
    \subsubsection{Results from primary reduction analysis}
    \label{sec:eclipse_depths}

    To joint-fit the three eclipse datasets only the eclipse model parameters ($T_0$, $P$, $a/R_*$, $R_p/R_*$, $b$, $f_p/f_*$, $\sqrt{e}\cos\omega$, $\sqrt{e}\sin\omega$) were shared between datasets for all methods, with all systematics and PSF parameters being separate for each dataset. For all joint-fits we cut the first 45 minutes to remove the strongest settling effects and fit the remaining settling with the exponential ramp model.
    
    The results from joint-fitting the light curves are shown in Table~\ref{tab:wn_GP_joint_fits}. While the results are all highly consistent, the pixel-fitting method tends to be a bit more conservative with larger uncertainties - which is similar to the simulations where there were only flux-conserving systematics present (i.e.  FCS scenario; see Section~\ref{sec:sim1}). We note that pixel-fitting while only accounting for flux-conserving systematics performs almost identically to aperture extraction without a GP, so any differences in the size of uncertainties are likely arising purely from the constraints on systematics which are not flux-conserved.
    
    \begin{table}
        \caption{Eclipse depths from the joint-fits of all three eclipses from the primary reduction analysis.}
        \centering
        \begin{tabular}{ll}
                \hline
                Method & Depth (ppm)\\
                \hline
            Ap. Ext. w/o GP & $272\pm40$\\
            Px-fit (only FCS) & $262\pm40$\\
            Ap. Ext. w/ GP & $273\pm43$\\
            Px-fit (common systematics) & $257\pm45$\\
            Px-fit (common+indep. sys.) & $253\pm55$\\
            Px-fit (common+shared indep. sys.) & $253\pm49$\\
                \hline
        \end{tabular}
        \label{tab:wn_GP_joint_fits}
    \end{table}
    
    \begin{table}
    \caption{Comparison of individual eclipse depths using aperture extraction from the primary reduction analysis.}
    \centering
    \begin{tabular}{lccccc}
    \toprule
    Model & $t_{cut}$ & E1    & E2 & E3 & Comb.   \\ \midrule
    \multirow{3}{*}{\texttt{L}}
    & 30 & 499$\pm$53 & 287$\pm$51 & 313$\pm$52 & 364$\pm$30 \\
& 45 & 446$\pm$53 & 273$\pm$52 & 316$\pm$52 & 344$\pm$30 \\
& 60 & 412$\pm$56 & 248$\pm$53 & 309$\pm$55 & 320$\pm$32 \\ \cmidrule(l){1-6}
    \multirow{3}{*}{\texttt{L+E}}
    & 30 & 305$\pm$71 & 235$\pm$65 & 291$\pm$61 & 276$\pm$38 \\
& 45 & 327$\pm$76 & 211$\pm$66 & 278$\pm$71 & 267$\pm$41 \\
& 60 & 324$\pm$76 & 228$\pm$76 & 244$\pm$80 & 266$\pm$45 \\ \cmidrule(l){1-6}
    \multirow{3}{*}{\texttt{L+GP}}
    & 30 & 389$\pm$126 & 284$\pm$64 & 312$\pm$60 & 309$\pm$41 \\
& 45 & 421$\pm$85 & 272$\pm$62 & 313$\pm$59 & 319$\pm$38 \\
& 60 & 404$\pm$70 & 249$\pm$60 & 307$\pm$62 & 312$\pm$37 \\ \cmidrule(l){1-6}
    \multirow{3}{*}{\texttt{L+E+GP}}
    & 30 & 306$\pm$79 & 244$\pm$75 & 288$\pm$69 & 279$\pm$43 \\
& 45 & 327$\pm$82 & 215$\pm$75 & 272$\pm$79 & 268$\pm$45 \\
& 60 & 321$\pm$85 & 226$\pm$80 & 243$\pm$88 & 262$\pm$49 \\
 \cmidrule(l){1-6}
    \end{tabular}
    \tablefoot{Each model included some combination of a linear slope (L), exponential ramp (E) and/or a GP. Initial time cut is denoted in minutes.}
    \label{tab:individual_eclipses}
    \end{table}

    We also tested fitting for the eclipse depth individually for each dataset. In this case, we assumed a circular orbit and used constraints on the central eclipse time $T_\mathrm{sec}$ from the other two eclipses as a prior on each individual fit as each eclipse alone gave poor constraints on the eclipse time. These priors are listed in Table~\ref{tab:T0_constraints}. We experimented with varying the initial time being cut ($t_\mathrm{cut}$) and the effect of fitting for an exponential ramp or using a GP. Due to the computational cost of the pixel-fitting method, we only tested these variations with aperture extraction. The results are presented in Table~\ref{tab:individual_eclipses} showing the eclipse depths individually for each eclipse as well as the combined result from the three eclipses calculated using error propagation (i.e. not a joint-fit).
    
    The individual eclipse depths show that fitting for an exponential ramp has a large effect on the result. Given the strong settling ramp seen at the beginning of the first eclipse (see Figure~\ref{fig:three_eclipses}), the eclipse depth may be biased if this ramp is not accounted for - even after cutting a full hour at the start. This would explain why the linear model without a GP (denoted `L' in the Model column) recovers very deep eclipses (>400ppm) inconsistent with the other two eclipses. The linear model with a GP (L+GP) performs slightly differently and recovers large uncertainties likely because the GP height scale is increasing to fit this initial settling ramp. As initial settling effects are not in line with a stationary Gaussian process, we would prefer not to use a GP to fit them and so our preferred choice of model is the linear with exponential ramp and a GP (L+E+GP) model. This performs very similarly to the linear with exponential ramp without a GP (L+E) model but including a GP is slightly more conservative as it can account for additional systematics which could potentially lie in these data. We see that both these models give eclipse depths that are all within $2\sigma$ of each other and varying the time cut between 30-60 minutes has minimal impact on the result.

 \subsubsection{Results from secondary reduction analysis}
 
The secondary analysis, conducted using an independent pipeline and the \texttt{juliet} package to fit the light curve, yields consistent results. Table\,\ref{tab:eclise_depth_sec} presents the eclipse depth results for various detrending methods in the combined fit. The combined eclipse depths remain consistent within 1$\sigma$ of those from the primary analysis, which used classic aperture photometry or pixel-fitting, except for the \texttt{L} and \texttt{L+E} models in the secondary analysis. However, these models are strongly disfavored, compared to the other models. The use of GPs inflates the uncertainty. 

 \begin{table}
    \caption{Comparison of jointly fitted eclipse depths for various methods from the secondary analysis.}
    \centering
    \begin{tabular}{lccc}
    \toprule
    Model & Extraction &  Comb. depth (ppm) &  ln$\mathcal{Z}$ \\ \midrule
        \multirow{2}{*}{\texttt{L}} & optimal & 423$\pm$32  & 17995 \\
         & classic& 283$\pm$33 & 17849 \\ \hline
        \multirow{2}{*}{\texttt{E}}  & optimal & 257$\pm$33 & 18048 \\
        & classic & 258$\pm$35 & 17869 \\ \hline
       \multirow{2}{*}{\texttt{L+E}} & optimal &  320$\pm$40 & 18012 \\
        & classic & 211$\pm$37 & 17845\\ \hline
        \multirow{2}{*}{\texttt{L+GP}} & optimal & 254$\pm$53  & 18026 \\
         & classic & 257$\pm$45 & 17841 \\ \hline
        \multirow{2}{*}{\texttt{L+E+GP}}& optimal &  235$\pm$70 & 18016 \\
        & classic  & 268$\pm$51& 17828\\ \hline 
    \end{tabular}
    \tablefoot{200 integrations are cut at the beginning of each observations -- equivalent to 37 minutes.}
    \label{tab:eclise_depth_sec}
    \end{table}

    \subsection{Including the masked cosmic ray pixels}
    
    \begin{figure}
        \includegraphics[width=\columnwidth]{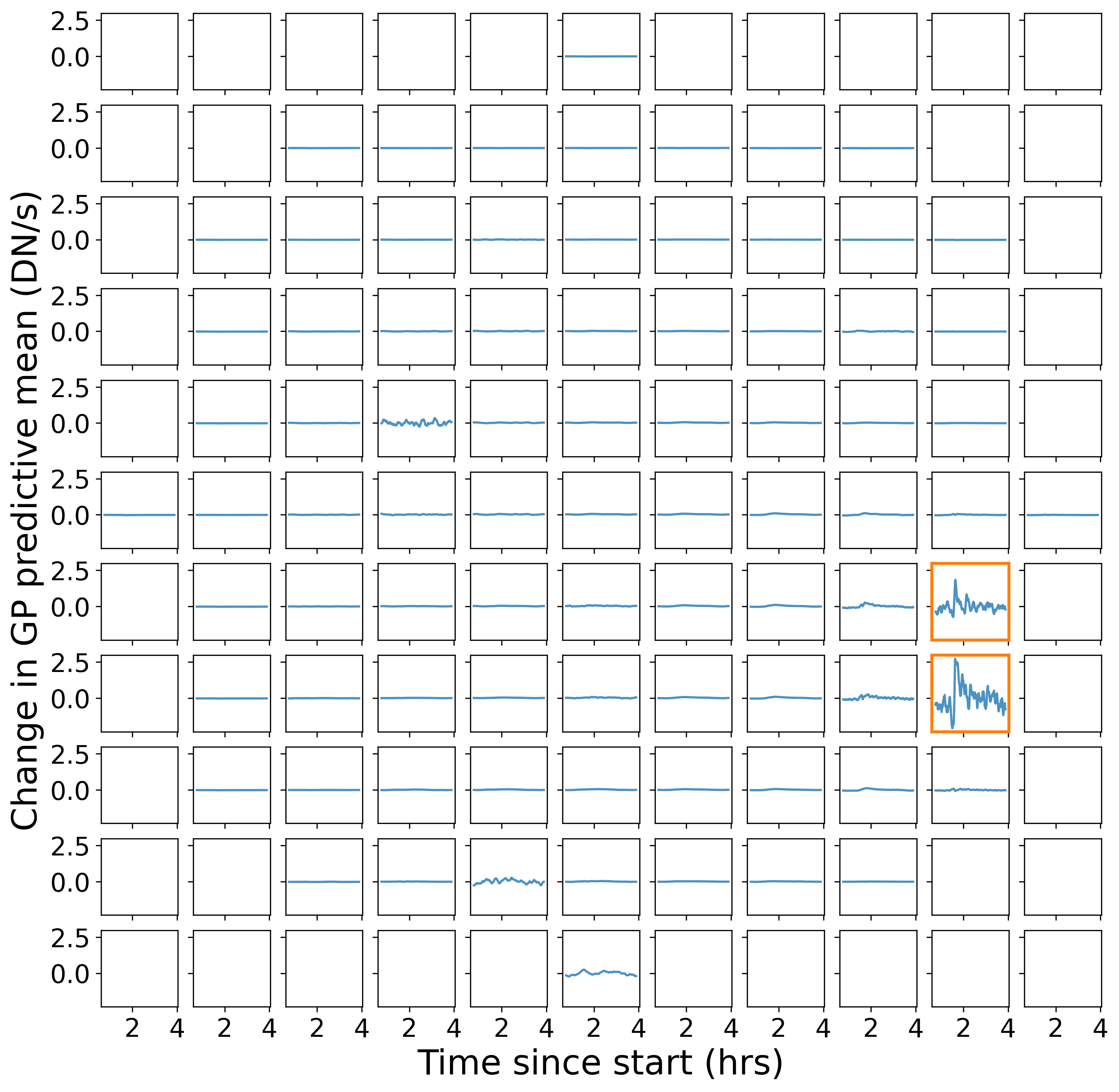}
        \caption{Difference in GP predictive mean from pixel-fitting the second eclipse due to fitting for independent pixel systematics. Pixels affected by the strong cosmic ray persistence effect from Figure~\ref{fig:real_px_lightcurves} are highlighted in orange. By fitting for independent pixel systematics the shape of this persistence effect is modelled by the GP and these pixels are also weighted away from.}
        \label{fig:cosmic_ray_visualised}
    \end{figure}
    
    \begin{table}
        \caption{Eclipse depths for the second eclipse either including or excluding the pixels containing a strong cosmic ray jump.}
        \centering
        \begin{tabular}{lll}
                \hline
                Cosmic ray pixels: & Excluded & Included\\
                \hline
            Ap. Ext. w/o GP & $211\pm66$ & $181\pm61$\\
            Px-fit (only FCS) & $195\pm73$ & $185\pm64$\\
            Ap. Ext. w/ GP & $215\pm75$ & $189\pm70$\\
            Px-fit (common systematics) & $194\pm80$ & $188\pm74$\\
            Px-fit (common+indep. sys.) & $208\pm99$ & $214\pm99$\\
            Px-fit (common+shared indep. sys.) & $204\pm85$ & $203\pm87$\\
                \hline
        \end{tabular}
        \label{tab:cosmic_ray}
    \end{table}
    
    One interesting test case for the pixel-fitting method is to include the pixels which appear to suffer a large persistence effect from a cosmic ray jump in the second eclipse (highlighted in Figure~\ref{fig:real_px_lightcurves}). These pixels were masked out of all pixel-fits and aperture extraction fits for the primary reduction analysis. The eclipse depths recovered including or excluding these pixels are given in Table~\ref{tab:cosmic_ray} for aperture extraction and pixel-fitting methods. We can see that including these pixels improves the constraints on the eclipse depth for most of the methods even though these pixels do not appear to contain clean data, and the aperture extraction methods also get their mean values shifted down by 25-30 ppm. However, the pixel-fitting methods which account for systematics in individual pixels (labelled `common+indep. sys' and `common+shared indep. sys.') are not affected by the inclusion of these pixels. This could be analogous to the simulations in the IPS scenario (Section~\ref{sec:sim2}), where the pixel-fitting method would weight away from pixels containing independent pixel systematics and by doing so the eclipse depth was recovered more precisely.

    We also experimented with using the pixel-fitting GP predictive mean to visualise the systematics in individual pixels. We calculated the GP predictive mean fitting the pixel light curves both accounting for independent pixel systematics or not accounting for them (i.e. setting all $h_\mathrm{IPS; (i, j)} = 0$). Getting the difference between these two GP means gives an idea of how the inclusion of accounting for independent pixel systematics changes the fit to the pixel light curves, potentially revealing some of the shape of these systematics being fit. This is shown in Figure~\ref{fig:cosmic_ray_visualised} and the pixels featuring the cosmic ray jump are highlighted.

    \subsection{Variations in detector settling}
    \label{sec:settling}

    We investigated potential explanations for the variations in detector settling seen at the start of our three LHS 1140c observations (shown in Figure~\ref{fig:three_eclipses}). It was noticed that the previous MIRI observation for the first eclipse was a calibration observation which was taking darks for MIRI/Imaging. This observation also shows a strong negative slope very different from the start of the other two eclipses. We note that this calibration observation ended >5 hours before the start of the first LHS 1140 observation (the last observation of any kind was carried out using NIRSpec/IFU). However, MIRI  typically leaves the previous filter it used in position until almost immediately before the next MIRI observation begins. In this case, the OPAQUE filter was kept in place until switching to the F1500W filter, leaving a long time for each pixel to settle without receiving flux. For the other two eclipses, the F1000W and F770W filters were left in place until right before the observation.
    
\begin{figure}
        \includegraphics[width=\columnwidth]{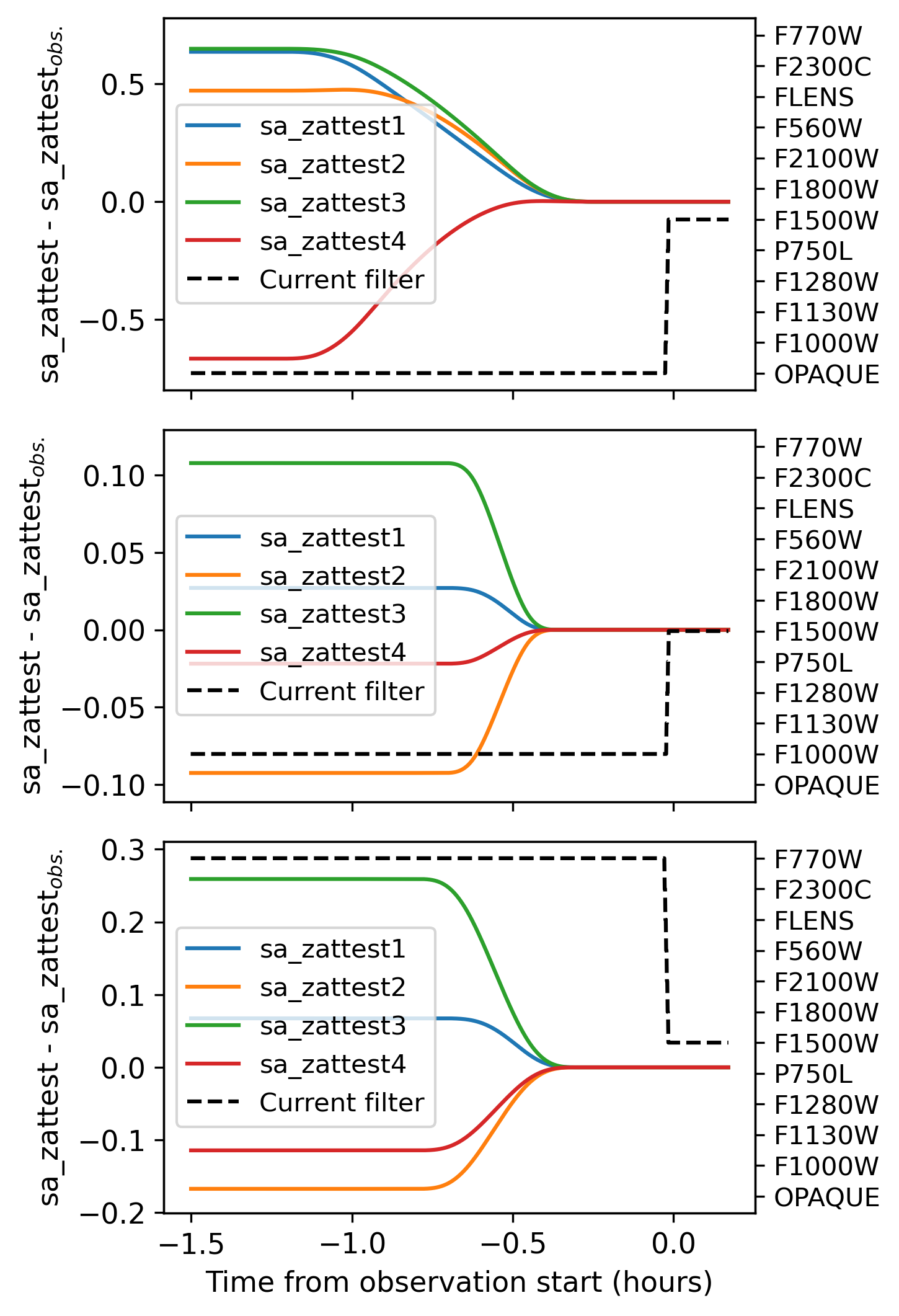}
        \caption{JWST mnemonics which track telescope orientation (\texttt{sa\_zattest}) overlaid with the mnemonic tracking current MIRI filter position. Left y-axis shows difference in telescope pointing parameters from our observations. Filter wheel position is plotted as black dashed line and corresponds to right y-axis, including any intermediate filters rotated through. In all three eclipses, coarse slewing towards the target finishes $\sim\!15$ minutes before the observation starts, leaving time for persistence effects from the last filter used.}
        \label{fig:mnemonics}
    \end{figure}

    We note that there is no shutter on MIRI, as the detector is continuously observing the sky and constantly resetting after each read; this means that the telescope will spend time observing LHS 1140 with the previous filter in the time after slewing to the target but before switching filter. It has already been noted that this change in filters can lead to persistence effects from different amounts of background flux in each filter \citep{Dicken2024}. However, the PSF of LHS 1140 observed through the different wavelength filters may also have sufficient time to produce persistence effects and affect the beginning of our observations. Examining this using the JWST engineering mnemonics for telescope orientation and MIRI filter position appears to support this possibility, with the telescope pointing in a similar direction to the observation period for about 10-15 minutes before the filter is switched to F1500W (see Figure~\ref{fig:mnemonics}).

\begin{figure}
        \includegraphics[width=\columnwidth]{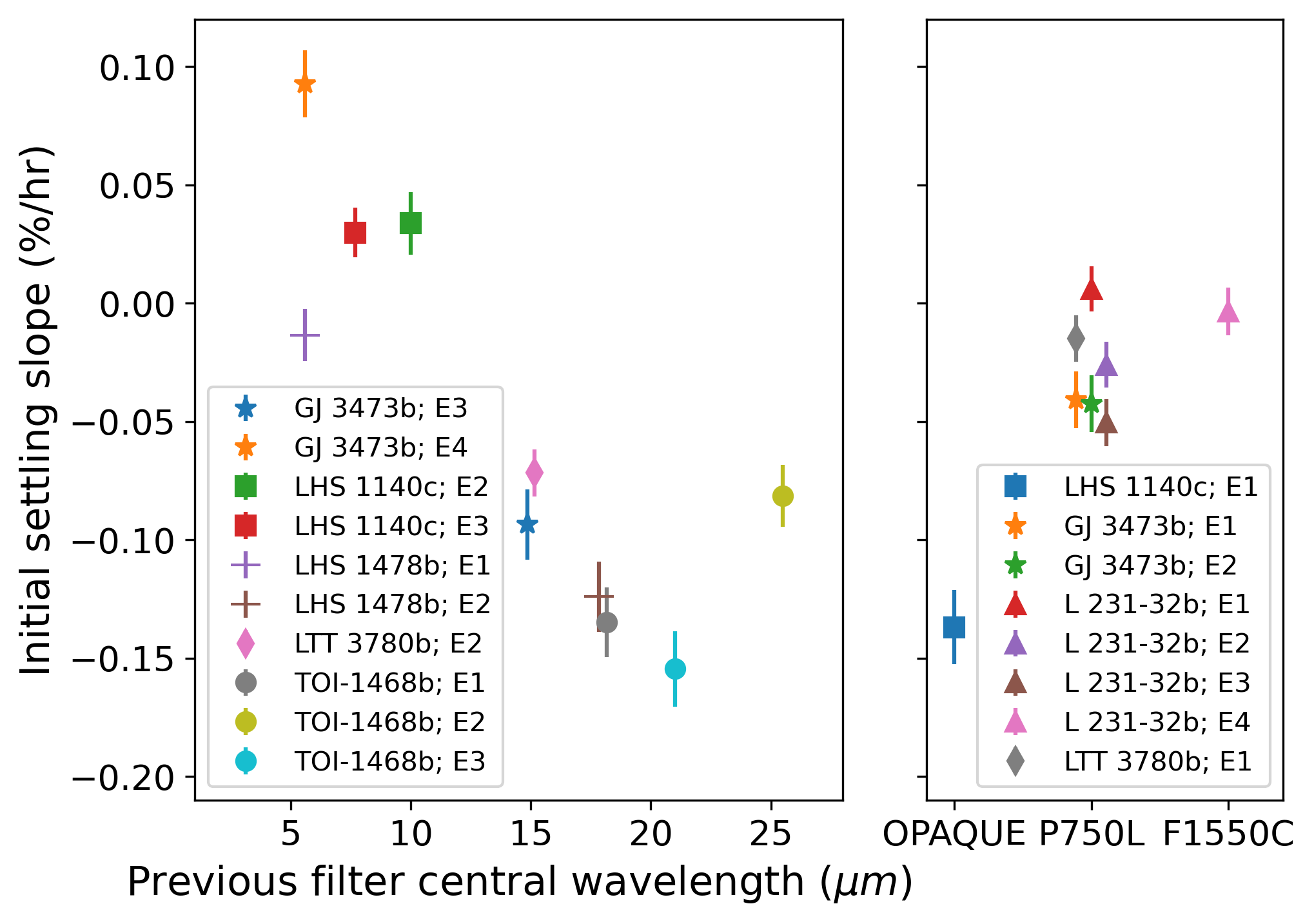}
        \caption{Possible trend in detector settling based on the previous filter used by MIRI. The mean and uncertainty in the slope of the first 30 minutes of the aperture extracted light curve is plotted for various Hot Rocks observations. Left plot shows previous imaging filters based on their central wavelength, the right plot includes the OPAQUE position for darks, the P750L prism for LRS and a coronagraphic filter. Observations which shared the same previous filters often display similar detector settling.}
        \label{fig:settling_by_filter}
    \end{figure}

    To test this idea, each of the observations in the Hot Rocks survey which used the SUB256 subarray were reduced in the same manner as described in Section~\ref{sec:reduction}. The four observations in the survey which used either the SUB64 or SUB128 subarrays would require a different background subtraction and occur on completely separate pixels on the detector so we restricted our comparison to just the SUB256 observations. After we had performed the same background subtraction and 5px radius aperture extraction as for the LHS 1140c eclipses, we took the first 30 minutes of each observation (clipping the first integration to avoid the `first exposure' effect) and fit a best-fit line to it. We then compare the slope of the first 30 minutes of each observation to the previous filter used. The previous filter used was found with the \texttt{IMIR\_HK\_FW\_CUR\_POS} engineering mnemonic, which provides the MIRI filter currently in position. In all cases, the observations are being switched to the F1500W filter. See Figure~\ref{fig:settling_by_filter} for a plot of these results.
    
    \begin{figure}
        \includegraphics[width=\columnwidth]{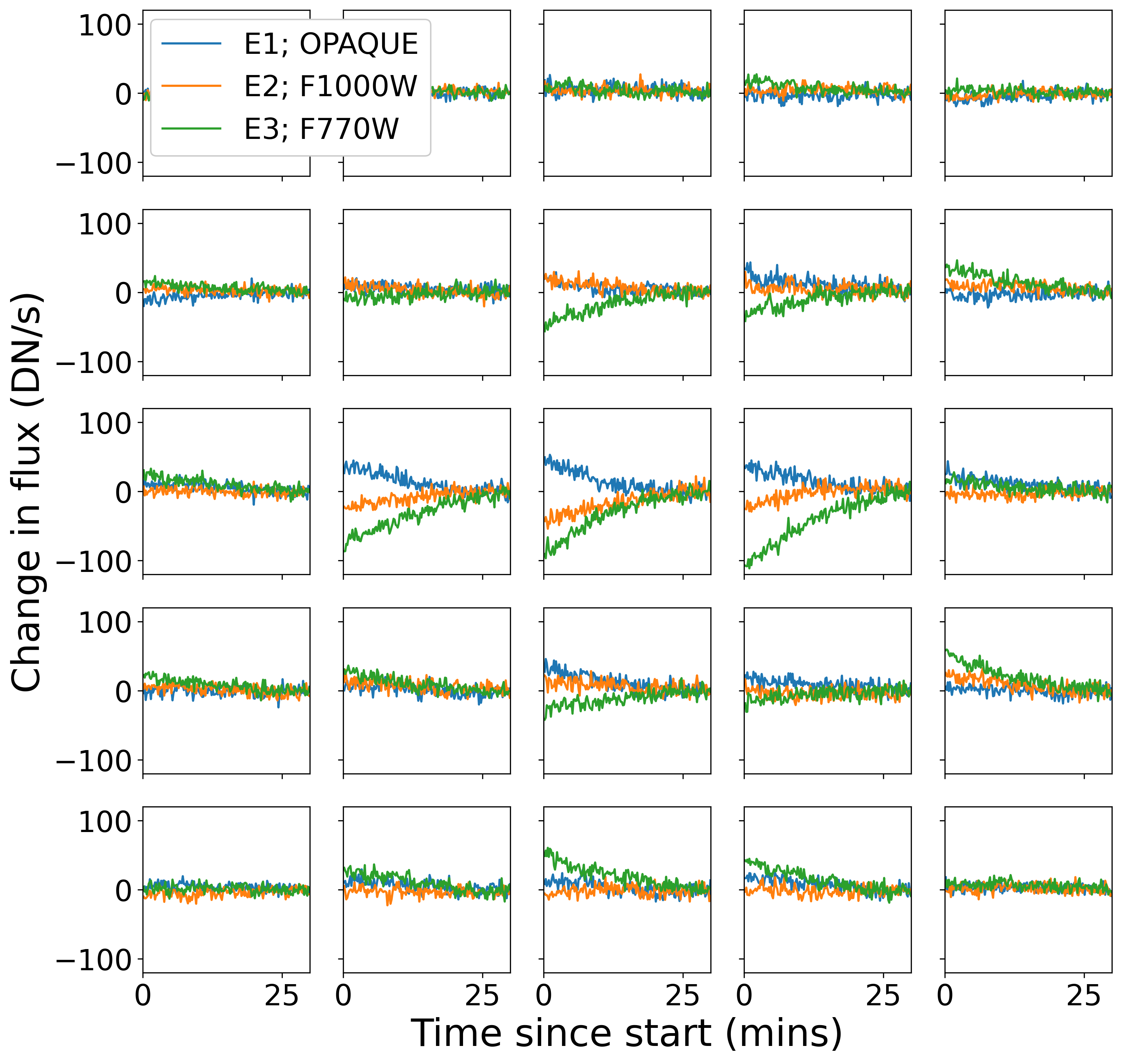}
        \caption{Central pixel light curves for the first 30 minutes of each LHS 1140c eclipse. Each plot shows the change in flux relative to the last ten integrations of the first 30 minutes. The legend specifies which previous filter was in place for each eclipse. Note the strong differences in the central pixel light curves, particularly between the first eclipse (which switched from the OPAQUE filter) and the third eclipse (which switched from the F770W filter).}
        \label{fig:LHS_initial}
    \end{figure}

    There appears to be a pattern where the observations which previously used short wavelength filters (i.e. F560W, F770W, F1000W) show flat or positive slopes at the start, while longer wavelength filters up to the F2000W filter show increasingly negative slopes. The F2550W filter doesn't quite match this trend but still shows a negative slope. The OPAQUE filter performs similarly to the long wavelength filters with a strong negative slope, while observations which previously had the P750L prism used for MIRI/LRS have slightly negative or flat slopes. The fourth eclipse of L 231-32b was the only observation which previously used a coronagraphic filter (F1550C), with this observation being consistent with a flat slope. We note that observations which both used the same previous filter tend to have quite consistent slopes, even when the targets being observed are different. This suggests that it may be possible to improve the consistency of the initial detector settling by switching filter earlier (e.g. before slewing or before guide star acquisition), which might enable better characterisation of detector settling and allow us to clip less data.

    However, the two observations which previously used the F560W filter show inconsistent settling, suggesting that there may be other important factors. For example, examining engineering mnemonics for these observations (not shown but similar to those for LHS 1140c in Figure~\ref{fig:mnemonics}) suggests that they differ significantly in terms of the time spent on target with the previous filter. The fourth eclipse of GJ 3473b spends $\sim$45 minutes on target with the F560W filter while the first eclipse of LHS 1478b only spends $\sim$10 minutes. This could explain the strong positive slope seen for the fourth eclipse of GJ 3473b, as it has more time to build up a significant persistence effect with the previous filter compared to the first eclipse of LHS-1478b.
    
    It is worth noting that if we avoid performing any background subtraction we recover similar slopes in the aperture extracted light curves. Most settling slopes are shifted by less than 0.01\%/hr if a background subtraction is not performed, with the most affected observation being the first LHS 1140c eclipse which is shifted 0.035\%/hr flatter (i.e. $-0.137\%/$hr to $-0.102\%$/hr). This implies that any trend is primarily driven by the strong settling effects seen in the central pixels of the PSF and not related to persistence effects in the background pixels. The strong settling of the central pixels is visualised in Figure~\ref{fig:LHS_initial}, which shows the pixel light curves centred on the PSF for the initial 30 minutes of each LHS 1140 observation. Consider that when the F770W filter is in position prior to the third observation, this filter should produce a $1.8$ times narrower PSF which has about seven times greater total flux compared to the F1500W filter (based on the filter bandpasses and BT-Settl models). This combination of a narrower PSF and a bandpass which collects more flux could likely explain the significant differences in the persistence effects seen on the central pixels compared to an observation which previously had the OPAQUE filter in and hence was exposed to very little light in the time immediately before the observation began. One caveat is that there is some time required to switch filter ($\approx\!8$s per filter position; \citealt{Wright_2015}), so the filters which are in between the previous filter and the F1500W filter could also introduce some persistence effects, although there is significantly less time for this to occur.

    \subsection{Pixel-fitting parameters recovered}
    \label{sec:detector_params}

\begin{figure*}
        \includegraphics[width=\textwidth]{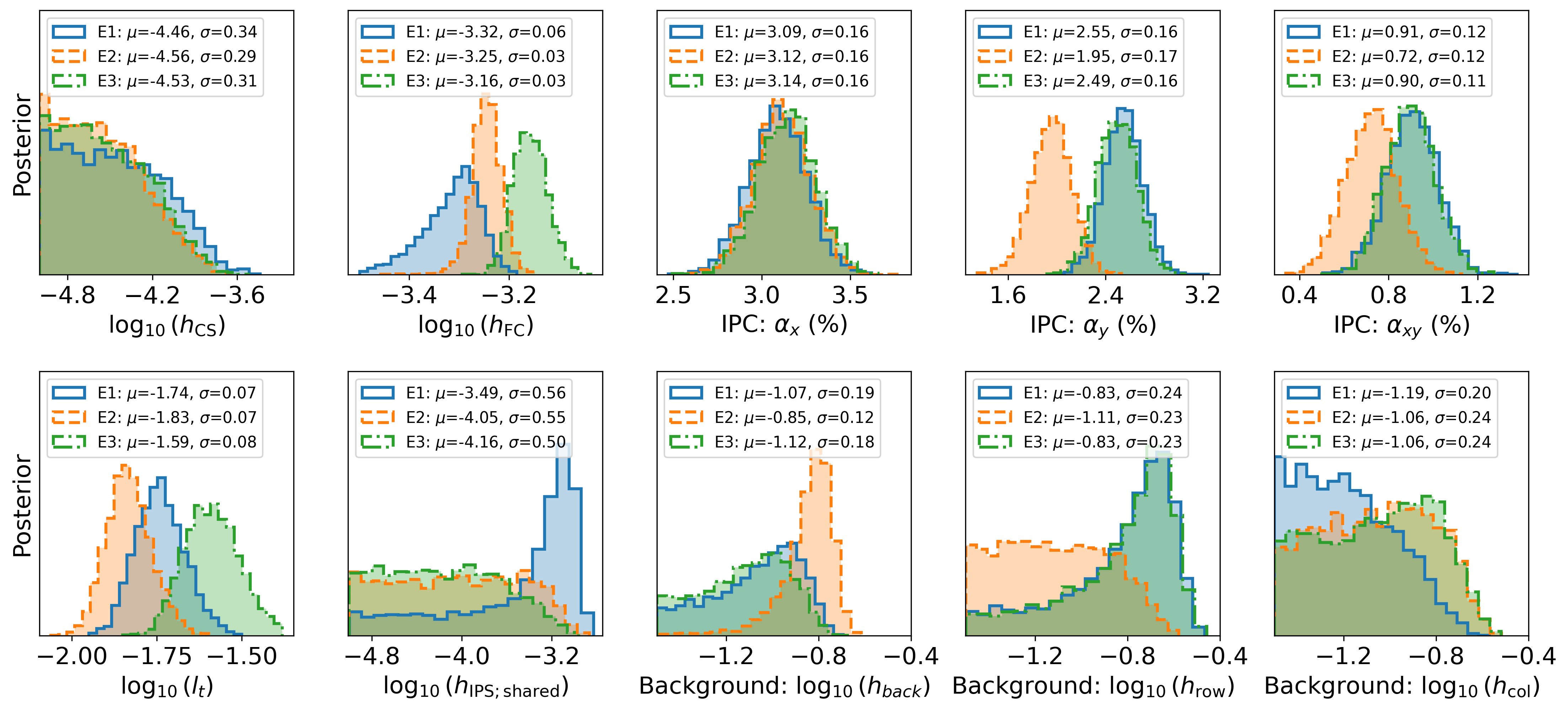}
        \caption{Marginal posterior distributions of various pixel-fitting systematics parameters for each eclipse. Values from joint-fit of three eclipses with independent systematics parameters for each eclipse. Common systematics and correlated background scatter were not strongly detected in any eclipse and were consistent with lower bounds. Flux-conserved systematics, interpixel capacitance and the time length scale of systematics show consistent values across each observation.}
        \label{fig:detector_param}
    \end{figure*}

    In addition to recovering the eclipse depth, the pixel-fitting method also recovers hyperparameters which quantify the level of different detector systematics. These hyperparameters have been described in Section~\ref{sec:kernel choice} as parameters of the kernel function, used to describe the correlation between different noise processes in the data. We have plotted the posteriors of some of these parameters in Figure~\ref{fig:detector_param}, all of which come from the joint-fit of the three eclipses using the pixel-fitting method which was accounting for both common systematics and independent systematics with a partially shared height scale. As stated earlier, while these datasets were being joint-fit, only the eclipse model parameters were being shared between datasets and the systematics parameters were independent for each dataset.
    
    It is notable that the systematics parameters are all consistent between each dataset. None of the datasets constrain the presence of common systematics (with height scale $h_\mathrm{CS}$) or found strong evidence of independent pixel systematics from the shared heightscale ($h_\mathrm{IPS; shared}$). The amplitude of the flux-conserving systematics ($h_\mathrm{FCS}$) -- which could be interpreted as the amplitude of the PSF changing shape over time -- appears to be well constrained. The correlations in noise arising from interpixel capacitance in the horizontal ($\alpha_x$), vertical ($\alpha_y$), and diagonal ($\alpha_{xy}$) directions are also tightly constrained and greater than zero. Combining these constraints through error propagation we get $\alpha_x = 3.12\pm0.09$\%, $\alpha_y = 2.35\pm0.09$\%, and $\alpha_{xy} = 0.84\pm0.07$\% - consistent with expected values \citep{Morrison_2023}.
    
    The length scale in time of the systematics, $l_t$, is also well constrained. However, there was a prior placed on each $l_t$ parameter taken from the posterior of the joint-fit accounting for common systematics but not independent pixel systematics. This was because the addition of independent pixel systematics could create a degeneracy between very gradual systematics in each pixel and the slope in each pixel being fit for in the mean function. As a result, the constraints on $l_t$ are more closely related to the timescale of the flux-conserving systematics and so could represent the rate at which the PSF is changing shape (in units of days).

    Finally, the parameters describing correlations in background scatter do not definitively constrain correlations in background noise either across all pixels ($h_\mathrm{back}$) or along the rows ($h_\mathrm{row}$) or columns ($h_\mathrm{col}$). The inclusion of fitting for correlations in noise along each row and column was largely a conservative choice as there are a number of row/column effects seen with MIRI and the lack of a detection does not rule these systematics out but instead puts an upper bound on their presence in these data. However, the background subtraction is not perfect so $h_\mathrm{back}$ almost certainly should be non-zero and all three eclipses do appear to favour the value being at least somewhat larger than the minimum prior bound of $h_\mathrm{back} = 10^{-1.5} \approx 0.032$ DN/s. By including more pixels or if our datasets had more integrations, this might provide the method more information to constrain the remaining level of background scatter in the data. This could be interesting as a way to compare different background subtraction strategies and identify which method best minimises the background scatter.

    \subsection{Absolute flux calibration}
    \label{sec:flux_cal}

    As our eclipse depth measurements correspond to the planet-to-star flux ratio, $f_p/f_*$, our measurement of the dayside planetary flux $f_p$ is sensitive to our models of flux from the host star across the F1500W filter bandpass. In addition, stellar models are required to determine the amount of radiation the dayside is exposed to, which will determine the expected flux emitted for each model. Inaccuracies in our stellar models may therefore affect our eclipse spectra models in two different ways. While we do not have spectra to examine the accuracy of these models across near to mid-infrared, we can use our observations to perform an absolute flux calibration and examine how closely the BT-Settl models match the observed flux across the F1500W bandpass.
    
    We performed the absolute flux calibration following the procedure used for the JWST Absolute Flux Calibration program \citep{flux_cal}. This involved running the standard JWST calibration pipeline but skipping the EMI correction step \textsc{emicorr} and turning off the after-jump flagging and shower detection steps of the \textsc{jump} step. We skipped the \textsc{photom} step in stage 2 and instead used the values given in \citet{flux_cal} to convert from DN/s to millijanskys (mJy). \textsc{photutils} was used for the centroiding and aperture extraction, with an aperture radius of $r = 5.69$px and background annulus radii of 8.63px and 11.45px - matching the radii used for the program. A conversion factor is used to convert the aperture extracted flux using a finite aperture to the total flux expected from an `infinite' aperture and this factor also accounts for the oversubtraction of background flux due to some of the PSF being included within the background annulus.
    
    We performed the flux calibration for each integration of our observations and calculated the mean flux. This resulted in flux values of $9.17\pm0.04$ mJy, $9.19\pm0.04$ mJy, and $9.17\pm0.04$ mJy for each eclipse respectively, with the uncertainties determined using the reported uncertainty in the calibration factor of $0.48\%$ (the uncertainties due to Poisson noise are negligible in comparison). In comparison, when we calculate the expected flux weighted across the F1500W bandpass using a grid of BT-Settl (CIFIST) models \citep{2011ASPC..448...91A}, we find an expected flux of $8.33\pm0.32$ mJy. The uncertainty in this value propagates uncertainties in the stellar distance $d = 14.9861\pm0.0153$ pc from Gaia DR3 \citep{gaia} and in the stellar radius, $R_*$, stellar effective temperature, $T_\mathrm{eff}$, and surface gravity, $\log(g)$, from \citet{2024ApJ...960L...3C}.
    
    This makes the measured flux $\sim\!2.6\sigma$ larger than the flux expected from the BT-Settl models after marginalising over $T_\mathrm{eff}$ and $\log(g)$ - or $(10.3\pm3.2)\%$ greater than the interpolated BT Settl model with $T_\mathrm{eff}$ and $\log(g)$ fixed to mean values\footnote{One caveat is that as we are scaling the BT-Settl models based on the stellar radius and system distance, it is possible that the BT-Settl models are accurate but for example the stellar radius value from \citet{2024ApJ...960L...3C} is inaccurate.}. If this increased stellar flux is real and only deviates from the BT-Settl models in a region around the F1500W bandpass, then we may expect this to roughly shift our bare rock and atmospheric eclipse models to be $\sim\!10\%$ shallower by affecting the $f_p/f_*$ ratio. In principle, this should not dramatically affect our conclusions given our $\sim\!16\%$ uncertainty in the measured eclipse depth (and in fact may provide stronger evidence on the lack of an atmosphere). However, this does not account for how stellar model inaccuracies may affect the radiative transfer and energy balance equations or how the BT-Settl models may deviate from the stellar spectrum at other wavelengths. We must therefore advise some caution in the interpretation of our eclipse models. We expect that future observations of M-dwarf spectra at near to mid-infrared will help us improve the accuracy of these models as there are currently significant challenges facing the modeling their spectra \citep{2016PhR...663....1S}.

    Our result of $\sim\!10\%$ increased flux is in contrast with the flux calibration results for TRAPPIST-1, as \citet{2023Natur.618...39G} reported a $13\%$ reduced flux compared to PHOENIX models, while \citet{Ducrot2024} and \citet{Ih2023} report only $\sim7\%$ reduced flux - both using SPHINX models. On the other hand, \citet{2023Natur.620..746Z} report an expected flux using SPHINX models which is consistent within $\sim\!1\%$ of the absolute flux calibrations. These works differ significantly in their methods of estimating flux from stellar models and in their choice of stellar parameters and stellar models used. However, all absolute flux calibrations performed were consistent within $\sim\!1\%$, although their methods differ from \citealt{flux_cal} and this work. We tested our own expected flux for TRAPPIST-1 using BT-Settl models and using the stellar parameters from \citet{Agol2021, gaia} and we obtained a flux of 2.507±0.077 mJy, which similar to the \citet{2023Natur.620..746Z} result is within 1-3\% of all reported absolute flux calibrations. Therefore, our flux estimates from BT-Settl models are consistent with the TRAPPIST-1 flux calibrations yet differ from our LHS 1140 flux calibrations.
    
    We may note that the calibration factor from the program was determined using much shorter duration observations often with only one or two integrations. Therefore these observations may have been affected by the reduced flux often seen in the first integration of each observation \citep{Morrison_2023}. If we calculate the flux using just the first integration of each eclipse, we do get reduced flux values of $9.10\pm0.04$ mJy, $9.09\pm0.04$ mJy and $9.09\pm0.04$ mJy for each eclipse, however these are still $\sim 2.3\sigma$ deeper than expected so this makes little difference.

    The absolute flux value can be used in combination with the eclipse depth to calculate the brightness temperature of LHS 1140c. This is performed by multiplying the eclipse depth, absolute flux value and the square of the ratio of the distance the system is away to the radius of the planet. This is compared to integrating Planck's law over the F1500W bandpass and finding a brightness temperature $T_B$ which corresponds to this level of emission. We performed this by calculating the emission from Planck's law over a range of brightness temperatures between 200K and 800K in 0.1K intervals and interpolating to invert emission to brightness temperature. Marginalising over the uncertainty in the eclipse depth from aperture extraction with a GP ($d = 273\pm43$ ppm) as well as the system distance and planetary radius, we recover a brightness temperature of $T_B = 561\pm44$K. This is close to the maximum brightness temperature we might expect from a (smooth) zero albedo bare rock with no heat redistibution of $T_{B,\mathrm{ max}} = 537\pm9$ K, while full heat redistibution even with zero albedo would be inconsistent with this value at only $T_{B,\mathrm{ full}} = 421\pm7$ K (and higher albedos would be even colder). However, this is only a simple calculation which does not take into account effects such as the greenhouse effect, scattering, or thermal beaming (discussed in Section~\ref{sec:5}). This calculation does have the benefit of not requiring the use of (potentially unreliable) stellar models while the atmospheric modelling described in Section~\ref{sec:5} requires the use of stellar models to account for the wavelength-dependent effects of absorption and emission as well as scattering.

\section{Modelling}
\label{sec:5}

    For the no atmosphere case, we can calculate the dayside temperature by balancing thermal emission and absorbed stellar radiation:
    \begin{equation}
        \epsilon T_\mathrm{day}^4 = T_*^4 \left(\frac{R_*}{a}\right)^2(1 - A_\mathrm{surf}) (f/4)
    \label{eq:no_atmo}
    \end{equation}
    where $T_*$ is the stellar effective temperature, $R_*$ is the stellar radius, $a$ is the semi-major axis, $A_\mathrm{surf}$ is the surface albedo (i.e. the Bond albedo of the surface), $\epsilon$ is the surface emissivity, and $f$ is the heat redistribution factor, where $f = 1/4$ represents full heat redistribution, while $f = 2/3$ represents zero heat redistribution \citep{Hansen2008}. This equation is adapted from \citet{heliossurfa} and assumes there is no extra heat emitted due to a high interior planetary temperature. For the bare rock case we set $f = 2/3$ as we expect insignificant heat redistribution without an atmosphere \citep{Joshi1997, Selsis2011, Kollf}. For all simulations, we assumed that the surface albedo was constant in wavelength, which implies $\epsilon = 1 - A_\mathrm{surf}$ (as emission and absorption can occur at different wavelengths this is not necessarily true otherwise). Emission was then calculated using Planck's law with the dayside temperature $T_\mathrm{day}$ and scaling by the emissivity $\epsilon$.
    
    \begin{figure*}
        \includegraphics[width=\textwidth]{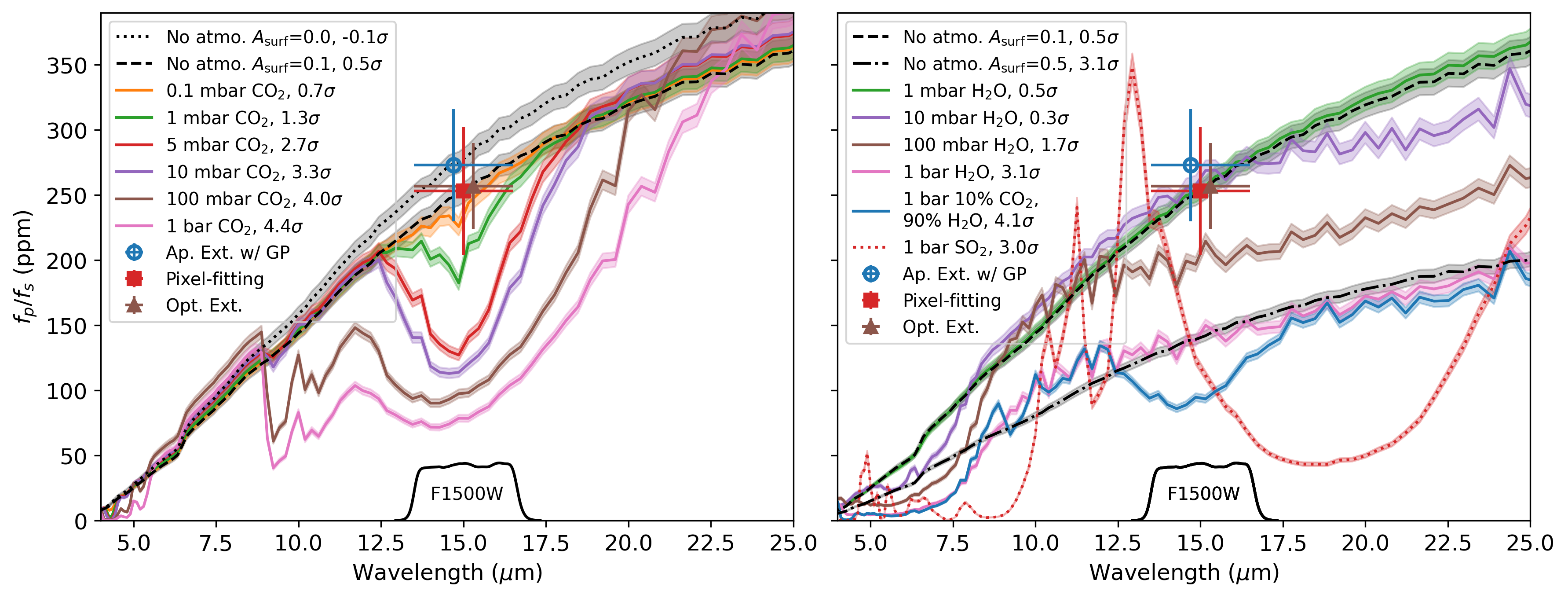}
        \caption{Atmospheric forward models compared to eclipse depth measurements using various analyses. Spectra are compared to eclipse depths recovered using aperture extraction with a GP, pixel-fitting or a secondary analysis fit using optimal extraction without a GP. Left: Emission spectra of bare rock and pure CO$_2$ models at varying surface pressures. Statistical significance given relative to joint-fitting aperture extracted light curves using GPs. Right: Similar plot for pure H$_2$O, pure SO$_2$ or H$_2$O/CO$_2$ mixture atmospheres. All atmospheric models account for heat redistribution using the f-factor parameterisation in \citet{Kollf} and use a surface albedo of A$_\mathrm{surf}$ = 0.1.}
        \label{fig:models}
    \end{figure*}
    
    We also added the reflected light component,
    \begin{equation}
        \frac{f_\mathrm{refl}}{f_*} = A_g \left(\frac{R_p}{a}\right)^2,
    \label{eq:reflection}
    \end{equation}
    where $A_g$ is the geometric albedo. For a Lambertian reflector we get $A_g = (2/3)A_\mathrm{surf}$ \citep{Heng2021}. Note for our target $(R_p/a)^2 \approx 4$ ppm so reflection is only a minor contribution.
    
    For the atmospheric cases, we used the 1D atmospheric model HELIOS \citep{heliosa, heliosb} with a solid surface bottom layer \citep{heliossurfa, heliossurfb} to calculate the planet’s temperature structure and its synthetic emission and reflection spectra. The temperature profile of a planet's atmosphere is primarily determined by the balance between radiative, convective processes and heat redistribution. The model considers well-mixed atmospheres with various gas compositions consisting of CO$_2$, H$_2$O, SO$_2$, N$_2$ and O$_2$. It employs k-distribution tables for opacities, integrating radiative fluxes over 386 spectral bands and 20 Gaussian points. Opacities are calculated using the HELIOS-K \citep{heliosk} code and using cross-sections for CO$_2$ from \citet{hitemp}, H$_2$O from \citet{waterlinelist}, SO$_2$ from \citet{SO2_linelist} and O$_2$ from \citet{O2_crosssection}. Additionally, we incorporate Rayleigh scattering cross-sections for H$_2$O,  CO$_2$, N$_2$, and O$_2$ \citep{Allens, sneep2005, thalman2014}, but did not include scattering for SO$_2$. For all atmospheric models, the surface albedo was fixed to $A_\mathrm{surf} = 0.1$.
    
    Heat redistribution by winds also plays an important role in determining the temperature structure of the atmosphere. The heat redistribution in HELIOS is controlled by the f-factor (the same $f$ parameter used in Equation~\ref{eq:no_atmo} for the no atmosphere case). The f-factor was approximated by the scaling equation from \citet{Kollf}. This analytical equation depends on the longwave optical depth at the surface, the surface pressure and the equilibrium temperature.
    
    Stellar spectra are taken from BT-Settl (CIFIST) models \citep{2011ASPC..448...91A} and linearly interpolated for the stellar parameters. We did not choose to adjust these stellar models based on the absolute flux calibration performed in Section~\ref{sec:flux_cal} so note that these stellar models may be underestimating flux by $\sim$10\% in the F1500W bandpass.

    Figure~\ref{fig:models} shows a range of models including pure CO$_2$, pure H$_2$O, pure SO$_2$, and a CO$_2$/H$_2$O mixture. Each model was run 100 times with the input planetary parameters $R_p$, $a$, $g_p$ (planetary surface gravity) and stellar parameters $R_*$, $T_\mathrm{eff}$, $\log(g)$ randomly drawn from their constraints listed in \citet{2024ApJ...960L...3C} to propagate the uncertainty in these parameters. The figure displays the emission spectra resulting from the modelling, with the shaded regions representing the 1$\sigma$ range of the computed values.

    We find that the eclipse depths from aperture photometry, pixel-fitting and the independent analysis using optimal extraction are all in excellent agreement with a low-albedo bare rock ($A_\mathrm{surf} \sim 0.1$). Higher albedo surfaces of $A_\mathrm{surf} = 0.5$ diverge at $>\!3\sigma$ (uncertainties given relative to the joint-fit using aperture extraction with GPs; $d = 273\pm43$ ppm). Extremely tenuous (<1 mbar) CO$_2$ or H$_2$O atmospheres naturally behave similarly to bare rocks and cannot be distinguished using these observations.

\begin{figure}
        \includegraphics[width=\columnwidth]{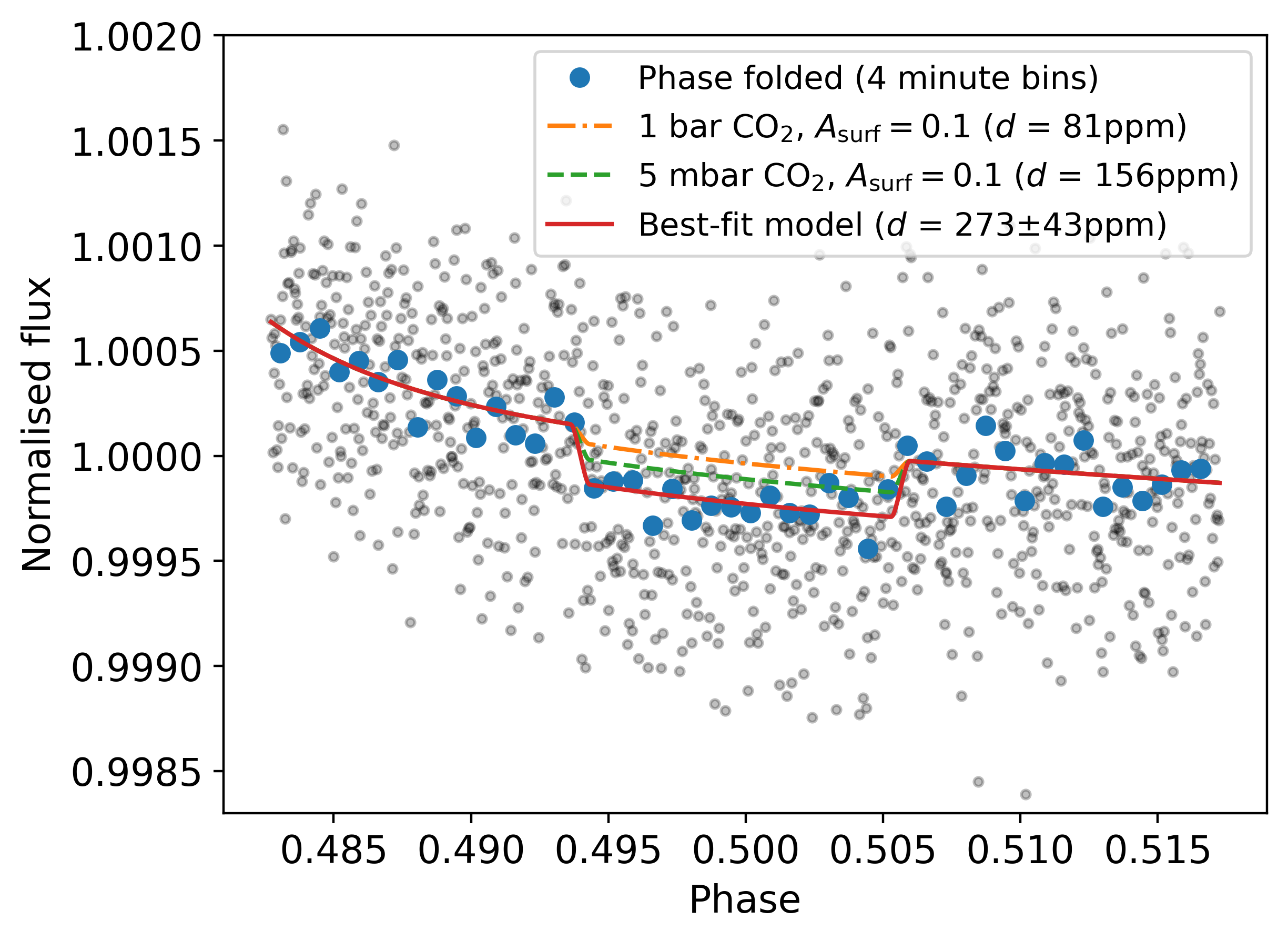}
        \caption{Phase-folded observations compared to model light curves of pure CO$_2$ atmospheres. Observations plotted without binning (black) and with binning (blue circles). In red is the phase folded best-fit eclipse model from joint-fitting the aperture extracted light curves with a GP. The orange and green dashed lines show expected eclipse depths for different pressure pure CO$_2$ atmospheres.}
        \label{fig:phase_folded}
    \end{figure}

    However, we do find that a pure CO$_2$ atmosphere with a surface pressure of only $P_s = 5$ mbar is discrepant with observations at $2.7\sigma$, making a Mars-like atmosphere ($\sim\!6$ mbar, 95\% CO$_2$) unlikely. These constraints are only weakly affected by our heat redistribution model as fixing $f = 2/3$ (i.e. no heat redistribution) has a negligible effect on the 5 mbar result and only slightly weakens the 100 mbar and 1 bar pure CO$_2$ results to be between $3-4\sigma$. This implies that our constraints on the lack of a thick pure CO$_2$ atmosphere largely derive from the lack of CO$_2$ absorption rather than from the lack of heat redistribution. We compare the 5 mbar and 1 bar pure CO$_2$ eclipse depths to our phase-folded light curve fit in Figure~\ref{fig:phase_folded} to show how far they diverge from the light curve data.
    
    Pure H$_2$O atmospheres are unlikely from a theoretical perspective as models suggest they are highly vulnerable to atmospheric escape for planets around M dwarfs \citep{Luger2015, 2019Natur.573...87K}. From our data, we cannot rule out tenuous 10 mbar models but we can rule out 1 bar H$_2$O atmospheres at $3.1\sigma$. In contrast to the pure CO$_2$ atmospheres, this result is completely dependent on our heat redistribution model and is not the result of a particular absorption feature. All atmospheric models plotted in Figure~\ref{fig:models} use the analytic f-factor parameterisation from \citep{Kollf}, but if we instead fix $f = 2/3$ then all pure H$_2$O atmospheres up to 1 bar are within 1$\sigma$ of our results (not plotted). We have also included a mixed 10\% CO$_2$, 90\% H$_2$O 1 bar atmosphere which appears to behave similarly to the pure H$_2$O 1 bar atmosphere but with the addition of a CO$_2$ absorption feature over the 15$\upmu$m region observed, making it less consistent with observations.

    As we may expect volcanically outgassed secondary atmospheres to include sulphur-based molecules such as SO$_2$, we also tested a range of pure SO$_2$ atmospheres. In reality, sulphur chemistry is complex and SO$_2$ is very photochemically sensitive; thus, more sophisticated photochemical modelling would need to be performed to understand what the dominant form of sulphur would be for this planet (see e.g. \citealt{Bello-Arufe_2025}). It is likely that a dense SO$_2$ atmosphere would require constant replenishment from volcanic outgassing, which given the low eccentricity of the planet ($e < 0.0172$ to 95\% confidence, see Appendix~\ref{app:eccentricity}) would not be expected from tidal forces, unlike suggestions for the L 98-59 system \citep{Seligman2024}. Noting these caveats, we recover that pure 1 bar SO$_2$ atmospheres are ruled out at $3\sigma$ when modelling heat redistribution (plotted in Figure~\ref{fig:models}); however, without any heat redistribution, this result would decrease to $2.3\sigma$. Atmospheres of 100 mbar or weaker are all within $2\sigma$ regardless of the effect of heat redistribution. We note that longer wavelengths such as those covered by the F1800W filter might provide significantly tighter constraints for thinner SO$_2$ atmospheres. We have included models of SO$_2$ in a CO$_2$ background in Appendix~\ref{app:mixed_SO2} which suggest that combinations of these molecules may produce particularly strong signatures which we can rule out down to 1 mbar. However, we again note that these models exclude any photochemical or geochemical modelling and we hope future work can examine to what significance physically plausible volcanically outgassed atmospheres are ruled out to.

    We also tested a pure 1 bar O$_2$ atmosphere but found it indistinguishable from a bare rock model of the same albedo and have therefore not included it in Figure~\ref{fig:models}. This makes sense as O$_2$ is only weakly absorbing at these wavelengths and our heat redistribution model predicts minimal heat redistribution as the atmosphere is very optically thin. We highlight that the atmospheres this method is sensitive to are not simply any atmospheres with high surface pressures, they must also be reasonably optically thick for significant heat redistribution to occur \citep{Kollf, Ih2023}. We may expect a similar result for pure N$_2$ atmospheres.

    In addition, we tested various concentrations of CO$_2$ in a background of N$_2$ at various surface pressures. The statistical significance of each model is listed in Table~\ref{tab:CO2_concentration}. We found that even low concentrations of CO$_2$ are ruled out with high confidence at sufficiently high surface pressure. For example, 100 ppm CO$_2$ in a 1 bar N$_2$ atmosphere is rejected at 3.7$\sigma$. We also ran each model with O$_2$ instead of N$_2$ and every result was affected by less than 0.2$\sigma$. This makes sense as these molecules are both weakly absorbing and are simply acting as inert background gases with similar mean molecular weights. This implies that an Earth-like atmosphere of 78\% N$_2$, 21\% O$_2$, $\sim$400 ppm CO$_2$ at 1 bar is also strongly rejected by our observations.
    
    \begin{table}
        \caption{Statistical significance in std. dev. of various atmospheres containing CO$_2$ with N$_2$ as a background gas.}
        \centering
        \begin{tabular}{lccc}
                \hline
                Pressure & 1 ppm CO$_2$ & 100 ppm CO$_2$ & 1000 ppm CO$_2$ \\
                \hline
            0.1 mbar & 0.6 & 0.6 & 0.6\\
            1 mbar & 0.6 & 0.6 & 0.7\\
            5 mbar & 0.6 & 0.7 & 0.8\\
            10 mbar & 0.6 & 0.7 & 0.9\\
            50 mbar & 0.6 & 1.0 & 1.9\\
            100 mbar & 0.7 & 1.5 & 2.7\\
            500 mbar & 0.9 & 3.1 & 4.0\\
            1 bar & 1.2 & 3.7 & 4.3\\
            5 bar & 2.7 & 4.5 & 4.8\\
            10 bar & 3.4 & 4.8 & 5.0\\
                \hline
        \end{tabular}
        \tablefoot{Results given relative to the aperture extraction joint-fit with a GP ($d$ = 273 $\pm$ 43 ppm). Assumes a surface albedo of $A_\mathrm{surf} = 0.1$.}
        \label{tab:CO2_concentration}
    \end{table}
    
    One caveat is that our bare rock models do not account for surface roughness, which can make a bare rock surface appear hotter at eclipse relative to a smooth spherical surface. This happens because the surfaces tilted towards the observer are also tilted towards its host star and are therefore hotter \citep{Hapke_1993, Coy2024}. This effect - referred to as "thermal beaming" - has been observed for both the Moon \citep{Wohlfarth2023} and Mercury \citep{Mercury1998}. It is also the focus of a Cycle 2 program to help characterise the surface of the hot rocky exoplanet LHS 3844b (GO 4008; \citealt{2023jwst.prop.4008Z}). The effect may not be very significant at 15$\upmu$m based on simulations of Mercury (a similar temperature planet) in \citet{Wohlfarth2023}, but more modelling work would need to be performed to confirm this.

\section{Discussion}
\label{sec:6}

    Our observations rule out a wide range of possible atmospheres on LHS 1140c. While we cannot rule out very optically thin atmospheres such as pure O$_2$ atmospheres (which were a possible outcome of atmospheric escape modelling for TRAPPIST-1b \citealt{Totton2024}), these atmospheres would be difficult to detect using any technique. The consistency with a bare rock despite our high sensitivity to CO$_2$ is in stark contrast to rocky planet atmospheres within our Solar System and also appears inconsistent with volcanically outgassed atmospheres. This work further backs up the finding in \citet{2024ApJ...970L...2C} that the $4\sigma$ discrepancy between transit depths measured with Spitzer and TESS (previously noted in \citealt{2024ApJ...960L...3C}) may be resolved by the transit depth being underestimated by the TESS observations and is unlikely to be due to CO$_2$ absorption.
    
    These constraints are very informative because the planet's low equilibrium temperature of $422\pm7$ K \citep{2024ApJ...960L...3C}, its relatively mild present-day radiation environment \citep{NUV_obs}, and the reported age of the system (>5 Gyr; \citealt{Dittmann2017}), make it one of the better known candidates to have an outgassed atmosphere or to have retained some of its primary atmosphere \citep{volcano_degas, novelphysics}. In contrast, our result is in line with the suggestion that historic levels of XUV fluence may be underestimated for mid-type M dwarfs such as LHS 1140, making atmospheric retention more difficult \citep{Pass2025}. As this star also hosts LHS 1140b -- a larger Super-Earth which likely still retains an atmosphere\footnote{There is still some discussion on the composition of LHS 1140b, see \citet{2024ApJ...970L...2C} and \citet{Damiano2024} for  evidence distinguishing rocky, water world, or cloudy mini-Neptune scenarios.} -- our new atmospheric constraints are expected to help make the LHS 1140 system an important benchmark for understanding atmospheric escape and the cosmic shoreline \citep{2017ApJ...843..122Z}.
    
    We have introduced a new technique, which can act as an alternative to aperture photometry or optimal extraction. Section~\ref{sec:3} demonstrates that our method presents a clear advantage when we have individual pixels contaminated by systematics, as it can weight away from them towards cleaner pixels. For the JWST observations, we see in the case of our second eclipse that the pixel-fitting method could weight away from two discrepant pixels apparently affected by a persistence effect from a cosmic ray. While the method has disadvantages, such as its increased computational cost and complexity to implement, it provides a new way to analyse and explore data at the pixel-level which can identify problematic pixels in an empirical way. It is also important to note that while our observations do not appear heavily impacted by systematics, an observation of another target in our survey (LHS 1478b) appears to be significantly impacted by the presence of systematics \citep{Prune2024}, which our method may be able to alleviate. The fact that our method is capable of distinguishing between systematics in individual pixels or across all pixels could also help to diagnose the cause of these systematics.
    
    Pixel-fitting therefore provides a powerful alternative to aperture photometry, but we do not claim it is a replacement for it. Highly flexible systematics models can come at the cost of providing more conservative results on clean data. This is a common trade-off when we analyse data for which we do not understand the extent of systematics present and the best approach is to use a range of methods varying in how conservative they are. By utilising both aperture photometry and pixel-fitting, we can test the robustness of our results and learn about potential systematics in our datasets. Having more independent techniques to analyse these data is particularly important given the decision for 500 hours of Director's Discretionary Time to be spent extending this technique to even more rocky exoplanets \citep{DDT}.
    
    The pixel-fitting method has huge potential as it could be applied to a wide range of existing time-series observations including spectroscopic time series. There is currently a move within exoplanet atmospheres research to analyse JWST spectroscopic time series at the native or pixel resolution, yet there is concern over how much correlation this introduces in the noise between neighbouring light curves, with standard analysis approaches typically ignoring correlations between light curves (see \citealt{Ih2021} and \citealt{Fortune2024} for how this can affect atmospheric retrievals). We note that MIRI/LRS uses the same detector as MIRI/Imaging (but on a subarray that is different from that of our observations); thus, many of the detector systematics seen in our observations (e.g. the flux-conserving systematics) might affect MIRI/LRS observations too. Future work extending pixel-fitting to spectroscopy could examine whether this anti-correlation between neighbouring pixel light curves could explain the large increase in scatter in transmission spectra noted when extracting spectroscopic light curves with one pixel-wide bins \citep{Bell2024}. Implementing pixel-fitting for spectroscopy would differ as the eclipse depth would not be shared across all pixels -- but only between pixels at the same wavelength. The computational cost could be a barrier to implementation as the spectral trace is spread across more pixels. Strategies such as binning in time and parallelising over multiple GPUs could help to mitigate this.

    Finally, we believe the relationship between the initial detector settling slope and the previous filter used by MIRI warrants further study as a significant fraction of MIRI TSOs are affected by this. Therefore, mitigating these effects could provide a serious improvement to the efficiency of the instrument. A strategy such as changing when the filter is moved into place and/or changing how MIRI is clocked in idle mode could have a large effect on the initial settling slope.

\section{Conclusions}
\label{sec:7}

    Our three eclipses of LHS 1140c using MIRI/Imaging at 15$\upmu$m have allowed us to rule out a wide range of possible atmospheres. We have found that LHS 1140c is consistent with a low-albedo bare rock and is unlikely to host an atmosphere capable of substantial heat redistribution or an atmosphere containing significant levels ($>$100 ppm) of CO$_2$. Using an absolute flux calibration, we measured its dayside brightness temperature to be $T_B = 561\pm44$ K, close to the maximum expected dayside temperature of $T_\text{day; max} = 537\pm9$ K. These results are fully consistent with the scenario that LHS 1140c lacks a substantial atmosphere. There are alternative possibilities, including dense atmospheres that are very optically thin, but these models may require some fine-tuning to explain the lack of strong absorbers such as CO$_2$. If LHS 1140c does indeed lack an atmosphere, then it could have strong implications for the cosmic shoreline model of atmospheric escape, as LHS 1140c is a super-Earth with one of the lowest instellations of any target this technique has been applied to so far (after TRAPPIST-1b and c; \citealt{2023Natur.618...39G, 2023Natur.620..746Z}). Given that the outer planet LHS 1140b is also a super-Earth, but one that is  likely to host an atmosphere \citep{Damiano2024, 2024ApJ...970L...2C}, this could place each planet on either side of the cosmic shoreline.

    We have identified a possible relationship in the strength and sign of the initial settling slope with the previous filter used by MIRI. This may help to inform ways to mitigate this effect in future. We have also demonstrated a new method of analysing time series photometry, which produces similar results to aperture photometry on these datasets. The benefits of this method include:

    \begin{itemize}
    \item identifying and weighting away from systematics arising in individual pixels, potentially leading to more precise and robust measurements.

    \item identifying whether systematics arise from individual pixels or are common to all pixels, helping to determine the source of the systematics.

    \item weighting towards higher S/N pixels, similar to optimal extraction.

    \item retrieving additional information about the detector, such as how the PSF behaves over time and the strength of interpixel capacitance.

    \item constraining information about the amplitude of background scatter remaining after the background subtraction.

    \end{itemize}

    Overall, we hope this paper provides greater insight into the systematics present in MIRI data and provides useful approaches for dealing with them. This may help to robustly characterise more exoplanets in the future.
    
\begin{acknowledgements}

    We are grateful to the anonymous referee for the careful
reading and helpful suggestions which improved this manuscript. We would like to thank Charles Cadieux for providing orbital constraints on the LHS 1140 system and Karl Gordon for his advice in performing the absolute flux calibration. We also thank Bibiana Prinoth for a careful reading of the manuscript and helping to improve the clarity of this work. We acknowledge the input from the following individuals to the GO 3730 proposal: Can Akin, Adam Burgasser, Chloe Fisher, Matthew Hooton, Andrea Guzmán Mesa, Anna Lueber, Nicholas Borsato, Bibiana Prinoth, Jens Hoeijmakers, Meng Tian, Mette Baungaard. Based on observations with the NASA/ESA/CSA James Webb Space Telescope. The data were obtained from the Mikulski Archive for Space Telescopes at the Space Telescope Science Institute, which is operated by the Association of Universities for Research in Astronomy, Incorporated, under NASA contract NAS 5-03127. Support for program number 3730 was provided through a grant from the STScI under NASA contract NAS5-03127. The raw data described here may be obtained from \url{https://dx.doi.org/10.17909/11g0-pd60}. The authors wish to acknowledge the Irish Centre for High-End Computing (ICHEC) for the provision of computational facilities and support. Many of the calculations in this work were performed on the astro01 system maintained by the Trinity Centre for High Performance Computing (Research IT). This system is co-funded by the School of Physics and by the Irish Research Council grant award IRCLA/2022/3788. NPG gratefully acknowledges support from Science Foundation Ireland and the Royal Society through a University Research Fellowship (URF\textbackslash R\textbackslash 201032). HDL and PCA acknowledge support from the Carlsberg Foundation, grant CF22-1254. JMM acknowledges support from the Horizon Europe Guarantee Fund, grant EP/Z00330X/1. NHA acknowledges support by the National Science Foundation Graduate Research Fellowship under Grant No. DGE1746891. EMV acknowledges financial support from the Swiss National Science Foundation (SNSF) Mobility Fellowship under grant no. P500PT\_225456/1. B.-O. D. acknowledges support from the Swiss State Secretariat for Education, Research and Innovation (SERI) under contract number MB22.00046. ADR acknowledges support from the Carlsberg Foundation, grant CF22-1548.

    We are grateful to the developers of \textsc{NumPy}, \textsc{SciPy}, \textsc{Matplotlib}, \textsc{Astropy}, \textsc{photutils}, \textsc{pandas}, \textsc{iPython}, \textsc{ArviZ}, \textsc{corner}, \textsc{JAX}, \textsc{PyMC}, \textsc{jaxoplanet} and \textsc{exoplanet} as these packages were used extensively in this work \citep{numpy, scipy, matplotlib, astropy, photutils, pandas, ipython, arviz_2019, corner, JAX, Salvatier2015, jaxoplanet, exoplanet}.
    
\end{acknowledgements}

\bibliographystyle{aa}
\bibliography{main}

\begin{thebibliography}{117}
\expandafter\ifx\csname natexlab\endcsname\relax\def\natexlab#1{#1}\fi

\bibitem[{{Agol} {et~al.}(2021){Agol}, {Dorn}, {Grimm}, {Turbet}, {Ducrot},
  {Delrez}, {Gillon}, {Demory}, {Burdanov}, {Barkaoui}, {Benkhaldoun},
  {Bolmont}, {Burgasser}, {Carey}, {de Wit}, {Fabrycky}, {Foreman-Mackey},
  {Haldemann}, {Hernandez}, {Ingalls}, {Jehin}, {Langford}, {Leconte},
  {Lederer}, {Luger}, {Malhotra}, {Meadows}, {Morris}, {Pozuelos}, {Queloz},
  {Raymond}, {Selsis}, {Sestovic}, {Triaud}, \& {Van Grootel}}]{Agol2021}
{Agol}, E., {Dorn}, C., {Grimm}, S.~L., {et~al.} 2021, \psj, 2, 1

\bibitem[{{Aigrain} \& {Foreman-Mackey}(2023)}]{2023ARA&A..61..329A}
{Aigrain}, S. \& {Foreman-Mackey}, D. 2023, \araa, 61, 329

\bibitem[{{Alam} {et~al.}(2025){Alam}, {Gao}, {Adams Redai}, {Wallack},
  {Wogan}, {Aguichine}, {Dattilo}, {Alderson}, {Batalha}, {Batalha}, {Kirk},
  {L{\'o}pez-Morales}, {Meech}, {Moran}, {Teske}, {Wakeford}, \&
  {Wolfgang}}]{Alam2025}
{Alam}, M.~K., {Gao}, P., {Adams Redai}, J., {et~al.} 2025, \aj, 169, 15

\bibitem[{{Allard} {et~al.}(2011){Allard}, {Homeier}, \&
  {Freytag}}]{2011ASPC..448...91A}
{Allard}, F., {Homeier}, D., \& {Freytag}, B. 2011, in Astronomical Society of
  the Pacific Conference Series, Vol. 448, 16th Cambridge Workshop on Cool
  Stars, Stellar Systems, and the Sun, ed. C.~{Johns-Krull}, M.~K. {Browning},
  \& A.~A. {West}, 91

\bibitem[{{Alonso}(2018)}]{eclipse_time_formula}
{Alonso}, R. 2018, in Handbook of Exoplanets, ed. H.~J. {Deeg} \& J.~A.
  {Belmonte}, 40

\bibitem[{{Astropy Collaboration} {et~al.}(2022){Astropy Collaboration},
  {Price-Whelan}, {Lim}, {Earl}, {Starkman}, {Bradley}, {Shupe}, {Patil},
  {Corrales}, {Brasseur}, {N{\"o}the}, {Donath}, {Tollerud}, {Morris},
  {Ginsburg}, {Vaher}, {Weaver}, {Tocknell}, {Jamieson}, {van Kerkwijk},
  {Robitaille}, {Merry}, {Bachetti}, {G{\"u}nther}, {Aldcroft},
  {Alvarado-Montes}, {Archibald}, {B{\'o}di}, {Bapat}, {Barentsen},
  {Baz{\'a}n}, {Biswas}, {Boquien}, {Burke}, {Cara}, {Cara}, {Conroy},
  {Conseil}, {Craig}, {Cross}, {Cruz}, {D'Eugenio}, {Dencheva}, {Devillepoix},
  {Dietrich}, {Eigenbrot}, {Erben}, {Ferreira}, {Foreman-Mackey}, {Fox},
  {Freij}, {Garg}, {Geda}, {Glattly}, {Gondhalekar}, {Gordon}, {Grant},
  {Greenfield}, {Groener}, {Guest}, {Gurovich}, {Handberg}, {Hart},
  {Hatfield-Dodds}, {Homeier}, {Hosseinzadeh}, {Jenness}, {Jones}, {Joseph},
  {Kalmbach}, {Karamehmetoglu}, {Ka{\l}uszy{\'n}ski}, {Kelley}, {Kern},
  {Kerzendorf}, {Koch}, {Kulumani}, {Lee}, {Ly}, {Ma}, {MacBride}, {Maljaars},
  {Muna}, {Murphy}, {Norman}, {O'Steen}, {Oman}, {Pacifici}, {Pascual},
  {Pascual-Granado}, {Patil}, {Perren}, {Pickering}, {Rastogi}, {Roulston},
  {Ryan}, {Rykoff}, {Sabater}, {Sakurikar}, {Salgado}, {Sanghi}, {Saunders},
  {Savchenko}, {Schwardt}, {Seifert-Eckert}, {Shih}, {Jain}, {Shukla}, {Sick},
  {Simpson}, {Singanamalla}, {Singer}, {Singhal}, {Sinha}, {Sip{\H{o}}cz},
  {Spitler}, {Stansby}, {Streicher}, {{\v{S}}umak}, {Swinbank}, {Taranu},
  {Tewary}, {Tremblay}, {de Val-Borro}, {Van Kooten}, {Vasovi{\'c}}, {Verma},
  {de Miranda Cardoso}, {Williams}, {Wilson}, {Winkel}, {Wood-Vasey}, {Xue},
  {Yoachim}, {Zhang}, {Zonca}, \& {Astropy Project Contributors}}]{astropy}
{Astropy Collaboration}, {Price-Whelan}, A.~M., {Lim}, P.~L., {et~al.} 2022,
  \apj, 935, 167

\bibitem[{{August} {et~al.}(2025){August}, {Buchhave}, {Diamond-Lowe},
  {Mendon{\c{c}}a}, {Gressier}, {Rathcke}, {Allen}, {Fortune}, {Jones}, {Meier
  Vald{\'e}s}, {Demory}, {Espinoza}, {Fisher}, {Gibson}, {Heng}, {Hoeijmakers},
  {Hooton}, {Kitzmann}, {Prinoth}, {Eastman}, \& {Barnes}}]{Prune2024}
{August}, P.~C., {Buchhave}, L.~A., {Diamond-Lowe}, H., {et~al.} 2025, \aap,
  695, A171

\bibitem[{Banerjee {et~al.}(2024)Banerjee, Barstow, Gressier, Espinoza, Sing,
  Allen, Birkmann, Challener, Crouzet, Haswell, Lewis, Lewis, \&
  Yang}]{Banerjee_2024}
Banerjee, A., Barstow, J.~K., Gressier, A., {et~al.} 2024, \apjl, 975, L11

\bibitem[{{Barber} {et~al.}(2006){Barber}, {Tennyson}, {Harris}, \&
  {Tolchenov}}]{waterlinelist}
{Barber}, R.~J., {Tennyson}, J., {Harris}, G.~J., \& {Tolchenov}, R.~N. 2006,
  \mnras, 368, 1087

\bibitem[{{Bell} {et~al.}(2024){Bell}, {Crouzet}, {Cubillos}, {Kreidberg},
  {Piette}, {Roman}, {Barstow}, {Blecic}, {Carone}, {Coulombe}, {Ducrot},
  {Hammond}, {Mendon{\c{c}}a}, {Moses}, {Parmentier}, {Stevenson},
  {Teinturier}, {Zhang}, {Batalha}, {Bean}, {Benneke}, {Charnay}, {Chubb},
  {Demory}, {Gao}, {Lee}, {L{\'o}pez-Morales}, {Morello}, {Rauscher}, {Sing},
  {Tan}, {Venot}, {Wakeford}, {Aggarwal}, {Ahrer}, {Alam}, {Baeyens},
  {Barrado}, {Caceres}, {Carter}, {Casewell}, {Challener}, {Crossfield},
  {Decin}, {D{\'e}sert}, {Dobbs-Dixon}, {Dyrek}, {Espinoza}, {Feinstein},
  {Gibson}, {Harrington}, {Helling}, {Hu}, {Iro}, {Kempton}, {Kendrew},
  {Komacek}, {Krick}, {Lagage}, {Leconte}, {Lendl}, {Lewis}, {Lothringer},
  {Malsky}, {Mancini}, {Mansfield}, {Mayne}, {Evans-Soma}, {Molaverdikhani},
  {Nikolov}, {Nixon}, {Palle}, {Petit dit de la Roche}, {Piaulet}, {Powell},
  {Rackham}, {Schneider}, {Steinrueck}, {Taylor}, {Welbanks}, {Yurchenko},
  {Zhang}, \& {Zieba}}]{Bell2024}
{Bell}, T.~J., {Crouzet}, N., {Cubillos}, P.~E., {et~al.} 2024, Nature
  Astronomy, 8, 879

\bibitem[{Bello-Arufe {et~al.}(2025)Bello-Arufe, Damiano, Bennett, Hu,
  Welbanks, MacDonald, Seligman, Sing, Tokadjian, Oza, \&
  Yang}]{Bello-Arufe_2025}
Bello-Arufe, A., Damiano, M., Bennett, K.~A., {et~al.} 2025, \apjl, 980, L26

\bibitem[{Bradbury {et~al.}(2018)Bradbury, Frostig, Hawkins, Johnson, Leary,
  Maclaurin, Necula, Paszke, Vander{P}las, Wanderman-{M}ilne, \& Zhang}]{JAX}
Bradbury, J., Frostig, R., Hawkins, P., {et~al.} 2018, {JAX}: composable
  transformations of {P}ython+{N}um{P}y programs

\bibitem[{Bradley {et~al.}(2024)Bradley, Sipőcz, Robitaille, Tollerud,
  Vinícius, Deil, Barbary, Wilson, Busko, Donath, Günther, Cara, Lim,
  Meßlinger, Burnett, Conseil, Droettboom, Bostroem, Bray, Bratholm, Jamieson,
  Ginsburg, Barentsen, Craig, Pascual, Rathi, Perrin, Morris, \&
  Perren}]{photutils}
Bradley, L., Sipőcz, B., Robitaille, T., {et~al.} 2024, astropy/photutils:
  1.13.0

\bibitem[{{Cadieux} {et~al.}(2024{\natexlab{a}}){Cadieux}, { Plotnykov},
  {Doyon}, {Valencia}, {Jahandar}, {Dang}, {Turbet}, {Fauchez}, {Cloutier},
  {Cherubim}, {Artigau}, {Cook}, {Edwards}, {Hallatt}, {Charnay}, {Bouchy},
  {Allart}, {Mignon}, {Baron}, {Barros}, {Benneke}, {Canto Martins}, {Cowan},
  {De Medeiros}, {Delfosse}, {Delgado-Mena}, {Dumusque}, {Ehrenreich},
  {Frensch}, {Gonz{\'a}lez Hern{\'a}ndez}, {Hara}, {Lafreni{\`e}re}, {Lo
  Curto}, {Malo}, {Melo}, {Mounzer}, {Passeger}, {Pepe}, {Poulin-Girard},
  {Santos}, {Sosnowska}, {Su{\'a}rez Mascare{\~n}o}, {Thibault}, {Vaulato},
  {Wade}, \& {Wildi}}]{2024ApJ...960L...3C}
{Cadieux}, C., { Plotnykov}, M., {Doyon}, R., {et~al.} 2024{\natexlab{a}},
  \apjl, 960, L3

\bibitem[{{Cadieux} {et~al.}(2024{\natexlab{b}}){Cadieux}, {Doyon},
  {MacDonald}, {Turbet}, {Artigau}, {Lim}, {Radica}, {Fauchez}, {Salhi},
  {Dang}, {Albert}, {Coulombe}, {Cowan}, {Lafreni{\`e}re}, {L'Heureux},
  {Piaulet-Ghorayeb}, {Benneke}, {Cloutier}, {Charnay}, {Cook},
  {Fournier-Tondreau}, {Plotnykov}, \& {Valencia}}]{2024ApJ...970L...2C}
{Cadieux}, C., {Doyon}, R., {MacDonald}, R.~J., {et~al.} 2024{\natexlab{b}},
  \apjl, 970, L2

\bibitem[{{Charbonneau} {et~al.}(2005){Charbonneau}, {Allen}, {Megeath},
  {Torres}, {Alonso}, {Brown}, {Gilliland}, {Latham}, {Mandushev}, {O'Donovan},
  \& {Sozzetti}}]{Charbonneau2005}
{Charbonneau}, D., {Allen}, L.~E., {Megeath}, S.~T., {et~al.} 2005, \apj, 626,
  523

\bibitem[{Chatterjee \& Pierrehumbert(2024)}]{novelphysics}
Chatterjee, R.~D. \& Pierrehumbert, R.~T. 2024, \apj, submitted,
  arXiv:2412.05188

\bibitem[{{Cox}(2000)}]{Allens}
{Cox}, A.~N. 2000, {Allen's astrophysical quantities}

\bibitem[{{Damiano} {et~al.}(2024){Damiano}, {Bello-Arufe}, {Yang}, \&
  {Hu}}]{Damiano2024}
{Damiano}, M., {Bello-Arufe}, A., {Yang}, J., \& {Hu}, R. 2024, \apjl, 968, L22

\bibitem[{{Deming} {et~al.}(2015){Deming}, {Knutson}, {Kammer}, {Fulton},
  {Ingalls}, {Carey}, {Burrows}, {Fortney}, {Todorov}, {Agol}, {Cowan},
  {Desert}, {Fraine}, {Langton}, {Morley}, \& {Showman}}]{2015ApJ...805..132D}
{Deming}, D., {Knutson}, H., {Kammer}, J., {et~al.} 2015, \apj, 805, 132

\bibitem[{{Demory} {et~al.}(2016){Demory}, {Gillon}, {de Wit}, {Madhusudhan},
  {Bolmont}, {Heng}, {Kataria}, {Lewis}, {Hu}, {Krick}, {Stamenkovi{\'c}},
  {Benneke}, {Kane}, \& {Queloz}}]{Demory2016}
{Demory}, B.-O., {Gillon}, M., {de Wit}, J., {et~al.} 2016, \nat, 532, 207

\bibitem[{{Dicken} {et~al.}(2024){Dicken}, {Mar{\'\i}n}, {Shivaei}, {Guillard},
  {Libralato}, {Glasse}, {Gordon}, {Cossou}, {Kavanagh}, {Temim}, {Flagey},
  {Klaassen}, {Rieke}, {Wright}, {Alberts}, {Azzollini},
  {{\'A}lvarez-M{\'a}rquez}, {Bouchet}, {Bright}, {Cracraft}, {Coulais},
  {Detre}, {Engesser}, {Fox}, {Gaspar}, {Gastaud}, {Glauser}, {Hines},
  {Kendrew}, {Labiano}, {Lagage}, {Lee}, {Law}, {Morrison}, {Noriega-Crespo},
  {Jones}, {Patapis}, {Scheithauer}, {Sloan}, \& {Tamas}}]{Dicken2024}
{Dicken}, D., {Mar{\'\i}n}, M.~G., {Shivaei}, I., {et~al.} 2024, \aap, 689, A5

\bibitem[{{Dittmann} {et~al.}(2017){Dittmann}, {Irwin}, {Charbonneau},
  {Bonfils}, {Astudillo-Defru}, {Haywood}, {Berta-Thompson}, {Newton},
  {Rodriguez}, {Winters}, {Tan}, {Almenara}, {Bouchy}, {Delfosse}, {Forveille},
  {Lovis}, {Murgas}, {Pepe}, {Santos}, {Udry}, {W{\"u}nsche}, {Esquerdo},
  {Latham}, \& {Dressing}}]{Dittmann2017}
{Dittmann}, J.~A., {Irwin}, J.~M., {Charbonneau}, D., {et~al.} 2017, \nat, 544,
  333

\bibitem[{{Ducrot} {et~al.}(2024){Ducrot}, {Lagage}, \& {Min}}]{Ducrot2024}
{Ducrot}, E., {Lagage}, P.-O., \& {Min}, M. 2024, in EGU General Assembly
  Conference Abstracts, EGU General Assembly Conference Abstracts, 20183

\bibitem[{Eastman {et~al.}(2013)Eastman, Gaudi, \& Agol}]{Eastman_2013}
Eastman, J., Gaudi, B.~S., \& Agol, E. 2013, \pasa, 125, 83

\bibitem[{{Emery} {et~al.}(1998){Emery}, {Sprague}, {Witteborn}, {Colwell},
  {Kozlowski}, \& {Wooden}}]{Mercury1998}
{Emery}, J.~P., {Sprague}, A.~L., {Witteborn}, F.~C., {et~al.} 1998, \icarus,
  136, 104

\bibitem[{{Espinoza}(2022)}]{espinoza_nestor_2022}
{Espinoza}, N. 2022, {TransitSpectroscopy}

\bibitem[{Espinoza {et~al.}(2019)Espinoza, Kossakowski, \&
  Brahm}]{Espinoza_2019}
Espinoza, N., Kossakowski, D., \& Brahm, R. 2019, \mnras, 490, 2262–2283

\bibitem[{Evans {et~al.}(2015)Evans, Aigrain, Gibson, Barstow, Amundsen,
  Tremblin, \& Mourier}]{Evans2015}
Evans, T.~M., Aigrain, S., Gibson, N., {et~al.} 2015, \mnras, 451, 680

\bibitem[{Foreman-Mackey(2016)}]{corner}
Foreman-Mackey, D. 2016, The Journal of Open Source Software, 1, 24

\bibitem[{{Foreman-Mackey} {et~al.}(2017){Foreman-Mackey}, {Agol},
  {Ambikasaran}, \& {Angus}}]{Mackey2017}
{Foreman-Mackey}, D., {Agol}, E., {Ambikasaran}, S., \& {Angus}, R. 2017, \aj,
  154, 220

\bibitem[{{Foreman-Mackey} {et~al.}(2021){Foreman-Mackey}, {Luger}, {Agol},
  {Barclay}, {Bouma}, {Brandt}, {Czekala}, {David}, {Dong}, {Gilbert},
  {Gordon}, {Hedges}, {Hey}, {Morris}, {Price-Whelan}, \& {Savel}}]{exoplanet}
{Foreman-Mackey}, D., {Luger}, R., {Agol}, E., {et~al.} 2021, {exoplanet:
  Gradient-based probabilistic inference for exoplanet data \& other
  astronomical time series}

\bibitem[{{Fortune} {et~al.}(2024){Fortune}, {Gibson}, {Foreman-Mackey},
  {Evans-Soma}, {Maguire}, \& {Ramkumar}}]{Fortune2024}
{Fortune}, M., {Gibson}, N.~P., {Foreman-Mackey}, D., {et~al.} 2024, \aap, 686,
  A89

\bibitem[{{Gaia Collaboration} {et~al.}(2021){Gaia Collaboration}, {Brown},
  {Vallenari}, {Prusti}, {de Bruijne}, {Babusiaux}, {Biermann}, {Creevey},
  {Evans}, {Eyer}, {Hutton}, {Jansen}, {Jordi}, {Klioner}, {Lammers},
  {Lindegren}, {Luri}, {Mignard}, {Panem}, {Pourbaix}, {Randich}, {Sartoretti},
  {Soubiran}, {Walton}, {Arenou}, {Bailer-Jones}, {Bastian}, {Cropper},
  {Drimmel}, {Katz}, {Lattanzi}, {van Leeuwen}, {Bakker}, {Cacciari},
  {Casta{\~n}eda}, {De Angeli}, {Ducourant}, {Fabricius}, {Fouesneau},
  {Fr{\'e}mat}, {Guerra}, {Guerrier}, {Guiraud}, {Jean-Antoine Piccolo},
  {Masana}, {Messineo}, {Mowlavi}, {Nicolas}, {Nienartowicz}, {Pailler},
  {Panuzzo}, {Riclet}, {Roux}, {Seabroke}, {Sordo}, {Tanga}, {Th{\'e}venin},
  {Gracia-Abril}, {Portell}, {Teyssier}, {Altmann}, {Andrae}, {Bellas-Velidis},
  {Benson}, {Berthier}, {Blomme}, {Brugaletta}, {Burgess}, {Busso}, {Carry},
  {Cellino}, {Cheek}, {Clementini}, {Damerdji}, {Davidson}, {Delchambre},
  {Dell'Oro}, {Fern{\'a}ndez-Hern{\'a}ndez}, {Galluccio}, {Garc{\'\i}a-Lario},
  {Garcia-Reinaldos}, {Gonz{\'a}lez-N{\'u}{\~n}ez}, {Gosset}, {Haigron},
  {Halbwachs}, {Hambly}, {Harrison}, {Hatzidimitriou}, {Heiter},
  {Hern{\'a}ndez}, {Hestroffer}, {Hodgkin}, {Holl}, {Jan{\ss}en}, {Jevardat de
  Fombelle}, {Jordan}, {Krone-Martins}, {Lanzafame}, {L{\"o}ffler}, {Lorca},
  {Manteiga}, {Marchal}, {Marrese}, {Moitinho}, {Mora}, {Muinonen}, {Osborne},
  {Pancino}, {Pauwels}, {Petit}, {Recio-Blanco}, {Richards}, {Riello},
  {Rimoldini}, {Robin}, {Roegiers}, {Rybizki}, {Sarro}, {Siopis}, {Smith},
  {Sozzetti}, {Ulla}, {Utrilla}, {van Leeuwen}, {van Reeven}, {Abbas}, {Abreu
  Aramburu}, {Accart}, {Aerts}, {Aguado}, {Ajaj}, {Altavilla}, {{\'A}lvarez},
  {{\'A}lvarez Cid-Fuentes}, {Alves}, {Anderson}, {Anglada Varela}, {Antoja},
  {Audard}, {Baines}, {Baker}, {Balaguer-N{\'u}{\~n}ez}, {Balbinot}, {Balog},
  {Barache}, {Barbato}, {Barros}, {Barstow}, {Bartolom{\'e}}, {Bassilana},
  {Bauchet}, {Baudesson-Stella}, {Becciani}, {Bellazzini}, {Bernet}, {Bertone},
  {Bianchi}, {Blanco-Cuaresma}, {Boch}, {Bombrun}, {Bossini}, {Bouquillon},
  {Bragaglia}, {Bramante}, {Breedt}, {Bressan}, {Brouillet}, {Bucciarelli},
  {Burlacu}, {Busonero}, {Butkevich}, {Buzzi}, {Caffau}, {Cancelliere},
  {C{\'a}novas}, {Cantat-Gaudin}, {Carballo}, {Carlucci}, {Carnerero},
  {Carrasco}, {Casamiquela}, {Castellani}, {Castro-Ginard}, {Castro Sampol},
  {Chaoul}, {Charlot}, {Chemin}, {Chiavassa}, {Cioni}, {Comoretto}, {Cooper},
  {Cornez}, {Cowell}, {Crifo}, {Crosta}, {Crowley}, {Dafonte}, {Dapergolas},
  {David}, {David}, {de Laverny}, {De Luise}, {De March}, {De Ridder}, {de
  Souza}, {de Teodoro}, {de Torres}, {del Peloso}, {del Pozo}, {Delbo},
  {Delgado}, {Delgado}, {Delisle}, {Di Matteo}, {Diakite}, {Diener},
  {Distefano}, {Dolding}, {Eappachen}, {Edvardsson}, {Enke}, {Esquej}, {Fabre},
  {Fabrizio}, {Faigler}, {Fedorets}, {Fernique}, {Fienga}, {Figueras},
  {Fouron}, {Fragkoudi}, {Fraile}, {Franke}, {Gai}, {Garabato},
  {Garcia-Gutierrez}, {Garc{\'\i}a-Torres}, {Garofalo}, {Gavras}, {Gerlach},
  {Geyer}, {Giacobbe}, {Gilmore}, {Girona}, {Giuffrida}, {Gomel}, {Gomez},
  {Gonzalez-Santamaria}, {Gonz{\'a}lez-Vidal}, {Granvik},
  {Guti{\'e}rrez-S{\'a}nchez}, {Guy}, {Hauser}, {Haywood}, {Helmi}, {Hidalgo},
  {Hilger}, {H{\l}adczuk}, {Hobbs}, {Holland}, {Huckle}, {Jasniewicz},
  {Jonker}, {Juaristi Campillo}, {Julbe}, {Karbevska}, {Kervella}, {Khanna},
  {Kochoska}, {Kontizas}, {Kordopatis}, {Korn}, {Kostrzewa-Rutkowska},
  {Kruszy{\'n}ska}, {Lambert}, {Lanza}, {Lasne}, {Le Campion}, {Le Fustec},
  {Lebreton}, {Lebzelter}, {Leccia}, {Leclerc}, {Lecoeur-Taibi}, {Liao},
  {Licata}, {Lindstr{\o}m}, {Lister}, {Livanou}, {Lobel}, {Madrero Pardo},
  {Managau}, {Mann}, {Marchant}, {Marconi}, {Marcos Santos}, {Marinoni},
  {Marocco}, {Marshall}, {Martin Polo}, {Mart{\'\i}n-Fleitas}, {Masip},
  {Massari}, {Mastrobuono-Battisti}, {Mazeh}, {McMillan}, {Messina},
  {Michalik}, {Millar}, {Mints}, {Molina}, {Molinaro}, {Moln{\'a}r},
  {Montegriffo}, {Mor}, {Morbidelli}, {Morel}, {Morris}, {Mulone}, {Munoz},
  {Muraveva}, {Murphy}, {Musella}, {Noval}, {Ord{\'e}novic}, {Orr{\`u}},
  {Osinde}, {Pagani}, {Pagano}, {Palaversa}, {Palicio}, {Panahi}, {Pawlak},
  {Pe{\~n}alosa Esteller}, {Penttil{\"a}}, {Piersimoni}, {Pineau}, {Plachy},
  {Plum}, {Poggio}, {Poretti}, {Poujoulet}, {Pr{\v{s}}a}, {Pulone}, {Racero},
  {Ragaini}, {Rainer}, {Raiteri}, {Rambaux}, {Ramos}, {Ramos-Lerate}, {Re
  Fiorentin}, {Regibo}, {Reyl{\'e}}, {Ripepi}, {Riva}, {Rixon}, {Robichon},
  {Robin}, {Roelens}, {Rohrbasser}, {Romero-G{\'o}mez}, {Rowell}, {Royer},
  {Rybicki}, {Sadowski}, {Sagrist{\`a} Sell{\'e}s}, {Sahlmann}, {Salgado},
  {Salguero}, {Samaras}, {Sanchez Gimenez}, {Sanna}, {Santove{\~n}a},
  {Sarasso}, {Schultheis}, {Sciacca}, {Segol}, {Segovia}, {S{\'e}gransan},
  {Semeux}, {Shahaf}, {Siddiqui}, {Siebert}, {Siltala}, {Slezak}, {Smart},
  {Solano}, {Solitro}, {Souami}, {Souchay}, {Spagna}, {Spoto}, {Steele},
  {Steidelm{\"u}ller}, {Stephenson}, {S{\"u}veges}, {Szabados}, {Szegedi-Elek},
  {Taris}, {Tauran}, {Taylor}, {Teixeira}, {Thuillot}, {Tonello}, {Torra},
  {Torra}, {Turon}, {Unger}, {Vaillant}, {van Dillen}, {Vanel}, {Vecchiato},
  {Viala}, {Vicente}, {Voutsinas}, {Weiler}, {Wevers}, {Wyrzykowski}, {Yoldas},
  {Yvard}, {Zhao}, {Zorec}, {Zucker}, {Zurbach}, \& {Zwitter}}]{gaia}
{Gaia Collaboration}, {Brown}, A.~G.~A., {Vallenari}, A., {et~al.} 2021, \aap,
  649, A1

\bibitem[{{Gehrels} {et~al.}(2004){Gehrels}, {Chincarini}, {Giommi}, {Mason},
  {Nousek}, {Wells}, {White}, {Barthelmy}, {Burrows}, {Cominsky}, {Hurley},
  {Marshall}, {M{\'e}sz{\'a}ros}, {Roming}, {Angelini}, {Barbier}, {Belloni},
  {Campana}, {Caraveo}, {Chester}, {Citterio}, {Cline}, {Cropper}, {Cummings},
  {Dean}, {Feigelson}, {Fenimore}, {Frail}, {Fruchter}, {Garmire}, {Gendreau},
  {Ghisellini}, {Greiner}, {Hill}, {Hunsberger}, {Krimm}, {Kulkarni}, {Kumar},
  {Lebrun}, {Lloyd-Ronning}, {Markwardt}, {Mattson}, {Mushotzky}, {Norris},
  {Osborne}, {Paczynski}, {Palmer}, {Park}, {Parsons}, {Paul}, {Rees},
  {Reynolds}, {Rhoads}, {Sasseen}, {Schaefer}, {Short}, {Smale}, {Smith},
  {Stella}, {Tagliaferri}, {Takahashi}, {Tashiro}, {Townsley}, {Tueller},
  {Turner}, {Vietri}, {Voges}, {Ward}, {Willingale}, {Zerbi}, \&
  {Zhang}}]{2004ApJ...611.1005G}
{Gehrels}, N., {Chincarini}, G., {Giommi}, P., {et~al.} 2004, \apj, 611, 1005

\bibitem[{{Gelman} \& {Rubin}(1992)}]{1992StaSc...7..457G}
{Gelman}, A. \& {Rubin}, D.~B. 1992, Statistical Science, 7, 457

\bibitem[{{Gibson}(2014)}]{Gibson2014}
{Gibson}, N.~P. 2014, \mnras, 445, 3401

\bibitem[{{Gibson} {et~al.}(2013){Gibson}, {Aigrain}, {Barstow}, {Evans},
  {Fletcher}, \& {Irwin}}]{Gibson2013}
{Gibson}, N.~P., {Aigrain}, S., {Barstow}, J.~K., {et~al.} 2013, \mnras, 428,
  3680

\bibitem[{{Gibson} {et~al.}(2012){Gibson}, {Aigrain}, {Roberts}, {Evans},
  {Osborne}, \& {Pont}}]{2012MNRAS.419.2683G}
{Gibson}, N.~P., {Aigrain}, S., {Roberts}, S., {et~al.} 2012, \mnras, 419, 2683

\bibitem[{{Glasse} {et~al.}(2015){Glasse}, {Rieke}, {Bauwens},
  {Garc{\'\i}a-Mar{\'\i}n}, {Ressler}, {Rost}, {Tikkanen}, {Vandenbussche}, \&
  {Wright}}]{background}
{Glasse}, A., {Rieke}, G.~H., {Bauwens}, E., {et~al.} 2015, \pasp, 127, 686

\bibitem[{{Gomes} \& {Ferraz-Mello}(2020)}]{modeleccentricity}
{Gomes}, G.~O. \& {Ferraz-Mello}, S. 2020, \mnras, 494, 5082

\bibitem[{{Gordon} {et~al.}(2022){Gordon}, {Rothman}, {Hargreaves}, {Hashemi},
  {Karlovets}, {Skinner}, {Conway}, {Hill}, {Kochanov}, {Tan}, {Wcis{\l}o},
  {Finenko}, {Nelson}, {Bernath}, {Birk}, {Boudon}, {Campargue}, {Chance},
  {Coustenis}, {Drouin}, {Flaud}, {Gamache}, {Hodges}, {Jacquemart}, {Mlawer},
  {Nikitin}, {Perevalov}, {Rotger}, {Tennyson}, {Toon}, {Tran}, {Tyuterev},
  {Adkins}, {Baker}, {Barbe}, {Can{\`e}}, {Cs{\'a}sz{\'a}r}, {Dudaryonok},
  {Egorov}, {Fleisher}, {Fleurbaey}, {Foltynowicz}, {Furtenbacher}, {Harrison},
  {Hartmann}, {Horneman}, {Huang}, {Karman}, {Karns}, {Kassi}, {Kleiner},
  {Kofman}, {Kwabia-Tchana}, {Lavrentieva}, {Lee}, {Long}, {Lukashevskaya},
  {Lyulin}, {Makhnev}, {Matt}, {Massie}, {Melosso}, {Mikhailenko}, {Mondelain},
  {M{\"u}ller}, {Naumenko}, {Perrin}, {Polyansky}, {Raddaoui}, {Raston},
  {Reed}, {Rey}, {Richard}, {T{\'o}bi{\'a}s}, {Sadiek}, {Schwenke},
  {Starikova}, {Sung}, {Tamassia}, {Tashkun}, {Vander Auwera}, {Vasilenko},
  {Vigasin}, {Villanueva}, {Vispoel}, {Wagner}, {Yachmenev}, \&
  {Yurchenko}}]{O2_crosssection}
{Gordon}, I.~E., {Rothman}, L.~S., {Hargreaves}, R.~J., {et~al.} 2022, \jqsrt,
  277, 107949

\bibitem[{{Gordon} {et~al.}(2025){Gordon}, {Sloan}, {Garcia Marin},
  {Libralato}, {Rieke}, {Aguilar}, {Bohlin}, {Cracraft}, {Decleir}, {Gaspar},
  {Kendrew}, {Law}, {Noriega-Crespo}, \& {Regan}}]{flux_cal}
{Gordon}, K.~D., {Sloan}, G.~C., {Garcia Marin}, M., {et~al.} 2025, \aj, 169, 6

\bibitem[{{Greene} {et~al.}(2023){Greene}, {Bell}, {Ducrot}, {Dyrek}, {Lagage},
  \& {Fortney}}]{2023Natur.618...39G}
{Greene}, T.~P., {Bell}, T.~J., {Ducrot}, E., {et~al.} 2023, \nat, 618, 39

\bibitem[{{Gressier} {et~al.}(2024){Gressier}, {Espinoza}, {Allen}, {Sing},
  {Banerjee}, {Barstow}, {Valenti}, {Lewis}, {Birkmann}, {Challener},
  {Manjavacas}, {Alves de Oliveira}, {Crouzet}, \& {Beck}}]{sulphur}
{Gressier}, A., {Espinoza}, N., {Allen}, N.~H., {et~al.} 2024, \apjl, 975, L10

\bibitem[{{Grimm} \& {Heng}(2015)}]{heliosk}
{Grimm}, S.~L. \& {Heng}, K. 2015, \apj, 808, 182

\bibitem[{{Hammond} {et~al.}(2025){Hammond}, {Guimond}, {Lichtenberg},
  {Nicholls}, {Fisher}, {Luque}, {Meier}, {Taylor}, {Changeat}, {Dang}, {Hay},
  {Herbort}, \& {Teske}}]{Hammond2024}
{Hammond}, M., {Guimond}, C.~M., {Lichtenberg}, T., {et~al.} 2025, \apjl, 978,
  L40

\bibitem[{{Hansen}(2008)}]{Hansen2008}
{Hansen}, B. M.~S. 2008, \apjs, 179, 484

\bibitem[{Hapke(1993)}]{Hapke_1993}
Hapke, B. 1993, Theory of Reflectance and Emittance Spectroscopy, Topics in
  Remote Sensing (Cambridge University Press)

\bibitem[{Harris {et~al.}(2020)Harris, Millman, van~der Walt, Gommers,
  Virtanen, Cournapeau, Wieser, Taylor, Berg, Smith, Kern, Picus, Hoyer, van
  Kerkwijk, Brett, Haldane, del R{\'{i}}o, Wiebe, Peterson,
  G{\'{e}}rard-Marchant, Sheppard, Reddy, Weckesser, Abbasi, Gohlke, \&
  Oliphant}]{numpy}
Harris, C.~R., Millman, K.~J., van~der Walt, S.~J., {et~al.} 2020, Nature, 585,
  357

\bibitem[{Hattori {et~al.}(2024)Hattori, Garcia, Murray, Dong, Dholakia, Degen,
  \& Foreman-Mackey}]{jaxoplanet}
Hattori, S., Garcia, L., Murray, C., {et~al.} 2024, {exoplanet-dev/jaxoplanet:
  Astronomical time series analysis with JAX}

\bibitem[{{Heng} {et~al.}(2021){Heng}, {Morris}, \& {Kitzmann}}]{Heng2021}
{Heng}, K., {Morris}, B.~M., \& {Kitzmann}, D. 2021, Nature Astronomy, 5, 1001

\bibitem[{Hoffman \& Gelman(2014)}]{Hoffman2011}
Hoffman, M.~D. \& Gelman, A. 2014, Journal of Machine Learning Research, 15,
  1593–1623

\bibitem[{{Horne}(1986)}]{optimal_extraction}
{Horne}, K. 1986, \pasp, 98, 609

\bibitem[{{Hu} {et~al.}(2024){Hu}, {Bello-Arufe}, {Zhang}, {Paragas},
  {Zilinskas}, {van Buchem}, {Bess}, {Patel}, {Ito}, {Damiano}, {Scheucher},
  {Oza}, {Knutson}, {Miguel}, {Dragomir}, {Brandeker}, \& {Demory}}]{55cancrie}
{Hu}, R., {Bello-Arufe}, A., {Zhang}, M., {et~al.} 2024, \nat, 630, 609

\bibitem[{Hunter(2007)}]{matplotlib}
Hunter, J.~D. 2007, Computing in Science \& Engineering, 9, 90

\bibitem[{{Ih} \& {Kempton}(2021)}]{Ih2021}
{Ih}, J. \& {Kempton}, E. M.~R. 2021, \aj, 162, 237

\bibitem[{{Ih} {et~al.}(2023){Ih}, {Kempton}, {Whittaker}, \&
  {Lessard}}]{Ih2023}
{Ih}, J., {Kempton}, E. M.~R., {Whittaker}, E.~A., \& {Lessard}, M. 2023,
  \apjl, 952, L4

\bibitem[{{Ingalls} {et~al.}(2012){Ingalls}, {Krick}, {Carey}, {Laine},
  {Surace}, {Glaccum}, {Grillmair}, \& {Lowrance}}]{Ingalls2012}
{Ingalls}, J.~G., {Krick}, J.~E., {Carey}, S.~J., {et~al.} 2012, in Society of
  Photo-Optical Instrumentation Engineers (SPIE) Conference Series, Vol. 8442,
  Space Telescopes and Instrumentation 2012: Optical, Infrared, and Millimeter
  Wave, ed. M.~C. {Clampin}, G.~G. {Fazio}, H.~A. {MacEwen}, \& J.~M.
  {Oschmann}, Jr., 84421Y

\bibitem[{Jensen {et~al.}(1995)Jensen, Kjærulff, \& Kong}]{blocked_Gibbs}
Jensen, C.~S., Kjærulff, U., \& Kong, A. 1995, International Journal of
  Human-Computer Studies, 42, 647

\bibitem[{Joshi {et~al.}(1997)Joshi, Haberle, \& Reynolds}]{Joshi1997}
Joshi, M., Haberle, R., \& Reynolds, R. 1997, Icarus, 129, 450

\bibitem[{{Kite} \& {Barnett}(2020)}]{volcano_degas}
{Kite}, E.~S. \& {Barnett}, M.~N. 2020, Proceedings of the National Academy of
  Science, 117, 18264

\bibitem[{{Knutson} {et~al.}(2008){Knutson}, {Charbonneau}, {Allen}, {Burrows},
  \& {Megeath}}]{spitzer_intrapixelsensitivity}
{Knutson}, H.~A., {Charbonneau}, D., {Allen}, L.~E., {Burrows}, A., \&
  {Megeath}, S.~T. 2008, \apj, 673, 526

\bibitem[{{Koll}(2022)}]{Kollf}
{Koll}, D. D.~B. 2022, \apj, 924, 134

\bibitem[{{Koll} {et~al.}(2019){Koll}, {Malik}, {Mansfield}, {Kempton}, {Kite},
  {Abbot}, \& {Bean}}]{Koll2019}
{Koll}, D. D.~B., {Malik}, M., {Mansfield}, M., {et~al.} 2019, \apj, 886, 140

\bibitem[{{Kreidberg} {et~al.}(2019){Kreidberg}, {Koll}, {Morley}, {Hu},
  {Schaefer}, {Deming}, {Stevenson}, {Dittmann}, {Vanderburg}, {Berardo},
  {Guo}, {Stassun}, {Crossfield}, {Charbonneau}, {Latham}, {Loeb}, {Ricker},
  {Seager}, \& {Vanderspek}}]{2019Natur.573...87K}
{Kreidberg}, L., {Koll}, D. D.~B., {Morley}, C., {et~al.} 2019, \nat, 573, 87

\bibitem[{{Krissansen-Totton} {et~al.}(2024){Krissansen-Totton}, {Wogan},
  {Thompson}, \& {Fortney}}]{Totton2024}
{Krissansen-Totton}, J., {Wogan}, N., {Thompson}, M., \& {Fortney}, J.~J. 2024,
  Nature Communications, 15, 8374

\bibitem[{Kumar {et~al.}(2019)Kumar, Carroll, Hartikainen, \&
  Martin}]{arviz_2019}
Kumar, R., Carroll, C., Hartikainen, A., \& Martin, O. 2019, Journal of Open
  Source Software, 4, 1143

\bibitem[{{Luger} \& {Barnes}(2015)}]{Luger2015}
{Luger}, R. \& {Barnes}, R. 2015, Astrobiology, 15, 119

\bibitem[{{Mahajan} {et~al.}(2024){Mahajan}, {Eastman}, \&
  {Kirk}}]{Mahajan2024}
{Mahajan}, A.~S., {Eastman}, J.~D., \& {Kirk}, J. 2024, \apjl, 963, L37

\bibitem[{{Malik} {et~al.}(2017){Malik}, {Grosheintz}, {Mendon{\c{c}}a},
  {Grimm}, {Lavie}, {Kitzmann}, {Tsai}, {Burrows}, {Kreidberg}, {Bedell},
  {Bean}, {Stevenson}, \& {Heng}}]{heliosa}
{Malik}, M., {Grosheintz}, L., {Mendon{\c{c}}a}, J.~M., {et~al.} 2017, \aj,
  153, 56

\bibitem[{{Malik} {et~al.}(2019{\natexlab{a}}){Malik}, {Kempton}, {Koll},
  {Mansfield}, {Bean}, \& {Kite}}]{heliossurfa}
{Malik}, M., {Kempton}, E. M.~R., {Koll}, D. D.~B., {et~al.}
  2019{\natexlab{a}}, \apj, 886, 142

\bibitem[{{Malik} {et~al.}(2019{\natexlab{b}}){Malik}, {Kitzmann},
  {Mendon{\c{c}}a}, {Grimm}, {Marleau}, {Linder}, {Tsai}, \& {Heng}}]{heliosb}
{Malik}, M., {Kitzmann}, D., {Mendon{\c{c}}a}, J.~M., {et~al.}
  2019{\natexlab{b}}, \aj, 157, 170

\bibitem[{{Mandel} \& {Agol}(2002)}]{mandelagol}
{Mandel}, K. \& {Agol}, E. 2002, \apj, 580, L171

\bibitem[{{Mansfield} {et~al.}(2019){Mansfield}, {Kite}, {Hu}, {Koll}, {Malik},
  {Bean}, \& {Kempton}}]{Mansfield2019}
{Mansfield}, M., {Kite}, E.~S., {Hu}, R., {et~al.} 2019, \apj, 886, 141

\bibitem[{{Meier Vald{\'e}s} {et~al.}(2025){Meier Vald{\'e}s}, {Demory},
  {Diamond-Lowe}, {Mendon{\c{c}}a}, {August}, {Fortune}, {Allen}, {Kitzmann},
  {Gressier}, {Hooton}, {Jones}, {Buchhave}, {Espinoza}, {Fisher}, {Gibson},
  {Heng}, {Hoeijmakers}, {Prinoth}, {Rathcke}, \& {Eastman}}]{Meier2025}
{Meier Vald{\'e}s}, E.~A., {Demory}, B.~O., {Diamond-Lowe}, H., {et~al.} 2025,
  \aap, accepted, arXiv:2503.19772

\bibitem[{{Ment} {et~al.}(2019){Ment}, {Dittmann}, {Astudillo-Defru},
  {Charbonneau}, {Irwin}, {Bonfils}, {Murgas}, {Almenara}, {Forveille}, {Agol},
  {Ballard}, {Berta-Thompson}, {Bouchy}, {Cloutier}, {Delfosse}, {Doyon},
  {Dressing}, {Esquerdo}, {Haywood}, {Kipping}, {Latham}, {Lovis}, {Newton},
  {Pepe}, {Rodriguez}, {Santos}, {Tan}, {Udry}, {Winters}, \&
  {W{\"u}nsche}}]{2019AJ....157...32M}
{Ment}, K., {Dittmann}, J.~A., {Astudillo-Defru}, N., {et~al.} 2019, \aj, 157,
  32

\bibitem[{{Moore} {et~al.}(2006){Moore}, {Ninkov}, \&
  {Forrest}}]{2006OptEn..45g6402M}
{Moore}, A.~C., {Ninkov}, Z., \& {Forrest}, W.~J. 2006, Optical Engineering,
  45, 076402

\bibitem[{{Morello}(2015)}]{Morello2015}
{Morello}, G. 2015, \apj, 808, 56

\bibitem[{{Morello} {et~al.}(2016){Morello}, {Waldmann}, \&
  {Tinetti}}]{Morello2016}
{Morello}, G., {Waldmann}, I.~P., \& {Tinetti}, G. 2016, \apj, 820, 86

\bibitem[{{Morello} {et~al.}(2014){Morello}, {Waldmann}, {Tinetti}, {Peres},
  {Micela}, \& {Howarth}}]{Morello2014}
{Morello}, G., {Waldmann}, I.~P., {Tinetti}, G., {et~al.} 2014, \apj, 786, 22

\bibitem[{{Morrison} {et~al.}(2023){Morrison}, {Dicken}, {Argyriou}, {Ressler},
  {Gordon}, {Regan}, {Cracraft}, {Rieke}, {Engesser}, {Alberts},
  {Alvarez-Marquez}, {Colbert}, {Fox}, {Gasman}, {Law}, {Garcia Marin},
  {G{\'a}sp{\'a}r}, {Guillard}, {Kendrew}, {Labiano}, {Laine},
  {Noriega-Crespo}, {Shivaei}, \& {Sloan}}]{2023PASP..135g5004M}
{Morrison}, J.~E., {Dicken}, D., {Argyriou}, I., {et~al.} 2023, \pasp, 135,
  075004

\bibitem[{Morrison {et~al.}(2023)Morrison, Dicken, Argyriou, Ressler, Gordon,
  Regan, Cracraft, Rieke, Engesser, Alberts, Alvarez-Marquez, Colbert, Fox,
  Gasman, Law, Marin, Gáspár, Guillard, Kendrew, Labiano, Laine,
  Noriega-Crespo, Shivaei, \& Sloan}]{Morrison_2023}
Morrison, J.~E., Dicken, D., Argyriou, I., {et~al.} 2023, \pasa, 135, 075004

\bibitem[{Neal(2003)}]{Neal2003}
Neal, R.~M. 2003, The Annals of Statistics, 31, 705

\bibitem[{pandas~development team(2020)}]{pandas}
pandas~development team, T. 2020, pandas-dev/pandas: Pandas

\bibitem[{{Park Coy} {et~al.}(2024){Park Coy}, {Ih}, {Kite}, {Koll},
  {Tenthoff}, {Bean}, {Weiner Mansfield}, {Zhang}, {Xue}, {Kempton},
  {Wolhfarth}, {Hu}, {Lyu}, \& {Wohler}}]{Coy2024}
{Park Coy}, B., {Ih}, J., {Kite}, E.~S., {et~al.} 2024, arXiv e-prints,
  arXiv:2412.06573

\bibitem[{{Pass} {et~al.}(2025){Pass}, {Charbonneau}, \&
  {Vanderburg}}]{Pass2025}
{Pass}, E.~K., {Charbonneau}, D., \& {Vanderburg}, A. 2025, \apjl, accepted,
  arXiv:2504.01182

\bibitem[{{Patel} {et~al.}(2024){Patel}, {Brandeker}, {Kitzmann}, {Petit dit de
  la Roche}, {Bello-Arufe}, {Heng}, {Meier Vald{\'e}s}, {Persson}, {Zhang},
  {Demory}, {Bourrier}, {Deline}, {Ehrenreich}, {Fridlund}, {Hu}, {Lendl},
  {Oza}, {Alibert}, \& {Hooton}}]{55cancrie2}
{Patel}, J.~A., {Brandeker}, A., {Kitzmann}, D., {et~al.} 2024, \aap, 690, A159

\bibitem[{P\'erez \& Granger(2007)}]{ipython}
P\'erez, F. \& Granger, B.~E. 2007, Computing in Science and Engineering, 9, 21

\bibitem[{{Perrin} {et~al.}(2014){Perrin}, {Sivaramakrishnan}, {Lajoie},
  {Elliott}, {Pueyo}, {Ravindranath}, \& {Albert}}]{webbpsf}
{Perrin}, M.~D., {Sivaramakrishnan}, A., {Lajoie}, C.-P., {et~al.} 2014, in
  Society of Photo-Optical Instrumentation Engineers (SPIE) Conference Series,
  Vol. 9143, Space Telescopes and Instrumentation 2014: Optical, Infrared, and
  Millimeter Wave, ed. J.~M. {Oschmann}, Jr., M.~{Clampin}, G.~G. {Fazio}, \&
  H.~A. {MacEwen}, 91433X

\bibitem[{{Rackham} {et~al.}(2018){Rackham}, {Apai}, \&
  {Giampapa}}]{2018ApJ...853..122R}
{Rackham}, B.~V., {Apai}, D., \& {Giampapa}, M.~S. 2018, \apj, 853, 122

\bibitem[{Rakitsch {et~al.}(2013)Rakitsch, Lippert, Borgwardt, \&
  Stegle}]{Rakitsch2013}
Rakitsch, B., Lippert, C., Borgwardt, K., \& Stegle, O. 2013, in Advances in
  Neural Information Processing Systems, Vol.~26

\bibitem[{{Redfield} {et~al.}(2024){Redfield}, {Batalha}, {Benneke}, {Biller},
  {Espinoza}, {France}, {Konopacky}, {Kreidberg}, {Rauscher}, \& {Sing}}]{DDT}
{Redfield}, S., {Batalha}, N., {Benneke}, B., {et~al.} 2024, arXiv e-prints,
  arXiv:2404.02932

\bibitem[{{Rigby} {et~al.}(2023){Rigby}, {Perrin}, {McElwain}, {Kimble},
  {Friedman}, {Lallo}, {Doyon}, {Feinberg}, {Ferruit}, {Glasse}, {Rieke},
  {Rieke}, {Wright}, {Willott}, {Colon}, {Milam}, {Neff}, {Stark}, {Valenti},
  {Abell}, {Abney}, {Abul-Huda}, {Acton}, {Adams}, {Adler}, {Aguilar}, {Ahmed},
  {Albert}, {Alberts}, {Aldridge}, {Allen}, {Altenburg},
  {{\'A}lvarez-M{\'a}rquez}, {Alves de Oliveira}, {Andersen}, {Anderson},
  {Anderson}, {Argyriou}, {Armstrong}, {Arribas}, {Artigau}, {Arvai},
  {Atkinson}, {Bacon}, {Bair}, {Banks}, {Barrientes}, {Barringer}, {Bartosik},
  {Bast}, {Baudoz}, {Beatty}, {Bechtold}, {Beck}, {Bergeron}, {Bergkoetter},
  {Bhatawdekar}, {Birkmann}, {Blazek}, {Blome}, {Boccaletti}, {B{\"o}ker},
  {Boia}, {Bonaventura}, {Bond}, {Bosley}, {Boucarut}, {Bourque}, {Bouwman},
  {Bower}, {Bowers}, {Boyer}, {Bradley}, {Brady}, {Braun}, {Breda},
  {Bresnahan}, {Bright}, {Britt}, {Bromenschenkel}, {Brooks}, {Brooks},
  {Brown}, {Brown}, {Brown}, {Bunker}, {Burger}, {Bushouse}, {Cale}, {Cameron},
  {Cameron}, {Canipe}, {Caplinger}, {Caputo}, {Cara}, {Carey}, {Carniani},
  {Carrasquilla}, {Carruthers}, {Case}, {Catherine}, {Chance}, {Chapman},
  {Charlot}, {Charlow}, {Chayer}, {Chen}, {Cherinka}, {Chichester}, {Chilton},
  {Chonis}, {Clampin}, {Clark}, {Clark}, {Coe}, {Coleman}, {Comber}, {Comeau},
  {Connolly}, {Cooper}, {Cooper}, {Coppock}, {Correnti}, {Cossou}, {Coulais},
  {Coyle}, {Cracraft}, {Curti}, {Cuturic}, {Davis}, {Davis}, {Dean}, {DeLisa},
  {deMeester}, {Dencheva}, {Dencheva}, {DePasquale}, {Deschenes}, {Hunor
  Detre}, {Diaz}, {Dicken}, {DiFelice}, {Dillman}, {Dixon}, {Doggett},
  {Donaldson}, {Douglas}, {DuPrie}, {Dupuis}, {Durning}, {Easmin}, {Eck},
  {Edeani}, {Egami}, {Ehrenwinkler}, {Eisenhamer}, {Eisenhower}, {Elie},
  {Elliott}, {Elliott}, {Ellis}, {Engesser}, {Espinoza}, {Etienne}, {Etxaluze},
  {Falini}, {Feeney}, {Ferry}, {Filippazzo}, {Fincham}, {Fix}, {Flagey},
  {Florian}, {Flynn}, {Fontanella}, {Ford}, {Forshay}, {Fox}, {Franz}, {Fu},
  {Fullerton}, {Galkin}, {Galyer}, {Garc{\'\i}a Mar{\'\i}n}, {Gardner},
  {Gardner}, {Garland}, {Garrett}, {Gasman}, {Gaspar}, {Gaudreau}, {Gauthier},
  {Geers}, {Geithner}, {Gennaro}, {Giardino}, {Girard}, {Giuliano},
  {Glassmire}, {Glauser}, {Glazer}, {Godfrey}, {Golimowski}, {Gollnitz},
  {Gong}, {Gonzaga}, {Gordon}, {Gordon}, {Goudfrooij}, {Greene}, {Greenhouse},
  {Grimaldi}, {Groebner}, {Grundy}, {Guillard}, {Gutman}, {Ha}, {Haderlein},
  {Hagedorn}, {Hainline}, {Haley}, {Hami}, {Hamilton}, {Hammel}, {Hansen},
  {Harkins}, {Harr}, {Hart}, {Hart}, {Hartig}, {Hashimoto}, {Haskins},
  {Hathaway}, {Havey}, {Hayden}, {Hecht}, {Heller-Boyer}, {Henriques}, {Henry},
  {Hermann}, {Hernandez}, {Hesman}, {Hicks}, {Hilbert}, {Hines}, {Hoffman},
  {Holfeltz}, {Holler}, {Hoppa}, {Hott}, {Howard}, {Howard}, {Hunter},
  {Hunter}, {Hurst}, {Husemann}, {Hustak}, {Ilinca Ignat}, {Illingworth},
  {Irish}, {Jackson}, {Jahromi}, {Jakobsen}, {James}, {James}, {Januszewski},
  {Jenkins}, {Jirdeh}, {Johnson}, {Johnson}, {Jones}, {Jones}, {Jones},
  {Jones}, {Jordan}, {Jordan}, {Jurczyk}, {Jurling}, {Kaleida}, {Kalmanson},
  {Kammerer}, {Kang}, {Kao}, {Karakla}, {Kavanagh}, {Kelly}, {Kendrew},
  {Kennedy}, {Kenny}, {Keski-kuha}, {Keyes}, {Kidwell}, {Kinzel}, {Kirk},
  {Kirkpatrick}, {Kirshenblat}, {Klaassen}, {Knapp}, {Knight}, {Knollenberg},
  {Koehler}, {Koekemoer}, {Kovacs}, {Kulp}, {Kumari}, {Kyprianou}, {La Massa},
  {Labador}, {Labiano}, {Lagage}, {Lajoie}, {Lallo}, {Lam}, {Lamb}, {Lambros},
  {Lampenfield}, {Langston}, {Larson}, {Law}, {Lawrence}, {Lee}, {Leisenring},
  {Lepo}, {Leveille}, {Levenson}, {Levine}, {Levy}, {Lewis}, {Lewis},
  {Libralato}, {Lightsey}, {Link}, {Liu}, {Lo}, {Lockwood}, {Logue}, {Long},
  {Long}, {Loomis}, {Lopez-Caniego}, {Lorenzo Alvarez}, {Love-Pruitt}, {Lucy},
  {Luetzgendorf}, {Maghami}, {Maiolino}, {Major}, {Malla}, {Malumuth},
  {Manjavacas}, {Mannfolk}, {Marrione}, {Marston}, {Martel}, {Maschmann},
  {Masci}, {Masciarelli}, {Maszkiewicz}, {Mather}, {McKenzie}, {McLean},
  {McMaster}, {Melbourne}, {Mel{\'e}ndez}, {Menzel}, {Merz}, {Meyett}, {Meza},
  {Miskey}, {Misselt}, {Moller}, {Morrison}, {Morse}, {Moseley}, {Mosier},
  {Mountain}, {Mueckay}, {Mueller}, {Mullally}, {Murphy}, {Murray}, {Murray},
  {Mustelier}, {Muzerolle}, {Mycroft}, {Myers}, {Myrick}, {Nanavati}, {Nance},
  {Nayak}, {Naylor}, {Nelan}, {Nickson}, {Nielson}, {Nieto-Santisteban},
  {Nikolov}, {Noriega-Crespo}, {O'Shaughnessy}, {O'Sullivan}, {Ochs}, {Ogle},
  {Oleszczuk}, {Olmsted}, {Osborne}, {Ottens}, {Owens}, {Pacifici}, {Pagan},
  {Page}, {Park}, {Parrish}, {Patapis}, {Paul}, {Pauly}, {Pavlovsky}, {Pedder},
  {Peek}, {Pena-Guerrero}, {Penanen}, {Perez}, {Perna}, {Perriello},
  {Phillips}, {Pietraszkiewicz}, {Pinaud}, {Pirzkal}, {Pitman}, {Piwowar},
  {Platais}, {Player}, {Plesha}, {Pollizi}, {Polster}, {Pontoppidan},
  {Porterfield}, {Proffitt}, {Pueyo}, {Pulliam}, {Quirt}, {Quispe Neira},
  {Ramos Alarcon}, {Ramsay}, {Rapp}, {Rapp}, {Rauscher}, {Ravindranath},
  {Rawle}, {Regan}, {Reichard}, {Reis}, {Ressler}, {Rest}, {Reynolds}, {Rhue},
  {Richon}, {Rickman}, {Ridgaway}, {Ritchie}, {Rix}, {Robberto}, {Robinson},
  {Robinson}, {Robinson}, {Rock}, {Rodriguez}, {Rodriguez Del Pino}, {Roellig},
  {Rohrbach}, {Roman}, {Romelfanger}, {Rose}, {Roteliuk}, {Roth}, {Rothwell},
  {Rowlands}, {Roy}, {Royer}, {Royle}, {Rui}, {Rumler}, {Runnels}, {Russ},
  {Rustamkulov}, {Ryden}, {Ryer}, {Sabata}, {Sabatke}, {Sabbi}, {Samuelson},
  {Sapp}, {Sappington}, {Sargent}, {Sauer}, {Scheithauer}, {Schlawin},
  {Schlitz}, {Schmitz}, {Schneider}, {Schreiber}, {Schulze}, {Schwab}, {Scott},
  {Sembach}, {Shanahan}, {Shaughnessy}, {Shaw}, {Shawger}, {Shay}, {Sheehan},
  {Shen}, {Sherman}, {Shiao}, {Shih}, {Shivaei}, {Sienkiewicz}, {Sing},
  {Sirianni}, {Sivaramakrishnan}, {Skipper}, {Sloan}, {Slocum}, {Slowinski},
  {Smith}, {Smith}, {Smith}, {Smith}, {Snyder}, {Soh}, {Sohn}, {Soto},
  {Spencer}, {Stallcup}, {Stansberry}, {Starr}, {Starr}, {Stewart},
  {Stiavelli}, {Straughn}, {Strickland}, {Stys}, {Summers}, {Sun}, {Sunnquist},
  {Swade}, {Swam}, {Swaters}, {Swoish}, {Taylor}, {Taylor}, {Te Plate}, {Tea},
  {Teague}, {Telfer}, {Temim}, {Thatte}, {Thompson}, {Thompson}, {Thomson},
  {Tikkanen}, {Tippet}, {Todd}, {Toolan}, {Tran}, {Trejo}, {Truong},
  {Tsukamoto}, {Tustain}, {Tyra}, {Ubeda}, {Underwood}, {Uzzo}, {Van Campen},
  {Vandal}, {Vandenbussche}, {Vila}, {Volk}, {Wahlgren}, {Waldman}, {Walker},
  {Wander}, {Warfield}, {Warner}, {Wasiak}, {Watkins}, {Weaver}, {Weilert},
  {Weiser}, {Weiss}, {Weissman}, {Welty}, {West}, {Wheate}, {Wheatley},
  {Wheeler}, {White}, {Whiteaker}, {Whitehouse}, {Whiteleather}, {Whitman},
  {Williams}, {Willmer}, {Willoughby}, {Wilson}, {Wirth}, {Wislowski}, {Wolf},
  {Wolfe}, {Wolff}, {Workman}, {Wright}, {Wu}, {Wu}, {Wymer}, {Yates},
  {Yeager}, {Yeates}, {Yerger}, {Yoon}, {Young}, {Yu}, {Zak}, {Zeidler},
  {Zhou}, {Zielinski}, {Zincke}, \& {Zonak}}]{JWST_mirrors}
{Rigby}, J., {Perrin}, M., {McElwain}, M., {et~al.} 2023, \pasp, 135, 048001

\bibitem[{{Rothman} {et~al.}(2010){Rothman}, {Gordon}, {Barber}, {Dothe},
  {Gamache}, {Goldman}, {Perevalov}, {Tashkun}, \& {Tennyson}}]{hitemp}
{Rothman}, L.~S., {Gordon}, I.~E., {Barber}, R.~J., {et~al.} 2010, \jqsrt, 111,
  2139

\bibitem[{Salvatier {et~al.}(2016)Salvatier, Wiecki, \&
  Fonnesbeck}]{Salvatier2015}
Salvatier, J., Wiecki, T.~V., \& Fonnesbeck, C. 2016, PeerJ Computer Science,
  2, e55

\bibitem[{{Seligman} {et~al.}(2024){Seligman}, {Feinstein}, {Lai}, {Welbanks},
  {Taylor}, {Becker}, {Adams}, {Morgan}, \& {Bergner}}]{Seligman2024}
{Seligman}, D.~Z., {Feinstein}, A.~D., {Lai}, D., {et~al.} 2024, \apj, 961, 22

\bibitem[{{Selsis} {et~al.}(2011){Selsis}, {Wordsworth}, \&
  {Forget}}]{Selsis2011}
{Selsis}, F., {Wordsworth}, R.~D., \& {Forget}, F. 2011, \aap, 532, A1

\bibitem[{{Shields} {et~al.}(2016){Shields}, {Ballard}, \&
  {Johnson}}]{2016PhR...663....1S}
{Shields}, A.~L., {Ballard}, S., \& {Johnson}, J.~A. 2016, \physrep, 663, 1

\bibitem[{{Sneep} \& {Ubachs}(2005)}]{sneep2005}
{Sneep}, M. \& {Ubachs}, W. 2005, \jqsrt, 92, 293

\bibitem[{{Spinelli} {et~al.}(2019){Spinelli}, {Borsa}, {Ghirlanda},
  {Ghisellini}, {Campana}, {Haardt}, \& {Poretti}}]{NUV_obs}
{Spinelli}, R., {Borsa}, F., {Ghirlanda}, G., {et~al.} 2019, \aap, 627, A144

\bibitem[{{Stevenson} {et~al.}(2012){Stevenson}, {Harrington}, {Fortney},
  {Loredo}, {Hardy}, {Nymeyer}, {Bowman}, {Cubillos}, {Bowman}, \&
  {Hardin}}]{BLISS_mapping}
{Stevenson}, K.~B., {Harrington}, J., {Fortney}, J.~J., {et~al.} 2012, \apj,
  754, 136

\bibitem[{Thalman {et~al.}(2014)Thalman, Zarzana, Tolbert, \&
  Volkamer}]{thalman2014}
Thalman, R., Zarzana, K., Tolbert, M., \& Volkamer, R. 2014, \jqsrt, 147,
  171–177

\bibitem[{{Tian} \& {Heng}(2024)}]{Tian2024}
{Tian}, M. \& {Heng}, K. 2024, \apj, 963, 157

\bibitem[{{Tulasi Ram} {et~al.}(2024){Tulasi Ram}, {Veenadhari}, {Dimri},
  {Bulusu}, {Bagiya}, {Gurubaran}, {Parihar}, {Remya}, {Seemala}, {Singh},
  {Sripathi}, {Singh}, \& {Vichare}}]{solarstorm2024}
{Tulasi Ram}, S., {Veenadhari}, B., {Dimri}, A.~P., {et~al.} 2024, Space
  Weather, 22, 2024SW004126

\bibitem[{Underwood {et~al.}(2016)Underwood, Tennyson, Yurchenko, Huang,
  Schwenke, Lee, Clausen, \& Fateev}]{SO2_linelist}
Underwood, D.~S., Tennyson, J., Yurchenko, S.~N., {et~al.} 2016, \mnras, 459,
  3890

\bibitem[{Virtanen {et~al.}(2020)Virtanen, Gommers, Oliphant, Haberland, Reddy,
  Cournapeau, Burovski, Peterson, Weckesser, Bright, {van der Walt}, Brett,
  Wilson, Millman, Mayorov, Nelson, Jones, Kern, Larson, Carey, Polat, Feng,
  Moore, {VanderPlas}, Laxalde, Perktold, Cimrman, Henriksen, Quintero, Harris,
  Archibald, Ribeiro, Pedregosa, {van Mulbregt}, \& {SciPy 1.0
  Contributors}}]{scipy}
Virtanen, P., Gommers, R., Oliphant, T.~E., {et~al.} 2020, Nature Methods, 17,
  261

\bibitem[{{Waldmann}(2012)}]{Waldmann1}
{Waldmann}, I.~P. 2012, \apj, 747, 12

\bibitem[{{Waldmann} {et~al.}(2013){Waldmann}, {Tinetti}, {Deroo}, {Hollis},
  {Yurchenko}, \& {Tennyson}}]{Waldmann2}
{Waldmann}, I.~P., {Tinetti}, G., {Deroo}, P., {et~al.} 2013, \apj, 766, 7

\bibitem[{{Wang} {et~al.}(2016){Wang}, {Hogg}, {Foreman-Mackey}, \&
  {Sch{\"o}lkopf}}]{2016PASP..128i4503W}
{Wang}, D., {Hogg}, D.~W., {Foreman-Mackey}, D., \& {Sch{\"o}lkopf}, B. 2016,
  \pasp, 128, 094503

\bibitem[{{Weiner Mansfield} {et~al.}(2024){Weiner Mansfield}, {Xue}, {Zhang},
  {Mahajan}, {Ih}, {Koll}, {Bean}, {Coy}, {Eastman}, {Kempton}, \&
  {Kite}}]{Mansfield2024}
{Weiner Mansfield}, M., {Xue}, Q., {Zhang}, M., {et~al.} 2024, \apjl, 975, L22

\bibitem[{{Whittaker} {et~al.}(2022){Whittaker}, {Malik}, {Ih}, {Kempton},
  {Mansfield}, {Bean}, {Kite}, {Koll}, {Cronin}, \& {Hu}}]{heliossurfb}
{Whittaker}, E.~A., {Malik}, M., {Ih}, J., {et~al.} 2022, \aj, 164, 258

\bibitem[{{Wohlfarth} {et~al.}(2023){Wohlfarth}, {W{\"o}hler}, {Hiesinger}, \&
  {Helbert}}]{Wohlfarth2023}
{Wohlfarth}, K., {W{\"o}hler}, C., {Hiesinger}, H., \& {Helbert}, J. 2023,
  \aap, 674, A69

\bibitem[{Wright {et~al.}(2015)Wright, Wright, Goodson, Rieke, Aitink-Kroes,
  Amiaux, Aricha-Yanguas, Azzollini, Banks, Barrado-Navascues,
  Belenguer-Davila, Bloemmart, Bouchet, Brandl, Colina, Örs Detre,
  Diaz-Catala, Eccleston, Friedman, García-Marín, Güdel, Glasse, Glauser,
  Greene, Groezinger, Grundy, Hastings, Henning, Hofferbert, Hunter, Jessen,
  Justtanont, Karnik, Khorrami, Krause, Labiano, Lagage, Langer, Lemke, Lim,
  Lorenzo-Alvarez, Mazy, McGowan, Meixner, Morris, Morrison, Müller, rgaard
  Nielson, Olofsson, O’Sullivan, Pel, Penanen, Petach, Pye, Ray, Renotte,
  Renouf, Ressler, Samara-Ratna, Scheithauer, Schneider, Shaughnessy,
  Stevenson, Sukhatme, Swinyard, Sykes, Thatcher, Tikkanen, van Dishoeck,
  Waelkens, Walker, Wells, \& Zhender}]{Wright_2015}
Wright, G.~S., Wright, D., Goodson, G.~B., {et~al.} 2015, \pasa, 127, 595

\bibitem[{{Zahnle} \& {Catling}(2017)}]{2017ApJ...843..122Z}
{Zahnle}, K.~J. \& {Catling}, D.~C. 2017, \apj, 843, 122

\bibitem[{{Zieba} {et~al.}(2023{\natexlab{a}}){Zieba}, {Hu}, {Kreidberg},
  {Morley}, {Tenthoff}, {Wohler}, \& {Wohlfarth}}]{2023jwst.prop.4008Z}
{Zieba}, S., {Hu}, R., {Kreidberg}, L., {et~al.} 2023{\natexlab{a}}, {The
  search for regolith on the airless exoplanet LHS 3844 b}, JWST Proposal.
  Cycle 2, ID. \#4008

\bibitem[{{Zieba} {et~al.}(2023{\natexlab{b}}){Zieba}, {Kreidberg}, {Ducrot},
  {Gillon}, {Morley}, {Schaefer}, {Tamburo}, {Koll}, {Lyu}, {Acu{\~n}a},
  {Agol}, {Iyer}, {Hu}, {Lincowski}, {Meadows}, {Selsis}, {Bolmont}, {Mandell},
  \& {Suissa}}]{2023Natur.620..746Z}
{Zieba}, S., {Kreidberg}, L., {Ducrot}, E., {et~al.} 2023{\natexlab{b}}, \nat,
  620, 746

\end{thebibliography}

\begin{appendix}

\section{Cosmic ray persistence effect}
\label{app:cosmic_ray}
\FloatBarrier
\begin{figure}
        \includegraphics[width=\columnwidth]{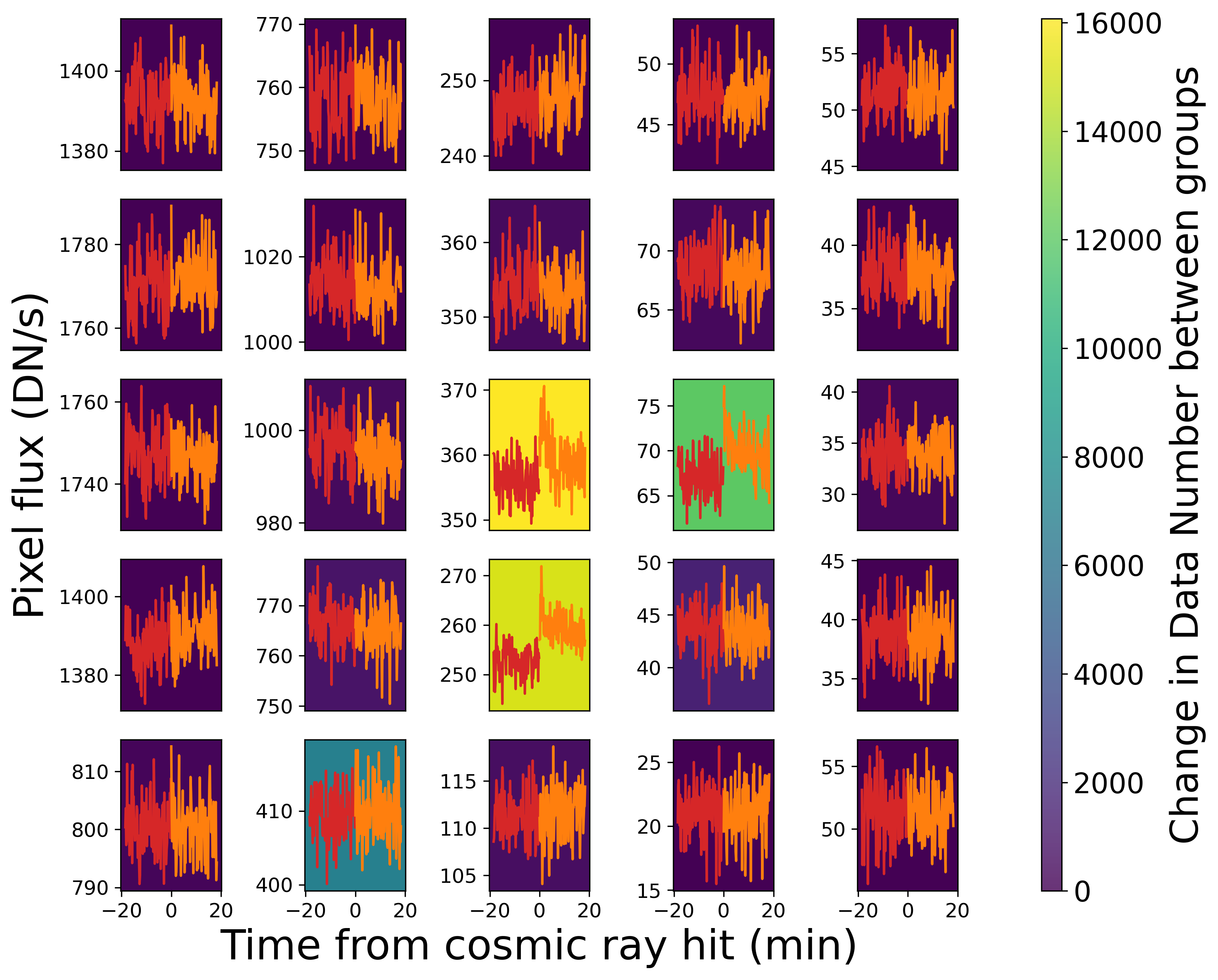}
        \caption{Pixel light curves centred on the pixels hit by a strong cosmic ray in the second eclipse. Light curves coloured in red before cosmic ray hit and in orange after. The background colour gives the change in data number between the groups before and after the cosmic ray hit for that pixel (after correcting for expected flux). Large data number jumps correspond to pixel light curves with strong persistence effects.}
        \label{fig:cosmic_ray_hit}
    \end{figure}
To verify that the persistence effect seen in Figure~\ref{fig:real_px_lightcurves} was caused by a cosmic ray, we took the uncalibrated `uncal' files for the second eclipse observation and calculated the change in data number between the groups before and after the cosmic ray hit. This occurred between groups 7 and 8 of the 524$^\mathrm{th}$ integration. To correct for the expected increase in data number due to flux from the target star, thermal background, etc. we subtracted the median frame of all integrations divided by the number of groups. This procedure had minimal effect as this expected increase in data number was far smaller (of the order of hundreds of data numbers) than the size of the jump from the cosmic ray ($\approx$16,000 data numbers). In Figure~\ref{fig:cosmic_ray_hit} we overlay this increase in data number between these two groups with the pixel light curves centred around this cosmic ray event. These pixel light curves are from the reduction described in Section~\ref{sec:reduction} which used the jump step correction and also performed an additional custom median outlier clipping step on the stage 1 and 2 processed data. These steps remove the significant jump in flux observed in many of these light curves in the integration containing the cosmic ray hit, but they do not perform any correction for this longer duration persistence artifact.

The large jump in data number of $\approx$16,000 DN seen in Figure~\ref{fig:cosmic_ray_hit} is larger than any other cosmic ray which hit within the 5px aperture during our three eclipse observations, suggesting these events are relatively rare. Examining pixels outside this aperture region which experience large (e.g. >5,000 DN) cosmic ray jumps also often correspond to similar but typically smaller persistence artifacts. To get a measure of the frequency of these events, we took the uncal files for all 22 observations in the Hot Rocks Survey and performed a similar test comparing the difference between successive groups across the full subarray (excluding the outermost rows and columns to avoid possible edge effects). Note this approach ignores cosmic rays which hit the first or last group which could also cause persistence effects. We found that jumps >5,000 DN occur at an averate rate of $\sim$0.0027 hits/(pixel hr) across each observation (excluding the eclipse of GJ-357b which was an outlier). Within a 5px radius aperture, we should expect jumps >5,000 DN every 4-7 hours with the rate varying between the observations, while hits >16,000 DN only occur every $\sim$70 hours. Therefore, smaller cosmic ray hits that could cause persistence effects may affect every observation or two, while effects as strong as Figure~\ref{fig:cosmic_ray_hit} should be quite rare. However, one exception is the eclipse of GJ 357b (Zgraggen et al., in prep), which was a huge outlier and for which hits >5,000 DN occurred every $\sim$13 minutes within a 5px radius aperture. This can likely be attributed to the observation being taken on 11$^\mathrm{th}$ May 2024, during one of the most intense Solar storms recorded in recent times \citep{solarstorm2024}.

In summary, persistence effects due to cosmic rays hitting individual pixels may be a widespread phenomenon for which the pixel-fitting method could be beneficial for weighting away from them. If these persistence effects could be modelled then it may be possible to directly fit them in the mean function of the pixel-fitting method (e.g. fitting an exponential ramp to the affected pixels) rather than weight away from them, potentially improving S/N. This could be more important if a strong cosmic ray were to affect one of the central pixels of the PSF or in the case of a high rate of cosmic ray hits as in the eclipse of GJ-357b.

\section{Effect of intrapixel sensitivity variations and flat fielding errors}
\label{app:intrapixel_testing}

As described in Section~\ref{sec:pixel_motivation}, there are multiple reasons why intrapixel sensitivity should be significantly reduced in MIRI time series compared to Spitzer. To confirm that it can be assumed to be negligble for MIRI, we performed tests to examine to what level intrapixel sensitivity could affect our observations. A related issue is flat fielding, while we did run the \textsc{flatfield} step in the JWST pipeline for the pixel-fitting analyses, any inaccuracies in the flat fielding would result in uncorrected for variations in sensitivity between different pixels (in contrast to intrapixel sensitivity which describes sensitivity variations within a pixel). We can examine both of these effects using simulated observations, similar to what has been performed for Spitzer \citep{Morello2015}.

\begin{figure*}
    \includegraphics[width=\textwidth]{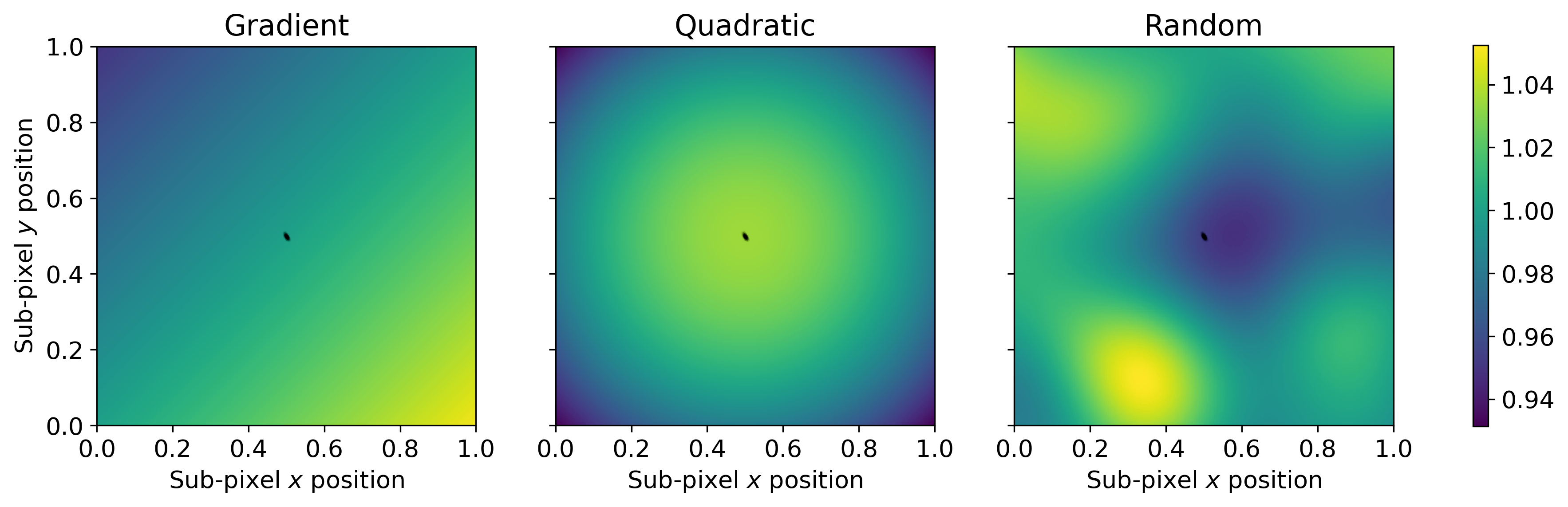}
    \caption{Various normalised sub-pixel sensitivity maps tested to examine intrapixel sensitivity. The colour scale gives the relative efficiency at each sub-pixel location. We visualise the level of pointing variation across an observation by plotting black dots for each centroid position in the first eclipse around the centre of each pixel map.}
    \label{fig:px_sensitivity}
\end{figure*}

All of our tests follow the same procedure. First, we generate a possible pixel sensitivity map. As we were unable to find any measurements of intrapixel sensitivity for MIRI in the literature, we tested various possibilities, shown in Figure~\ref{fig:px_sensitivity}. We considered a "gradient" map in which the pixel surface smoothly varies in sensitivity with a constant gradient from a relative efficiency of 95\% in the top left corner to 105\% in the bottom right corner. We also considered a "quadratic" map where the efficiency varied as a function of the square of the distance from the centre of the pixel. The normalised map has an efficiency ranging from 103\% at the pixel centre to 93\% at the edges. Our `random' sensitivity map was generated using a  GP with a 2D squared-exponential kernel function with correlation a function of the $x$ and $y$ position within a pixel. The height scale for this GP produced sensitivity variations between 95\% to 105\% relative efficiency. Each map was generated as a grid of $(250, 250)$ points within each pixel. We then copied this map for a $(15, 15)$ grid of pixels, creating an overall $(3750, 3750)$ sub-pixel resolution sensitivity map simulating 225 total pixels.

In addition to testing the scenario where all pixels share the same sensitivity map, we also generated maps which used randomised gradient maps, quadratic maps and random 2D GP maps which were independent for each pixel. These maps also exhibit $\sim$10\% variations in efficiency across each pixel. To test flat fielding errors, we generated $(250, 250)$ sub-pixel maps which had constant efficiency across each surface, but we varied this constant for each pixel with a standard deviation of 1\% and centred around 100\%. The MIRI reference file for flat fielding exhibits $\sim$1\% level variations between pixels, so if we run the \textsc{flatfield} step then we are modelling these maps being incorrect with a standard deviation of 1\%.

After generating all of these possible sensitivity maps, we generated a symmetric 2D Gaussian profile with the same FWHM as the MIRI F1500W PSF (FWHM = 4.436 pixels; \citealt{Dicken2024}). We used a 2D Gaussian profile as it can easily be calculated to arbitrary sub-pixel resolution and roughly matches the inner core of the MIRI PSF. To get a realistic level of pointing stability, we used the $x$ and $y$ centroid values from the first observation, which were calculated by fitting a 2D Gaussian profile to the PSF for each integration (see Section~\ref{sec:ap_ext_fitting} for more details). This likely overestimates the level of telescope pointing variation as this includes random scatter in fitting for the centroid position on top of the actual pointing instability. We discarded the centroid values from the first 45 minutes as these showed strong settling effects, likely arising from different pixels settling at different rates (see Section~\ref{sec:settling}). For each of the centroid values, we shifted the location of our 2D Gaussian profile by these $x$ and $y$ values and then multiplied our shifted Gaussian profile by each sub-pixel sensitivity map. We also did this for a completely uniform sensitivity map for reference. We then binned up these sub-pixel resolution values to pixel resolution and performed aperture extraction with a 5px radius aperture and using the centroid values which were used to shift the Gaussian profile. The resulting aperture extracted light curves were then subtracted from the light curves generated using a uniform map, and the standard deviation of this difference was calculated. This gives us a measure of how much additional noise is produced from the non-uniform pixel surfaces. Since we did not simulate any noise, the aperture extracted light curve using the uniform map was almost completely noise-free ($<$0.1 ppm RMS). We also examined the difference made to each pixel light curve to understand how this may affect the pixel-fitting method. In this case, we subtracted the pixel-resolution light curves generated with each map from the values using the uniform map. Calculating the standard deviation for each pixel light curve then gave us a measure of how much additional noise would be present in each pixel light curve for each map.

\begin{table}
    \caption{Simulated noise introduced by various (sub)-pixel sensitivity maps for both aperture extraction and for the pixel light curve most affected.}
    \centering
    \begin{tabular}{lcc}
        \hline
        Sensitivity map & Ap. Ext. (ppm) & Max pixel (ppm)\\
        \hline
        Flat field errors & 2.43 & 245.1 \\
        Gradient (shared) & 0.07 & 207.1 \\
        Quadratic (shared) & 0.00 & 48.2 \\
        Random (shared) & 0.02 & 100.6 \\
        Gradient (indep.) & 0.26 & 115.5 \\
        Quadratic (indep.) & 0.01 & 17.1 \\
        Random (indep.) & 0.56 & 222.9 \\
        \hline
    \end{tabular}
    \label{tab:intrapixel_table}
\end{table}

Our results were that all of the intrapixel sensitivity maps and flat fielding errors were completely negligible, never producing greater than a couple ppm of additional noise in our aperture extracted light curves (see Table~\ref{tab:intrapixel_table}). The pointing variation shows minimal correlation in time so most of this variation behaves like white noise and is tiny compared to the level of photometric noise in our observations ($\sim$800 ppm). We note that the noise would be added in quadrature to photon noise, so even 40 ppm of noise from intrapixel sensitivity would only increase the total aperture extracted noise by $\sim1$ ppm in combination with 800 ppm of photon noise. These effects are therefore more than an order of magnitude below the level that would impact our observations, meaning that significantly more extreme sub-pixel sensitivity variations would be needed to have any noticeable effect. While the maximum effect in a particular pixel could be a couple hundred ppm (see the `Max pixel (ppm)' column in Table~\ref{tab:intrapixel_table}), this was largely only in the outer regions of the PSF with the lowest S/N and the photon noise in any individual pixel for the real observations is never better than 3000 ppm even in the central pixels, so this effect should also be negligible for the pixel-fitting method. The exact values in Table~\ref{tab:intrapixel_table} would vary if different random draws of sensitivity maps were generated, but as the effect is so far below the level of photon noise it appears unlikely this would ever have a significant impact.

It is worth repeating that the Si:As detector technology used for MIRI is expected to show very minor intrapixel sensitivity, so extreme variations across each pixel surface is expected to be highly unlikely. This is similar to what was found with the Si:As detectors on Spitzer, which did not display the same level of intrapixel sensitivity as the shorter wavelength $3.6\upmu$m and $4.5\upmu$m channels which used InSb detectors \citep{spitzer_intrapixelsensitivity}. The narrower PSF widths from the shorter wavelength filters (e.g. F560W) may show slightly more variation, although rerunning the same tests with a reduced FWHM = 1.882px matching that of the F560W filter (and setting the aperture radius to match this FWHM), we still find that all maps produce $<20$ ppm of noise. This demonstrates that the exquisite pointing stability of JWST is enough to result in minimal intrapixel sensitivity, even if the pixels do demonstrate ~10\% level variations in sensitivity across their surfaces. Therefore, it appears unlikely that intrapixel sensitivity or flat fielding errors have any significant effect on time series observations with MIRI/Imaging.

\section{Eclipse timing and eccentricity constraints}
\label{app:eccentricity}

Given that the time of primary transit is already tightly constrained by transit observations, and our observations are the first to detect the eclipse of LHS 1140c, this should permit a tight constraint on the eccentricity parameter $\sqrt{e}\cos\omega$\footnote{And in principle on other stellar parameters such as the stellar density (e.g. see \citealt{Mahajan2024}), although we do not explore this within this work}. This is because the time gap from primary to secondary transit $\Delta t$ (ignoring light delay) is strongly dependent on $\cos\omega$. This time gap is given in \citet{eclipse_time_formula} as:
\begin{equation}
\Delta t = P\left[\frac{1}{2} + \frac{\left(1 + \frac{1}{\sin^2{i}}\right)e\cos\omega}{\pi}\right],
\label{eq:eclipse_timing}
\end{equation}
where $i$ is the orbital inclination ($89.80^{+0.14}_{-0.19}$ $\degree$ for our target; \citealt{2024ApJ...960L...3C}). A negative value of $\cos\omega$ therefore leads to the eclipse occurring earlier than expected (relative to a zero eccentricity orbit) while a positive value would result in a later than expected time.

We marginalise over $\sqrt{e}\sin\omega$ too, although our observations provide little improvement to the constraint on this value as it mostly affects the duration of the transit and eclipse and the existing transit data already constrains this very well. We also account for light delay in our eclipse model as this adds an approximate 22 second delay to the expected eclipse time.

\begin{figure}
    \includegraphics[width=\columnwidth]{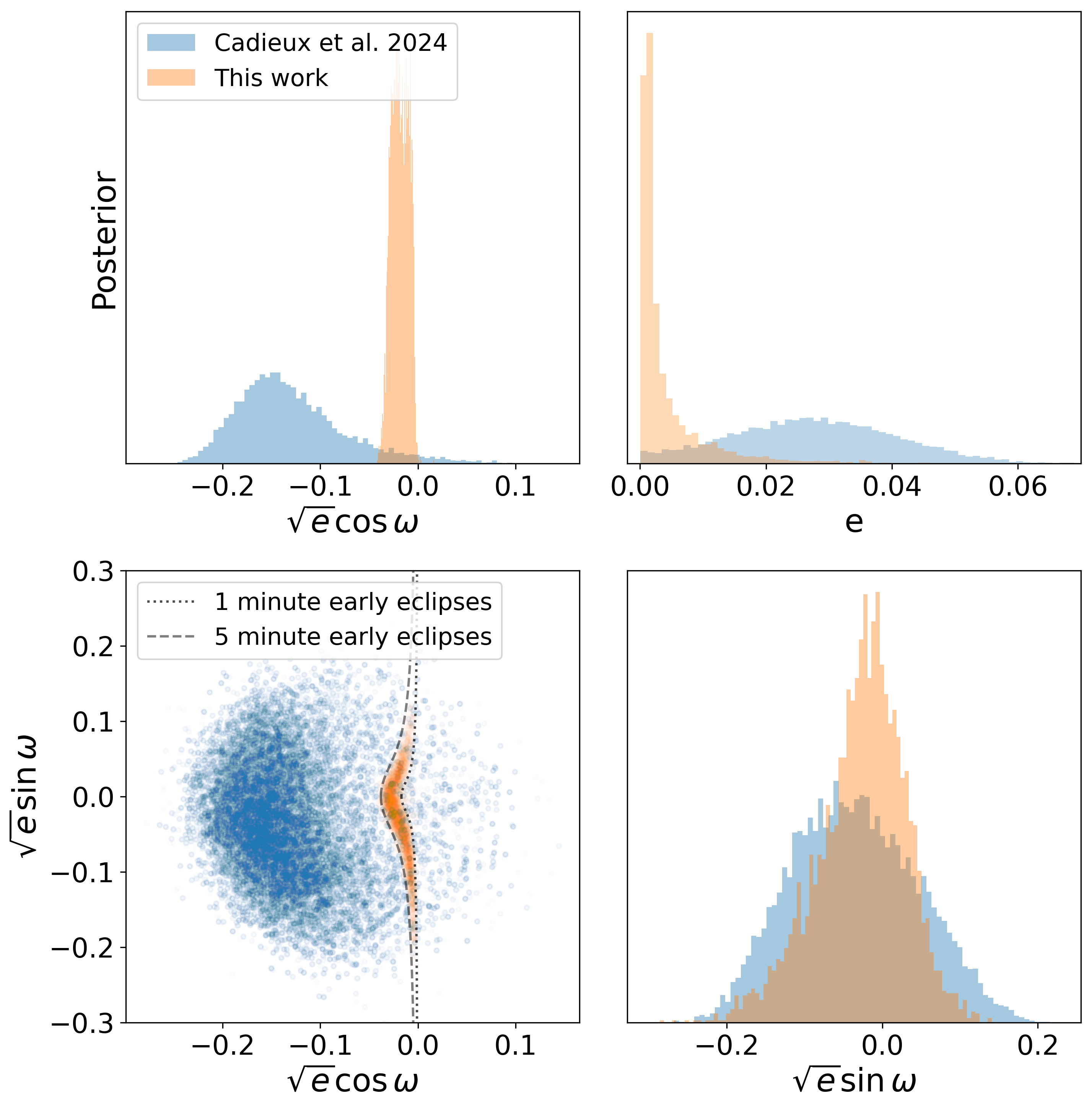}
    \caption{Comparison of eccentricity constraints between \citet{2024ApJ...960L...3C} (in blue; C. Cadieux, priv. comm.) and this work (in orange). The posterior from \citet{2024ApJ...960L...3C} was used as the prior for our eccentricity parameters. The additional constraint from the eclipse time significantly narrows the range of $\sqrt{e}\cos\omega$ (top left), which helps tighten the eccentricity constraint much closer to zero (top right). In the bottom left plot, we see that the new joint-posterior of $\sqrt{e}\cos\omega$ and $\sqrt{e}\sin\omega$ is largely constrained to lie between values which correspond to 1 to 5 minute early eclipses (relative to a circular orbit). The constraints on $\sqrt{e}\sin\omega$ are mostly unchanged from the prior values however (bottom right).}
    \label{fig:eccentricity}
\end{figure}

The eccentricity is constrained to be closer to zero than the previous analysis from \citet{2024ApJ...960L...3C}. This is highlighted by Figure~\ref{fig:eccentricity}, which shows that $\sqrt{e}\cos\omega$ is constrained much tighter compared to the posterior from \citet{2024ApJ...960L...3C} (C. Cadieux, priv. comm.), while $\sqrt{e}\sin\omega$ is relatively unaffected. Note the posterior from \citet{2024ApJ...960L...3C} was used as the priors on $\sqrt{e}\cos\omega$ and $\sqrt{e}\sin\omega$ in our analysis as our observations alone would not put good constraints on $\sqrt{e}\sin\omega$, while removing the prior on $\sqrt{e}\cos\omega$ has negligible impact. While we have constrained the eccentricity to be very small ($e < 0.0172$ to 95\% confidence), the observations are quite inconsistent with a zero eccentricity orbit - as can be seen in the scatter plot of $\sqrt{e}\sin\omega$ against $\sqrt{e}\cos\omega$. This is because all three eclipses favour the eclipse occurring slightly early - as shown in Appendix~\ref{app:eclipse_timing} - resulting in a negative value of $\sqrt{e}\cos\omega$. By using Equation~\ref{eq:eclipse_timing} with our posterior values, we recover that the eclipse occurs $2.8\pm0.9$ minutes early relative to a circular orbit. Our very small eccentricity value is consistent with dynamical modelling which predicts a short circularisation timescale for both LHS 1140b and LHS 1140c \citep{modeleccentricity}.

\subsection{Individual eclipse time constraints}
\label{app:eclipse_timing}

\begin{figure}
    \includegraphics[width=\columnwidth]{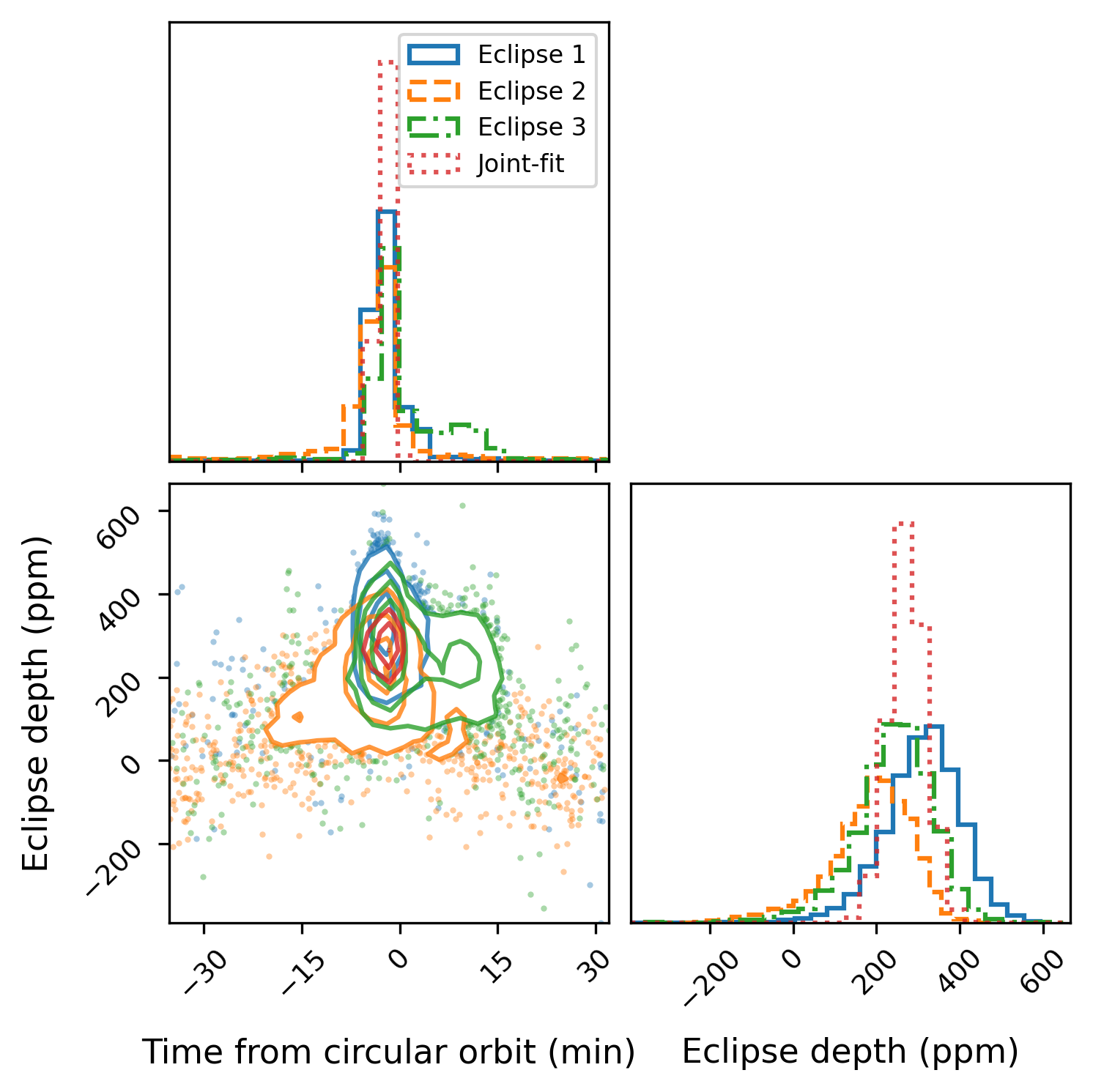}
    \caption{Posterior of central eclipse time (relative to expected time for a circular orbit) versus eclipse depth from individual eclipse fits and a joint-fit. Each dataset was fit using aperture extraction with a GP and the values from the joint-fit also using aperture extraction with a GP are overplotted in red. All individual fits are highly consistent with the joint-fit.}
    \label{fig:consistency}
\end{figure}

We also tested fitting each eclipse completely independently using aperture extraction with a GP, as opposed to Section~\ref{sec:eclipse_depths} where we used eclipse times from the other two eclipses to constrain each eclipse time. Figure~\ref{fig:consistency} shows the constraints on the central eclipse times (given relative to a circular orbit) and eclipse depths for each of the three eclipses fit independently. We see that while each individual dataset does not strongly detect an eclipse, the highest probability central eclipse times all peak close to the expected time for a circular orbit and the eclipse depths are all closely matching. This consistency offers strong evidence that our observed eclipses are real and not the result of systematics.

In addition to testing a fully independent fit to each of the three datasets, we also tested a joint-fit with a shared eclipse depth but separate eclipse times (using aperture extraction with GPs). This could be thought of as a joint-fit which assumes a circular orbit but accounts for transit timing variations (TTVs), although these are probably unlikely given the large separation between the two known planets in the system. Unfortunately, due to the low S/N of each individual eclipse, the central eclipse time is not strongly constrained by each individual dataset and so a uniform prior was chosen bounded within 30 minutes either side of the expected eclipse time for a circular orbit. We found that all three eclipses were favoured to occur a couple of minutes before the expected time with zero eccentricity, which explains the negative $\sqrt{e}\cos\omega$ constraint when joint-fitting eccentricity parameters to all three datasets.

\begin{figure}
        \includegraphics[width=\columnwidth]{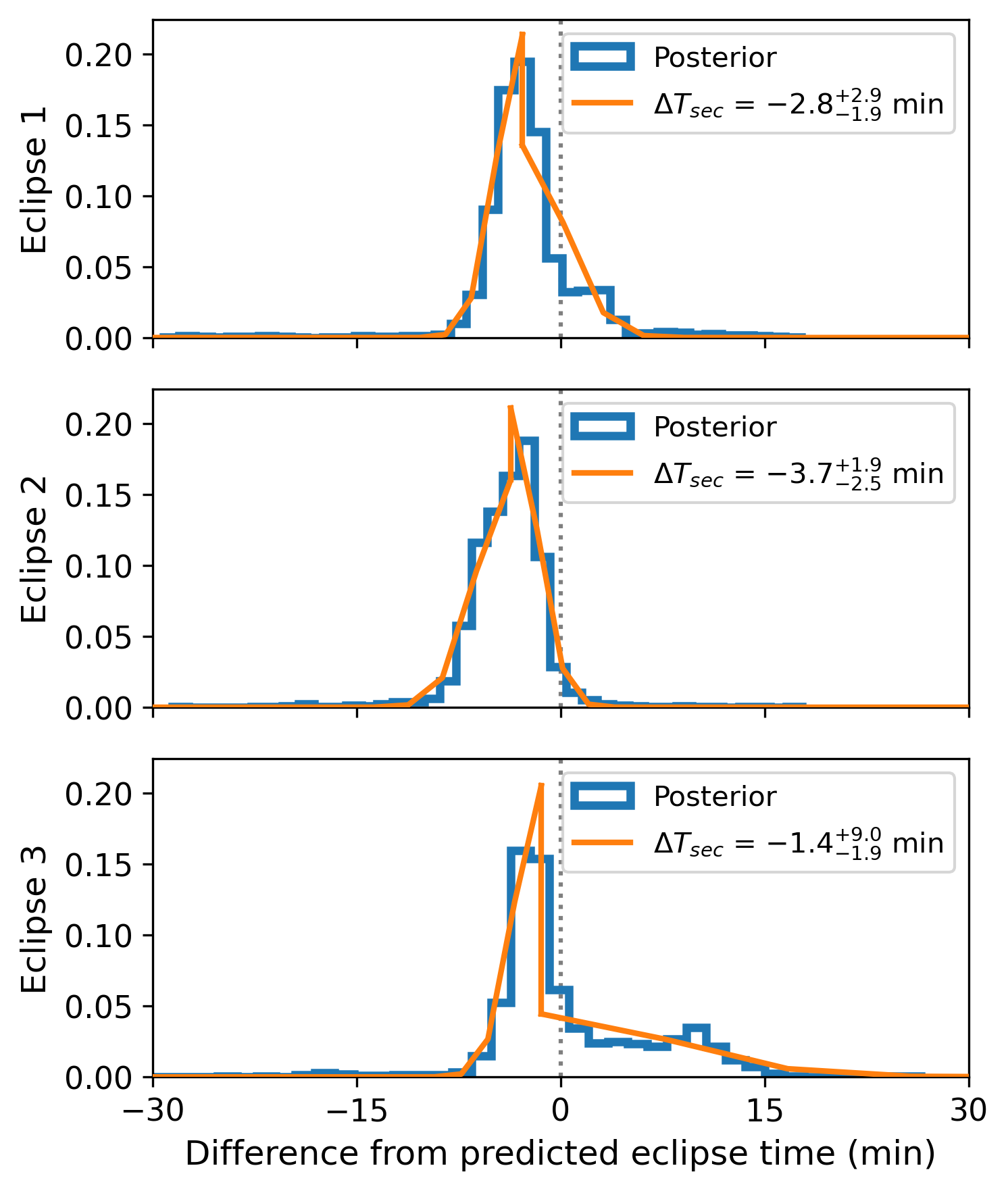}
        \caption{Central eclipse time posteriors from a joint-fit of all three eclipses using a shared eclipse depth but separate eclipse times. These posteriors were approximated using the median and $1 \sigma$ upper and lower uncertainties (highlighted in orange) and used to constrain the eclipse times when fitting individual eclipses. The predicted eclipse time was calculated using values in Table~\ref{tab:priors} but assumes a circular orbit and has been adjusted for light delay. All observations favour the eclipse to occur slightly early.}
        \label{fig:T0_priors}
    \end{figure}

Table~\ref{tab:T0_constraints} gives the central eclipse time constraints for each eclipse from this joint-fit without adjusting for light delay. Figure~\ref{fig:T0_priors} plots a histogram of the posterior for each of the central eclipse times. We note that Table~\ref{tab:T0_constraints} approximates these posteriors using the median value and upper and lower Gaussian uncertainties. The accuracy of these approximations is visualised by overlaying the probability distributions they produce in orange over the true posteriors in blue. The posteriors deviate substantially from symmetric Gaussian distributions but the different upper and lower uncertainties appear to do a reasonable job at tracing the true posterior.
    
We note that the constraints in Table~\ref{tab:T0_constraints} were used as priors when fitting individual eclipse depths as described in Section~\ref{sec:eclipse_depths} and Appendix~\ref{app:extra_eclipse_depths}. The way this was performed was by using the central eclipse time constraints from the two eclipses not being fit as priors on the expected eclipse time for the eclipse which was being fit (after adding/subtracting the correct number of periods between them). This helped constrain the individual eclipse fits which otherwise would have struggled to constrain the central eclipse time from a single dataset.

\begin{table}
    \caption{Central eclipse time constraints from each eclipse.}
    \centering
    \begin{tabular}{lccc}
        \hline
          & T$_\mathrm{sec}$ (BJD TDB - 2460276)\\
        \hline
        Eclipse 1 & $0.3716_{-0.0013}^{+0.0020}$\\
        Eclipse 2 & $223.2694_{-0.0017}^{+0.0013}$\\
        Eclipse 3 & $234.6048_{-0.0013}^{+0.0062}$\\
        \hline
    \end{tabular}
    \label{tab:T0_constraints}
\end{table}

\section{Tables of priors for both reduction analyses}
\label{app:prior_tables}

The choice of priors chosen for various parameters in the primary and secondary reductions analyses are included in Tables~\ref{tab:priors} and \ref{tab:priors:juliet}.
\begin{table}
    \caption{Priors used for eclipse parameters in the primary reduction analysis.}
    \centering
    \begin{tabular}{lll}
        \hline
        Parameter & Distribution & Prior \\
        \hline
        $f_p/f_*$ & Uniform & [-$\infty$, $+\infty$]\\
        $T_0$ & Normal & $2460304.70775\pm0.00004$\\
        $P$ & Normal & $3.777940\pm0.000002$\\
        $a/R_*$& Normal & $27.1_{-0.3}^{+0.2}$\\
        $R_p/R_*$ & Normal & $0.05312\pm0.00028$\\
        $b$ & Trunc. Normal & $0.09\pm0.06$ \& $b > 0$\\
        $\!\sqrt{e}\cos\omega$ & Normal & $-0.14_{-0.04}^{+0.06}$\\
        $\!\sqrt{e}\sin\omega$ & Normal & $-0.04_{-0.08}^{+0.09}$\\
        \hline
    \end{tabular}\newline
    \tablefoot{Values taken from \citet{2024ApJ...970L...2C} for $T_0$, $a/R_*$, $R_p/R_*$, $b$ and \citet{2024ApJ...960L...3C} for $P$, $\sqrt{e}\cos\omega$ and $\sqrt{e}\sin\omega$.}
    \label{tab:priors}
\end{table}

\begin{table}
\caption{Priors used for the secondary analysis fit.}
\centering
\begin{threeparttable}
    \begin{tabular}{lll}
        \hline
        Parameter & Distribution & Prior \\
        \hline
        $f_p/f_*$ & Uniform & [-0.01, 0.01]\\
        $T_0$ & Normal & (2460304.70775, 0.00004)\\
        $P$ & Fixed & $3.777940$ \\
        $a/R_*$ & Normal & (27.1, 0.1)\\
        $R_p/R_*$ & Fixed & 0.05312\\
        $b$ & Trunc. Normal & (0.09, 0.1) \& $b > 0$\\
        $\!\sqrt{e}\cos\omega$ & Normal & (-0.14, 0.1)\\
        $\!\sqrt{e}\sin\omega$ & Normal & (-0.04, 0.1)\\
        $m_{\text{flux}}$\tnote{a} & Normal & (0.0, 0.1) \\
        $\sigma$\tnote{b} & LogUniform & (10, 10000) \\
        \hline
        \multicolumn{3}{l}{Linear (L)} \\
        $\theta$ & Uniform & [-100,100] \\
        \multicolumn{3}{l}{Exponential (E)} \\
        $A$ & Uniform & [-1,1] \\
        $\tau$ & Uniform & [-1,1] \\
        \multicolumn{3}{l}{Gaussian Process (GP)} \\
        $GP_{\xi}$  & LogUniform & ($10^{-8}$, $10^{-2}$) \\
        $GP_{\rho}$ & LogUniform & ($10^{-8}$, $10^{2}$) \\            
        \hline
    \end{tabular}
    \begin{tablenotes}\footnotesize
        \item[a] Out-of-transit flux.
        \item[b] White noise (ppm).
    \end{tablenotes}
\end{threeparttable}
\tablefoot{Values taken from \citet{2024ApJ...970L...2C} for $T_0$, $a/R_*$, $R_p/R_*$, $b$ and \citet{2024ApJ...960L...3C} for $P$, $\sqrt{e}\cos\omega$ and $\sqrt{e}\sin\omega$. We indicate the priors and distributions for the parameters of the detrending methods. Each parameter is fitted independently for each observation.}
\label{tab:priors:juliet}
\end{table}

\FloatBarrier
\section{Atmospheres containing both SO$_2$ and CO$_2$}
\label{app:mixed_SO2}
\begin{figure}
    \includegraphics[width=\columnwidth]{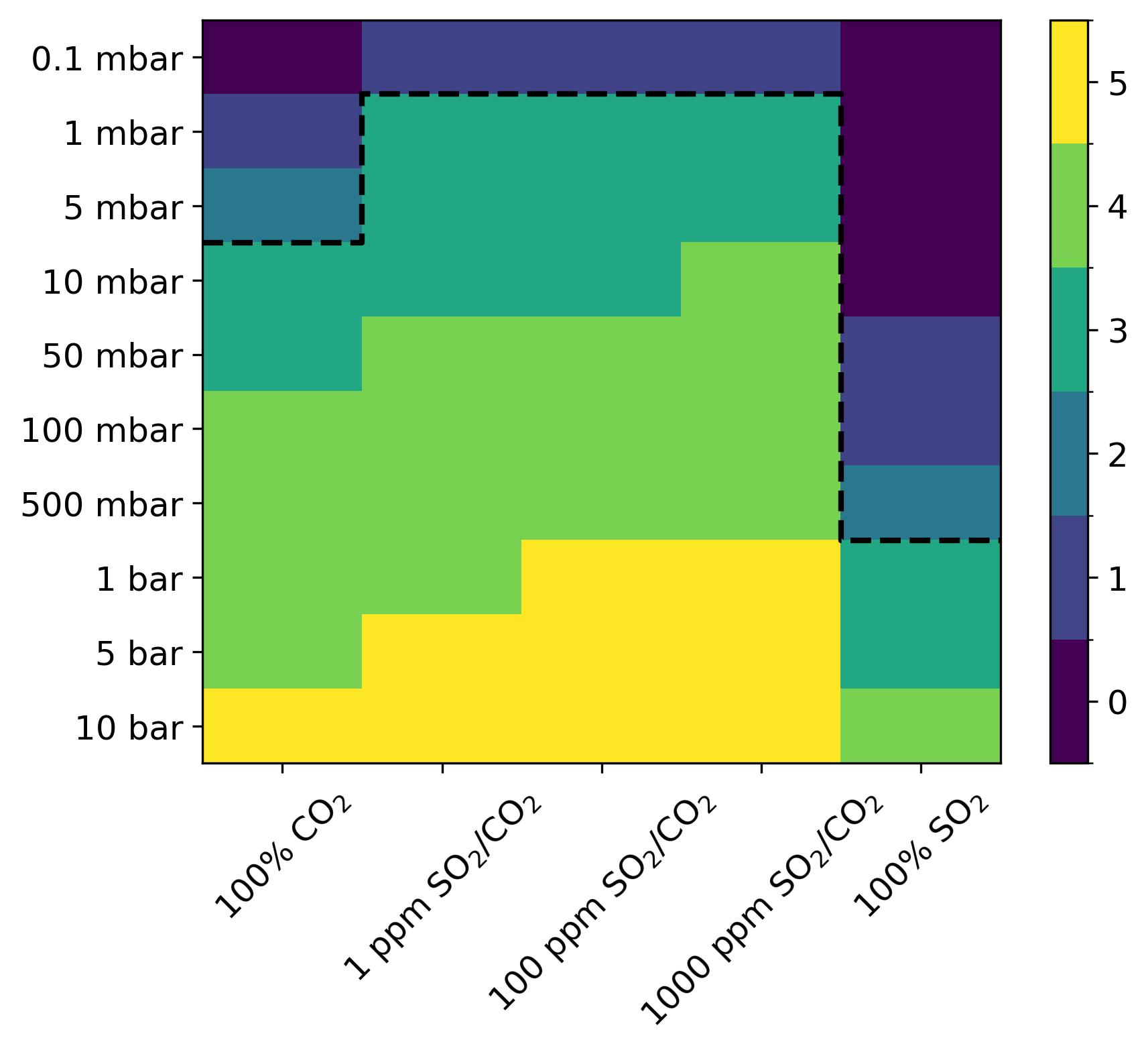}
    \caption{Minimum significance different atmospheres are ruled out to based on surface pressure and composition. Pressures below the black dotted line are all ruled out to at least $3\sigma$. These models suggest various combinations of SO$_2$ and CO$_2$ are ruled out down to pressures of only 1 mbar, stronger than the results for either pure CO$_2$ or pure SO$_2$.}
    \label{fig:mixed_SO2}
\end{figure}

In addition to the models described in Section~\ref{sec:5}, we tested various concentrations on SO$_2$ in a CO$_2$ background. These are meant to represent possible secondary atmospheres which may arise from volcanic outgassing. We refer back to the limitations of these models discussed in Section~\ref{sec:5} and we recommend that future modelling work be conducted to understand how photochemistry, atmospheric escape and realistic levels of outgassing may affect physically plausible secondary atmospheres.

Our models suggest that the combination of SO$_2$ at concentrations between 1 ppm - 1000 ppm in a CO$_2$ background are ruled out to very low pressures of just 1 mbar to at least $3\sigma$. Figure~\ref{fig:mixed_SO2} summarises these constraints and highlights which compositions and pressures are ruled out at $\ge3\sigma$. We note that these models all include heat redistribution as modelled by the analytic parameterisation in \citet{Kollf}. These results are intriguing as they suggest that eclipse photometry at 15$\upmu$m may be particularly sensitive to volcanic atmospheres. The addition of small concentrations of SO$_2$ appears to make the CO$_2$ absorption feature more prominent in the model spectra, likely due to the effect of SO$_2$ on the temperature-pressure profile of the atmosphere.

\section{Allan deviation plots}
\label{app:allan_deviation}

We took the last 1,024 integrations of each observation (cutting the first $\sim$44 minutes) and subtracted our best-fit exponential ramp and linear model from our aperture extracted light curve fits (which were joint-fit using a GP). We then calculated the standard deviation of these residuals (referred to as the RMS) at different binnings increasing in powers of two - starting from not binning at all and up to 256 points per bin. We also performed the same operation for 10,000 independent draws of white noise with standard deviation scaled to match the real data and each with 1,024 points. This was in order to examine the expected level of deviation of uncorrelated white noise from the expected $1/\sqrt{N}$ for $N$ binned points. We may expect sufficiently strong correlated noise to cause the RMS value to increase faster than $1/\sqrt{N}$ as $N$ increases.

We see that for all three observations our Allan deviation plots do not increase any faster than $1/\sqrt{N}$. We do see some deviation less than $1/\sqrt{N}$, although this is not indicative of correlated noise and it does not deviate beyond the $3\sigma$ boundaries of the simulated white noise draws. It is however possible that this effect is related to an imperfect best-fit model subtracted from the data i.e. the exponential ramp and linear slope fit is off, resulting in some anti-correlation between datapoints at either end of the dataset.

We note that while our residuals are consistent with white noise according to this Allan deviation test, this should not be used to assume that these observations lack any correlated noise. For example, we can perform a similar test for each white noise draw but adding time-correlated systematics from an exponential kernel with a timescale of 30 minutes and amplitude of 50ppm. In this case, we find that only $\sim$3\% of these simulations will increase above $1/\sqrt{N}$ by >3$\sigma$ for at least one choice of binning. This is similar to the $\sim$1\% of cases that the white noise draws will deviate by >3$\sigma$ for at least one choice of binning (as multiple binnings are being tested for each draw this is larger than the expected $\sim$0.14\% of deviations if testing at any specific binning). While we could use a lower threshold of 2$\sigma$ for detecting correlated noise, note that $\sim$12\% of white noise draws will increase more than expected by >2$\sigma$ for at least one binning, so this would result in many false positive detections of correlated noise. Also we could increase the number of binnings tested to increase the detection rate of correlated noise, but this would also likely increase the rate of false positives for uncorrelated noise. As the amplitude of correlated noise increases, the detection rate at >3$\sigma$ will increase. For 100 ppm of correlated noise, we detect these systematics $\sim$22\% of the time. 150 ppm of correlated noise is detected $\sim$54\% of the time and 200 ppm detected 78\% of the time. Therefore, quite significant levels of correlated noise can be added and will still frequently follow the $1/\sqrt{N}$ trend for uncorrelated noise. This means that it is always important to experiment with methods which can account for the effect of correlated noise when the signal being measured is so small ($\sim$250 ppm).
\begin{figure}
        \includegraphics[width=\columnwidth]{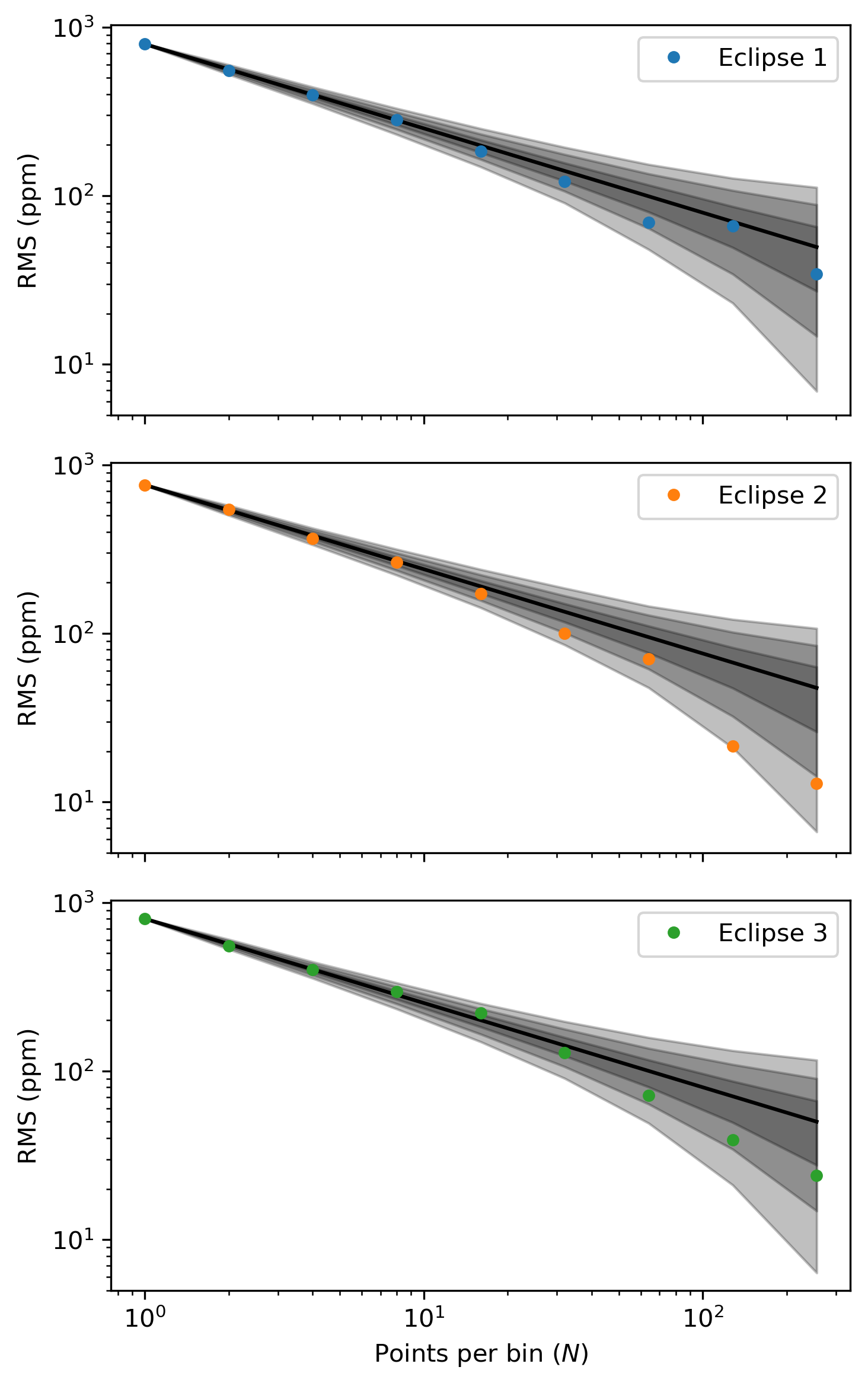}
        \caption{Allan deviation plot for each eclipse. The black line denotes the $1/\sqrt{N}$ trend expected for uncorrelated noise, while the shaded regions denote the 1$\sigma$, 2$\sigma$ and 3$\sigma$ confidence intervals for each particular binning calculated using 10,000 white noise draws. We note that only 75\% of the white noise draws stay within the 2$\sigma$ intervals for all bin sizes tested. We see that there can be a particularly large range of deviation from $1/\sqrt{N}$ for large bin sizes even for actual white noise.}
        \label{fig:allan_deviation}
\end{figure}
    
    \section{Individual eclipse depths for different pixel-fits}
    \label{app:extra_eclipse_depths}
    
    \begin{table*}
    \caption{Eclipse depths for individual eclipse depth fits using different pixel-fitting assumptions with aperture extraction results included for reference.}
        \centering
        \label{tab:px_tests}
        \begin{tabular}{lllll}
                \hline
                Method & E1 & E2 & E3 & Combined\\
                \hline
            Ap. Ext. w/o GP  & 327$\pm$76 & 211$\pm$66 & 278$\pm$71 & 267$\pm$41 \\
            Ap. Ext. w/ GP & 327$\pm$82 & 215$\pm$75 & 272$\pm$79 & 268$\pm$45 \\
            Px-fit (only FCS systematics) & 310$\pm$72 & 195$\pm$73 & 261$\pm$70   & 256$\pm$41 \\
Px-fit (varying PSF width, only FCS systematics) & 316$\pm$73 & 188$\pm$67 & 282$\pm$67   & 259$\pm$40 \\
Px-fit (common systematics) & 304$\pm$88 & 194$\pm$80 & 258$\pm$77   & 249$\pm$47 \\
Px-fit (common+indep. sys.) & 283$\pm$101 & 208$\pm$99 & 246$\pm$93   & 245$\pm$56 \\
Px-fit (8.5px; common+indep. sys.) & 289$\pm$97 & 203$\pm$93 & 243$\pm$89   & 243$\pm$54 \\
Px-fit (common+shared indep. sys.) & 294$\pm$95 & 204$\pm$85 & 257$\pm$78   & 249$\pm$49 \\
Px-fit (8.5px; common+shared indep. sys.) & 291$\pm$94 & 203$\pm$81 & 256$\pm$75   & 247$\pm$47 \\
                \hline
        \end{tabular}
    \end{table*}
    
    \begin{table*}
        \caption{Eclipse depths for individual eclipse depth fits using aperture extraction with different reductions.}
        \centering
        \label{tab:reduction_tests}
        \begin{tabular}{lllll}
                \hline
                Method & E1 & E2 & E3 & Combined\\
                \hline
            Reference & 323$\pm$82 & 213$\pm$73 & 276$\pm$78  & 266$\pm$45 \\
Reference, 4px rad. & 329$\pm$83 & 213$\pm$83 & 270$\pm$80  & 271$\pm$47 \\
Reference, 4.5px rad. & 319$\pm$83 & 211$\pm$75 & 277$\pm$76  & 266$\pm$45 \\
Reference, 5.5px rad. & 332$\pm$82 & 219$\pm$73 & 282$\pm$75  & 273$\pm$44 \\
Reference, 6.0px rad. & 317$\pm$83 & 219$\pm$76 & 301$\pm$78  & 276$\pm$46 \\
Default pipeline & 315$\pm$83 & 216$\pm$74 & 262$\pm$76  & 261$\pm$45 \\
Default jump step & 323$\pm$84 & 217$\pm$73 & 276$\pm$76  & 267$\pm$45 \\
Skip jump step & 325$\pm$85 & 216$\pm$72 & 271$\pm$77  & 265$\pm$45 \\
Skip emicorr step & 323$\pm$83 & 211$\pm$76 & 272$\pm$75  & 266$\pm$45 \\
Skip flat-fielding & 318$\pm$83 & 214$\pm$73 & 275$\pm$77  & 265$\pm$45 \\
Include all groups & 328$\pm$79 & 207$\pm$72 & 264$\pm$74  & 263$\pm$43 \\
                \hline
        \end{tabular}
    \end{table*}

    We tested a number of individual eclipse pixel-fits including varying some of the choices of the method. These are listed in Table~\ref{tab:px_tests}. For example, the PSF is modelled using the mean function and is always allowed to interpolate to shift its $x$ and $y$ position for each integration, but the PSF was not allowed to change width in previous fits. We tested allowing the PSF to change width by linearly scaling it along the x and y-axis for each integration, introducing two additional parameters to be fit per integration. We found this had negligible impact on all three eclipses, although we just tested it with only flux-conserved systematics accounted for. The constraints appeared to be slightly tighter from this run, possibly because the PSF model is fitting the data slightly more accurately.

    While running pixel-fitting including the various components of pixel systematics as chosen in Section~\ref{sec:3} and \ref{sec:4} such as including common systematics and independent pixel systematics, we also tested including more than just the 81 pixels which are 5px away from the centre of the PSF. We chose a larger radius of 8.5px to examine how this affects the results. Similar to optimal extraction, the method should be able to weight away from lower flux pixels so adding more pixels should always either tighten the constraints or leave them unchanged. We found that increasing from 5px away to 8.5px away (225 total pixels) had a negligible impact on the result other than slightly tightening the eclipse depth constraints. This was at a substantial computational cost as for each pixel we fit for the baseline flux $F_\mathrm{oot}$ and slope in flux $T_\mathrm{grad}$, white noise amplitude $\sigma$, and height scale of independent pixel systematics $h_\mathrm{IPS}$ (unless there is a shared height scale). Therefore this required fitting for more than 500 extra parameters and resulted in little change in the result. However, it is reassuring that our choice of pixels to include only had a small effect on the result. We note that for these pixel-fitting analyses a slightly larger convergence criterion of $\hat{r} < 1.02$ was used (instead of $\hat{r} < 1.01$ for the results presented in Section~\ref{sec:4}) in order to reduce computational costs.
    
    \subsection{Testing different reduction steps}
    \label{app:diff_reductions}

    We also tested varying individual reduction choices one-by-one to examine if any choice has a large effect. We applied aperture extraction with a GP to each individual eclipse in all cases. The `reference' reduction is the primary reduction as described in Section~\ref{sec:reduction}. To reiterate it here, it uses the standard JWST calibration pipeline except it modifies the jump rejection threshold to $7\sigma$ instead of the default of $4\sigma$, removes both the first and last group instead of just the last group and aperture extracts the light curve using a 5px radius circular aperture. For stage 2 we only skipped the photom step, although this should have no effect on the eclipse depth as this just linearly scales the light curve and we normalise the light curves which will cancel out this correction.

    We tested varying the radius aperture between 4px and 6px instead of 5px but it had minimal effect on the eclipse depth as can be seen in Table~\ref{tab:reduction_tests}. We also examined just running stage 1 and 2 of the pipeline with its default settings, using the \textsc{jump} step with the $4\sigma$ default threshold or skipping the \textsc{jump} step entirely. In addition, we tested skipping the \textsc{emicorr} step which corrects for electromagnetic interference in the raw data, skipping the flat-fielding step in stage 2 as well as including all groups (i.e. skipping the steps which exclude the first and last groups). We did not find any dramatic difference for any of the various reductions we tested which suggests our results are not highly sensitive to these reduction choices.

\section{Autocorrelation of pixel light curves}\label{app:anticorr}

\begin{figure*}
    \includegraphics[width=\textwidth]{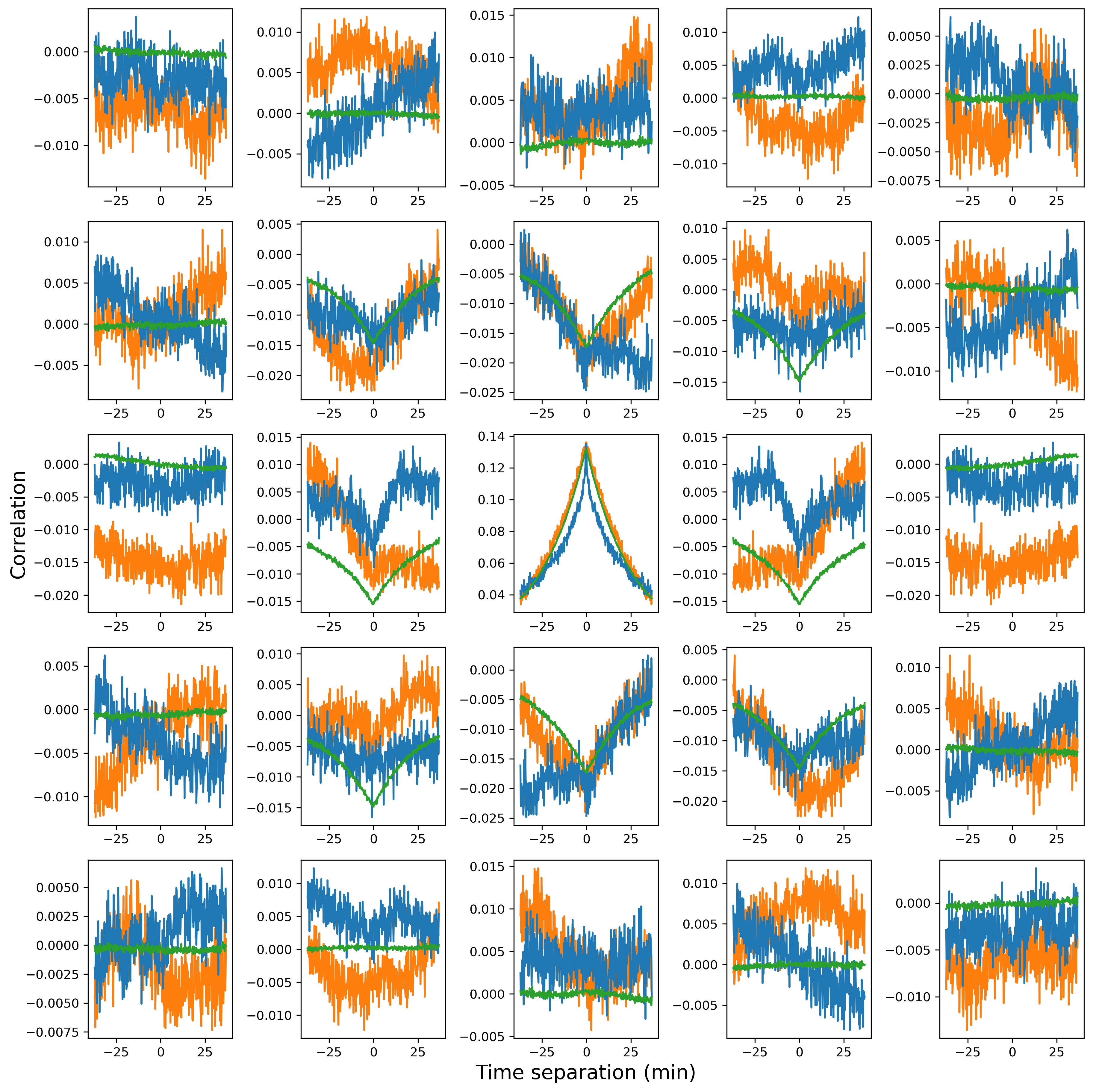}
    \caption{Autocorrelation of pixel light curves averaged across all the three eclipse datasets in blue, as well as autocorrelation of simulated pixel light curves for each eclipse in orange. The green is the averaged autocorrelation of 100 simulated eclipse datasets using our chosen kernel function. The autocorrelation of the real data is largely consistent with the general trend of the simulated data for three eclipses.}
    \label{fig:autocorr}
\end{figure*}

We can autocorrelate the pixel light curves for each eclipse to empirically inform our choice of kernel function. In practice this is somewhat complicated to implement as our pixel time series can be thought of as a 3D data cube with two spatial dimensions and a time dimension (X, Y, t). Figure~\ref{fig:autocorr} attempts to examine correlations in this space. In the centre of the (5, 5) subplot is the autocorrelation of each pixel light curve with itself as a function of separation in time, averaged together across all pixels included within the pixel fit (5px from the PSF centre) and averaged across all three eclipse datasets. This is plotted in blue when performed on the real JWST data. To examine our kernel function choice we also took our best-fit model from pixel-fitting the real dataset (accounting for both common systematics and independent pixel systematics) and generated random draws of noise for each of the three real datasets using the covariance matrix generated by our kernel function. We could then autocorrelate this noise in the same manner as for the real data which was then plotted in orange. Finally, instead of performing this once for each eclipse and averaging together, the autocorrelation plotted in green was recovered by generating noise 100 times for each eclipse dataset and averaging together the autocorrelation, giving a better idea for how the kernel function converges for a lot of synthetic observations.

For the plot immediately left or right of the centre plot, the autocorrelation was taken between horizontally adjacent pixels as a function of separation in time. Similarly the pixels above and below the central plot are for vertically adjacent pixels and so on for diagonally adjacent pixels. The outer plots are all for pixels that are separated with one pixel gap between them i.e. two pixels apart. We note that for our kernel choice, we expect neighbouring pixels to show an anti-correlation in time as we expect them to be flux-conserved. This is clear in the green light curves are there has been one hundred covariance matrix draws to average over. For both the blue and orange light curves they only have three eclipse datasets to average over so they may only show moderate evidence of an anti-correlation. However there does appear to be some sign of this anticorrelation, particularly in the adjacent neighbouring pixels where the blue light curves show signs of the V-shape seen in the green light curves. The pixels separated by more than one pixel show such a weak correlation in all cases that there is no evidence of anything to adjust in our kernel function for pixels at this separation.

The central subplot shows that our simulated noise does not quite match the correct kernel function as the real autocorrelation in blue is slightly sharper than either of the simulated autocorrelation. As this is the autocorrelation in time of pixels with themselves, this plot is for zero pixel separation and therefore the autocorrelation should just trace the kernel function in time (note we have excluded the point at zero separation in pixels and time which trivially gives a correlation of one). Our choice of the exponential kernel is able to match a lot of the sharpness of the autocorrelation of the real data however and we can clearly see that a smoother kernel such as the squared exponential kernel would be a less accurate match.

\section{Corner plots}\label{app:corner}

\begin{figure*}
    \includegraphics[width=\textwidth]{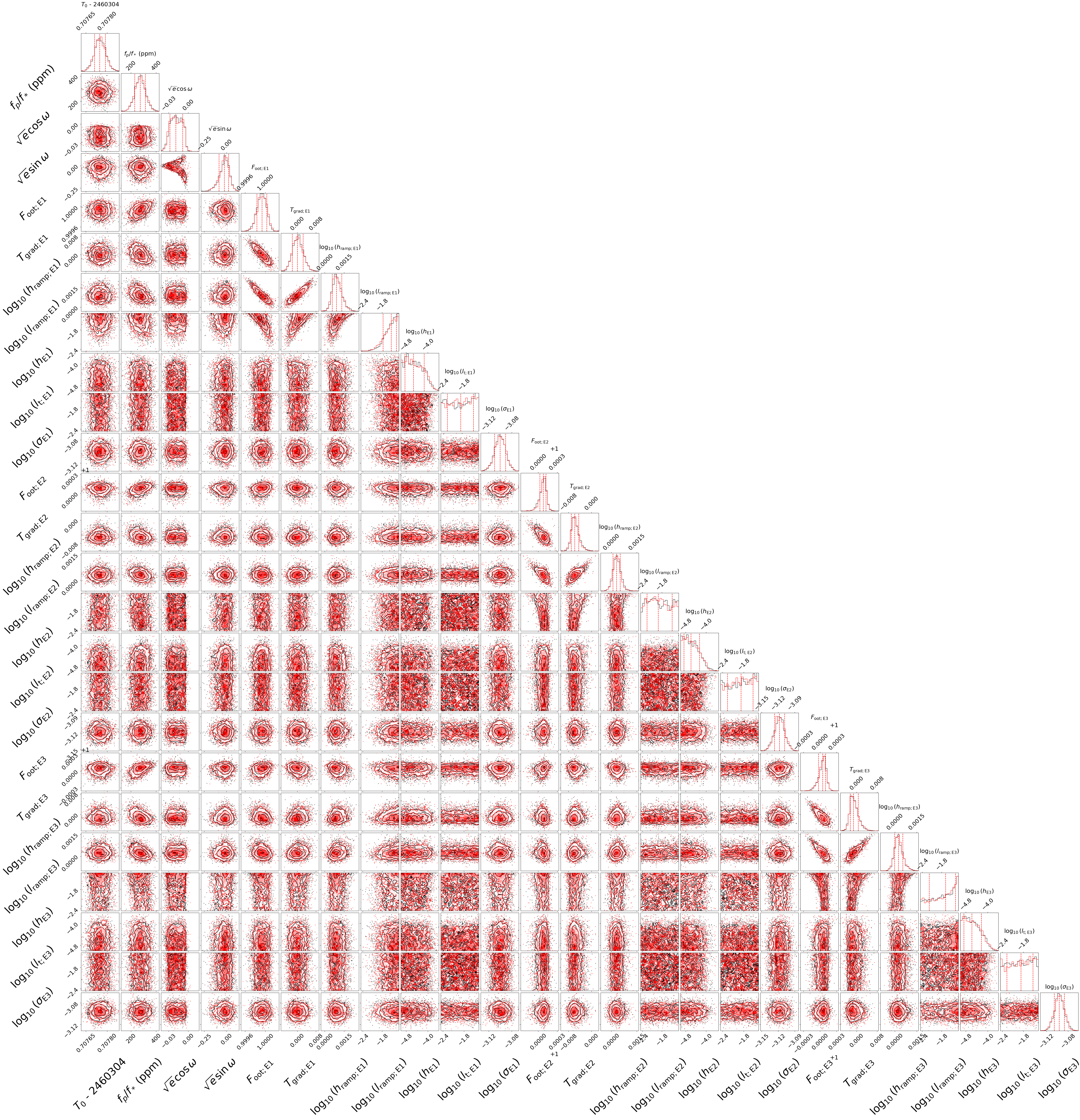}
    \caption{Corner plot of the joint-fit of aperture extracted light curves with GPs. The eclipse depth $f_p/f_*$ can be seen to be slightly correlated with many of the parameters which describe the baseline flux or detector settling in each eclipse. The slope in baseline flux is also somewhat degenerate with the initial detector settling slope for each eclipse. For many of the hyperparameters a log-uniform prior was used and so the logarithm of these parameters is plotted. Note the unusual relationship between the two eccentricity parameters which is explained in Appendix~\ref{app:eccentricity}.}
    \label{fig:corner_GP}
\end{figure*}

\begin{figure*}
    \includegraphics[width=\textwidth]{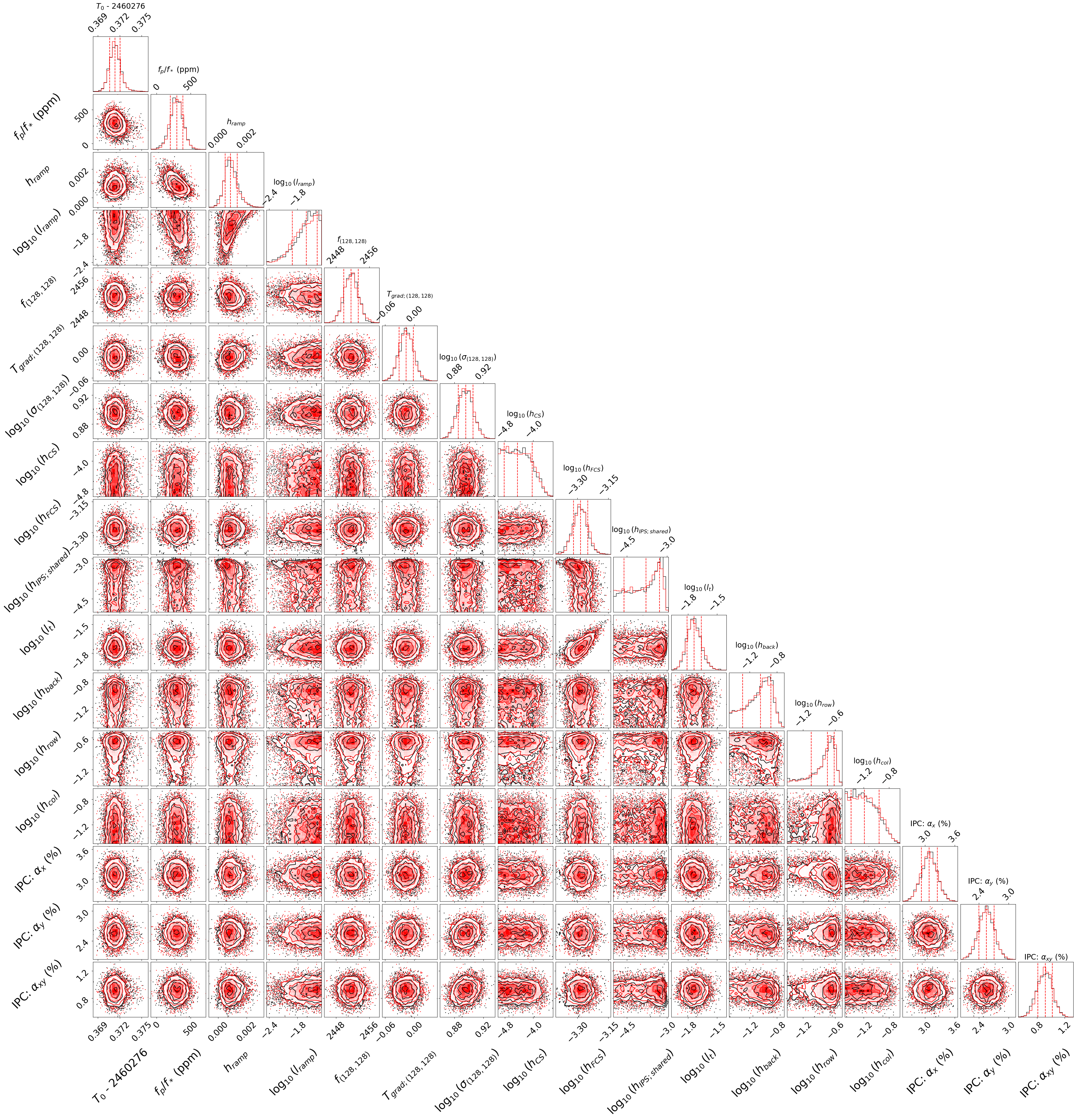}
    \caption{Corner plot of the first eclipse with the pixel-fitting method. Many parameters appear independent but we still find the eclipse depth is correlated with the detector settling parameters $h_\mathrm{ramp}$ and $l_\mathrm{ramp}$, similar to Figure~\ref{fig:corner_GP}. In addition, when the shared height scale of independent pixel systematics $h_\mathrm{IPS; shared}$ is high, the height scale of flux-conserving systematics $h_\mathrm{FCS}$ is reduced. This makes sense as the "bumps" seen in each pixel light curve in Figure~\ref{fig:real_px_lightcurves} could in principle be fit as either flux-conserved systematics or independent systematics, but the model does clearly prefer to fit a non-zero flux-conserving systematics height scale $h_\mathrm{FCS}$ while $h_\mathrm{IPS; shared}$ is consistent with zero. This means the model favours that these "bumps" are flux-conserving and not independent.}
    \label{fig:corner_e1}
\end{figure*}

\begin{figure*}
    \includegraphics[width=\textwidth]{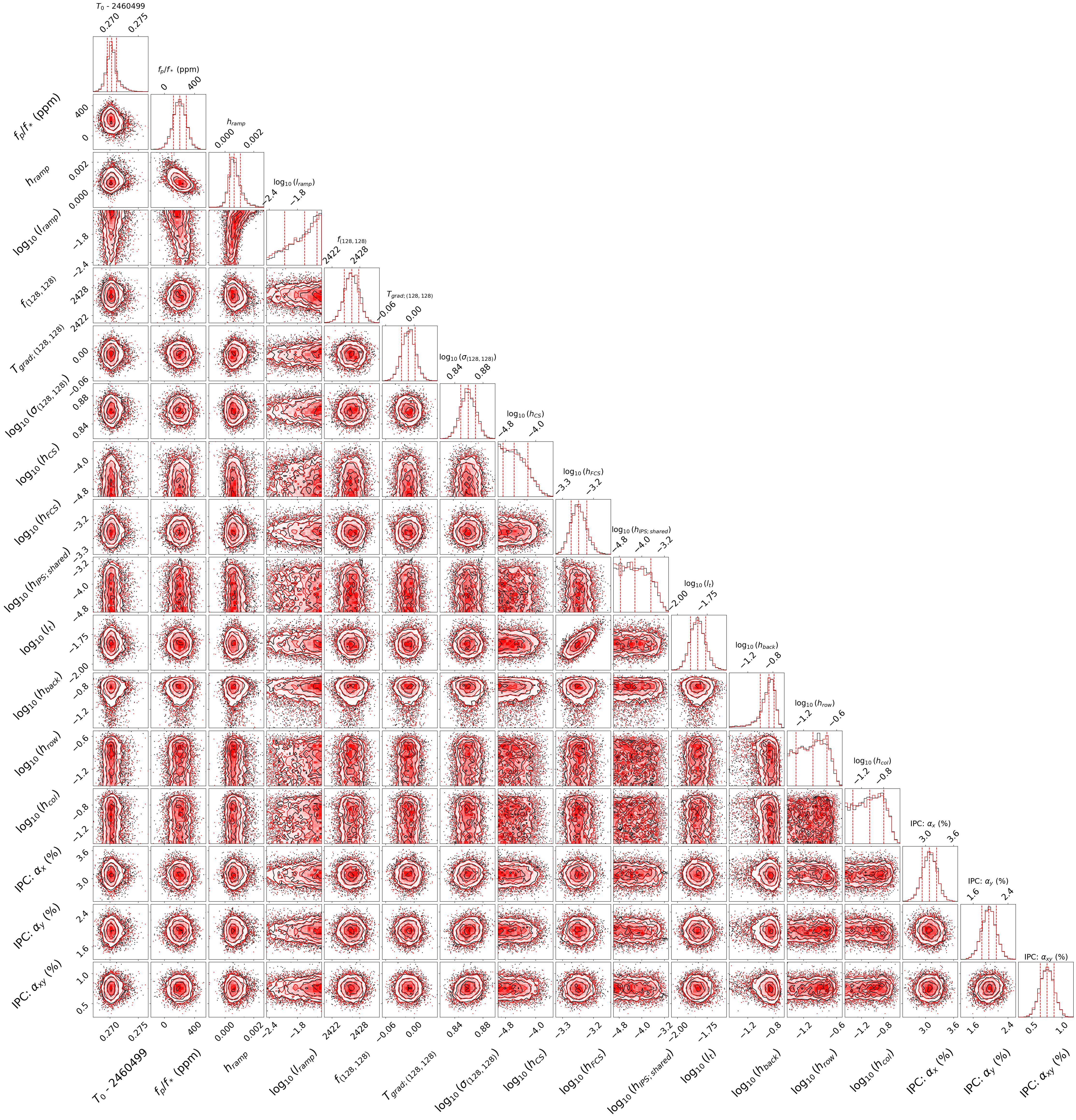}
    \caption{Same as Figure~\ref{fig:corner_e1} but for the second eclipse.}
    \label{fig:corner_e2}
\end{figure*}

\begin{figure*}
    \includegraphics[width=\textwidth]{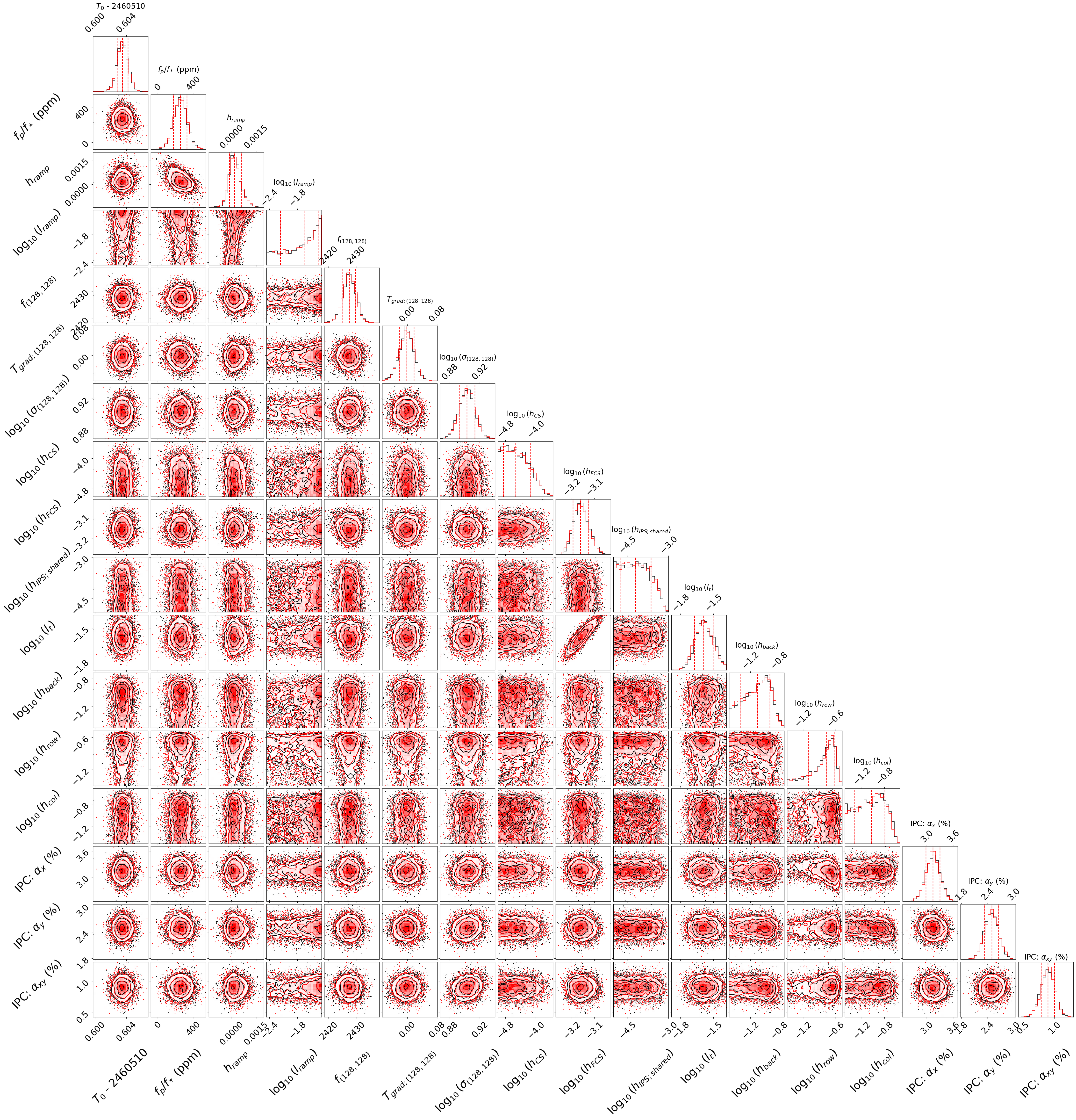}
    \caption{Same as Figure~\ref{fig:corner_e1} but for the third eclipse.}
    \label{fig:corner_e3}
\end{figure*}

We have included the corner plot featuring all parameters from the joint-fit of all three eclipses using aperture extraction with GPs. This is in Figure~\ref{fig:corner_GP}. In addition, we have a corner plot from each of the individual pixel-fits applied to each eclipse in Figures~\ref{fig:corner_e1}, ~\ref{fig:corner_e2} and ~\ref{fig:corner_e3}. These are all using the method which included common systematics and a shared independent pixel systematics height scale for some pixels. As there are approximately 2,300 parameters in each of these fits, only a subset of the parameters were included, with none of the changes in PSF position included (two parameters per integration) and only the parameters for the central pixel of the PSF included for parameters which are fit individually for each pixel.

\end{appendix}

\end{document}